\documentclass[a4paper,11pt]{article}
\pdfoutput=1

\usepackage{jheppub}

\usepackage{subfig}
\usepackage{xspace}
\usepackage[countmax]{subfloat}
\usepackage{slashed}

\usepackage{comment}

\usepackage{bbm}

\newcommand{\nbar}{{\bar n}}

\newcommand{\eventtwo}{\texttt{EVENT2}}

\newcommand{\ecf}[2]{e_{#1}^{(#2)}}

\newcommand{\ecflp}[2]{\tilde e_{#1}^{(#2)}}
\newcommand{\ecfop}[2]{\mathbf{E_{#1}}^{(#2)}}

\def\log{\text{log}}

\def\be{\begin{equation}}
\def\ee{\end{equation}}

\newcommand{\nn}{\nonumber}

\def\zcut{z_{\text{cut}}}

\newcommand{\Dobs}[2]{D_{#1}^{(#2)}}

\def\nbar{\bar n}

\DeclareRobustCommand{\Sec}[1]{Sec.~\ref{#1}}
\DeclareRobustCommand{\Secs}[2]{Secs.~\ref{#1} and \ref{#2}}
\DeclareRobustCommand{\App}[1]{App.~\ref{#1}}

\DeclareRobustCommand{\Fig}[1]{Fig.~\ref{#1}}
\DeclareRobustCommand{\Figs}[2]{Figs.~\ref{#1} and \ref{#2}}
\DeclareRobustCommand{\Eq}[1]{Eq.~(\ref{#1})}
\DeclareRobustCommand{\Eqs}[2]{Eqs.~(\ref{#1}) and (\ref{#2})}
\DeclareRobustCommand{\Ref}[1]{Ref.~\cite{#1}}
\DeclareRobustCommand{\Refs}[1]{Refs.~\cite{#1}}

\newcommand{\pythia}[1]{\textsc{Pythia\xspace #1}}
\newcommand{\madgraph}[1]{\textsc{MadGraph5\xspace #1}}
\newcommand{\fastjet}[1]{\textsc{FastJet\xspace #1}}

\newcommand{\vincia}[1]{\textsc{Vincia\xspace #1}}

\usepackage{color}
\definecolor{darkblue}{rgb}{0,0,0.5}
\definecolor{darkred}{rgb}{0.5,0,0}
\definecolor{purple}{rgb}{0.5,0.0,0.5}

\title{Factorization and Resummation for Groomed Multi-Prong Jet Shapes}

\author[1]{Andrew J. Larkoski,}
\affiliation[1]{Physics Department, Reed College, Portland, OR 97202, USA}
\author[2,3]{Ian Moult}
\affiliation[2]{Berkeley Center for Theoretical Physics, University of California, Berkeley, CA 94720, USA}
\affiliation[3]{Theoretical Physics Group, Lawrence Berkeley National Laboratory, Berkeley, CA 94720, USA}
\author[4]{and Duff Neill}
\affiliation[4]{Theoretical Division, MS B283, Los Alamos National Laboratory, Los Alamos, NM 87545, USA}

\emailAdd{larkoski@reed.edu}
\emailAdd{ianmoult@lbl.gov}
\emailAdd{duff.neill@gmail.com}

\abstract{Observables which distinguish boosted topologies from QCD jets are playing an increasingly important role at the Large Hadron Collider (LHC). These observables are often used in conjunction with jet grooming algorithms, which reduce contamination from both theoretical and experimental sources. In this paper we derive factorization formulae for groomed multi-prong substructure observables, focusing in particular on the groomed $D_2$ observable, which is used to identify boosted hadronic decays of electroweak bosons at the LHC. Our factorization formulae allow systematically improvable calculations of the perturbative $D_2$ distribution and the resummation of logarithmically enhanced terms in all regions of phase space using renormalization group evolution. They include a novel factorization for the production of a soft subjet in the presence of a grooming algorithm, in which clustering effects enter directly into the hard matching. We use these factorization formulae to draw robust conclusions of experimental relevance regarding the universality of the $D_2$ distribution in both $e^+e^-$ and $pp$ collisions. In particular, we show that the only process dependence is carried by the relative quark vs. gluon jet fraction in the sample, no non-global logarithms from event-wide correlations are present in the distribution, hadronization corrections are controlled by the perturbative mass of the jet, and all global color correlations are completely removed by grooming, making groomed $D_2$ a theoretically clean QCD observable even in the LHC environment. We compute all ingredients to one-loop accuracy, and present numerical results at next-to-leading logarithmic accuracy for $e^+e^-$ collisions, comparing with parton shower Monte Carlo simulations. Results for $pp$ collisions, as relevant for phenomenology at the LHC, are presented in a companion paper \cite{Larkoski:2017iuy}. 
}

\begin{document} 
\maketitle

\section{Introduction}\label{sec:intro}

The efficient and robust identification of hadronically-decaying boosted electroweak bosons is a problem of fundamental importance at Run 2 and into the future of the Large Hadron Collider (LHC) physics programme. The identification of boosted hadronically-decaying $H/W/Z$ bosons requires two key ingredients: observables for discriminating multi-prong jet substructure, and a method for removing contamination within the jets.  There has been substantial progress in both of these aspects of jet physics over the past few years.\footnote{For a recent review of theory and machine learning aspects of jet substructure, see \cite{Larkoski:2017jix}.}$^,$\footnote{A summary of studies from the LHC using jet substructure can be found at \url{https://twiki.cern.ch/twiki/bin/view/AtlasPublic} and \url{http://cms-results.web.cern.ch/cms-results/public-results/publications/}. For the Run 2 experimental status, see the performance documentation in \cite{ATLAS-CONF-2017-064,ATLAS-CONF-2016-039,ATL-PHYS-PUB-2017-009,ATLAS-CONF-2016-035,ATLAS-CONF-2017-062,ATLAS-CONF-2017-063,CMS-DP-2015-043,CMS-PAS-JME-15-002,CMS-DP-2016-070,CMS-DP-2015-038,CMS:2017wyc}.}   One of the most powerful observables for discrimination of two-prong substructure is $D_2$ \cite{Larkoski:2014gra,Larkoski:2015kga}, based on the $n$-point energy correlation functions \cite{Larkoski:2013eya}.  $D_2$ parametrically separates jets with one- and two-prongs, and has been used extensively during Run 2 by ATLAS \cite{ATLAS-CONF-2015-035,Aad:2015rpa,Aaboud:2016okv,Aaboud:2016trl,Aaboud:2016qgg,Aaboud:2017zfn,Aaboud:2017ahz,Aaboud:2017eta,Aaboud:2017itg,Aaboud:2017ecz}.  Because of the high-luminosity environment of the LHC, the robust measurement of the substructure of jets requires methods for removing contamination from underlying event, pile-up, or other sources that are mostly uncorrelated with the hard scattering.  Of these so-called jet grooming techniques, the modified mass drop (mMDT) \cite{Dasgupta:2013ihk,Dasgupta:2013via} and soft drop \cite{Larkoski:2014wba} groomers are theoretically most well-understood.  Soft drop groomed mass distributions of jets initiated both by light QCD partons and by hadronically decaying $Z$ bosons were measured recently by CMS \cite{CMS-PAS-JME-14-002}.

Many theory advancements in understanding these observables and techniques have been made recently.  The groomed jet mass and several related single prong observables have been calculated, both for QCD jets and electroweak bosons, up to next-to-leading logarithm (NLL) \cite{Dasgupta:2013ihk,Dasgupta:2013via,Larkoski:2014wba,Dasgupta:2015yua}, and very recently for top jets \cite{Hoang:2017kmk}.  \Refs{Frye:2016okc,Frye:2016aiz} presented the first jet substructure calculation to next-to-next-to-leading logarithmic accuracy (NNLL) matched to fixed-order at ${\cal O}(\alpha_s^2)$ for the mMDT or soft drop groomed jet mass.   A number of two-prong observables, both ungroomed \cite{Dasgupta:2015lxh} and groomed \cite{Dasgupta:2016ktv,Salam:2016yht}, have been computed at leading-logarithmic accuracy (LL), and the observable $D_2$ was calculated for QCD jets and boosted $Z$ bosons produced in $e^+e^-$ collisions to next-to-leading logarithmic accuracy (NLL) \cite{Larkoski:2015kga}. These calculations have helped to put the understanding of jet substructure observables on firmer theoretical footing, and have inspired a number of new jet substructure techniques currently used at the LHC.

In this paper, we present the first theory calculations to NLL accuracy for groomed jets on which two-prong observables like $D_2$ are measured, and present a systematically improvable framework for their calculation. This is achieved using factorization formulae for groomed two-prong jet observables derived using the techniques of soft-collinear effective theory (SCET) \cite{Bauer:2000yr,Bauer:2001ct,Bauer:2001yt,Bauer:2002nz,Rothstein:2016bsq}. These allow for the resummation of logarithmically enhanced terms in all regions of phase space using the renormalization group evolution of field theoretic operators, significantly simplifying higher order calculations, and also provide operator definitions of non-perturbative effects.  We will illustrate our framework by performing a calculation of the $D_2$ observable on jets produced in $e^+e^-$ collisions, although our approach is more general, and could be applied to related observables, such as $N_2$ \cite{Moult:2016cvt}. In a companion paper \cite{Larkoski:2017iuy} we present $D_2$ distributions for mMDT/soft drop groomed jets produced in $pp$ collisions for processes of phenomenological relevance for jet substructure at the LHC.

Schematically, the soft drop groomer steps through the clustering history of a jet and removes those branches in the jet which fail the requirement
\begin{equation}\label{eq:sddef}
\frac{\min[E_i,E_j]}{E_i+E_j}> \zcut \left(\frac{\theta_{ij}}{R}\right)^\beta\,.
\end{equation}
Here, $E_i$ is the energy of branch $i$, $\theta_{ij}$ is the angle between branches $i$ and $j$, and $R$ is the jet radius.  $\zcut$ and $\beta$ are parameters of soft drop; in this paper, we will only consider $\beta = 0$, for which soft drop coincides with the mMDT groomer (Due to their equivalence we will use their names interchangeably.). Typical values of $\zcut$ are $\zcut\approx 0.1$. When \Eq{eq:sddef} is satisfied, the procedure terminates.  On jets that have been groomed with soft drop, we will measure both the jet mass and the observable $D_2$.\footnote{Without a mass cut, $D_2$ is not infrared and collinear safe, but is Sudakov safe \cite{Larkoski:2013paa,Larkoski:2015lea}. It can therefore be calculated in resummed perturbation theory.}  We will work in the formal limit
\begin{equation}\label{eq:scalinglimit}
\frac{m_J^2}{E_J^2} \ll \zcut \ll 1\,,
\end{equation}  
where $m_J$ is the groomed jet mass and $E_J$ is the jet energy.  This limit is both relevant for phenomenology and vital for theoretical simplicity.  \Eq{eq:scalinglimit} is satisfied for electroweak scale masses, TeV scale jet energies, and $\zcut = 0.1$.  By working in the regime of \Eq{eq:scalinglimit}, only collinear emissions pass the soft drop requirement, no non-global logarithms arising from color-connections to the rest of the event \cite{Dasgupta:2001sh} contribute to the shape of the distribution, and corrections from hadronization are significantly reduced.  This will enable us to make precise and robust predictions for the distribution of $D_2$ as measured on these jets.

In fact, the factorization formulae that we derive for the mMDT/soft dropped $D_2$ exhibit a universality, and the resulting $D_2$ distribution is largely independent of the jet energy.  From the factorization formulae, we will show that:
\begin{itemize}
\item the leading non-perturbative corrections are suppressed by powers of the groomed jet mass, and are independent of the jet energy;
\item the endpoint of the distribution is set by the grooming parameter $\zcut$, and is independent of the jet mass and energy;
\item perturbative power corrections are suppressed by $m_J^2/(\zcut E_J^2)\ll 1$ throughout the distribution;
\item the quark or gluon flavor of a jet is well-defined at leading power in $\zcut \ll 1$. 
\end{itemize}
These properties imply that the full mMDT/soft drop $D_2$ is only very weakly dependent on the jet mass $m_J$ and the jet energy $E_J$. This is a desirable property, especially for experimental applications. The explicit design of observables that are independent of mass and $p_T$ (or energy) cuts has been a subject of recent research \cite{Dolen:2016kst,Moult:2016cvt}.  Due to the final property, quark and gluon jet distributions can be individually extracted from $D_2$ distributions of jets produced in different processes.

The outline of this paper is as follows.  In \Sec{sec:obs}, we define the mMDT and soft drop groomers appropriate for both jets produced in $e^+e^-$ and $pp$ collisions, and define the energy correlation functions and the ratio observable $D_2$.  In \Sec{sec:factee}, we will review the previously published factorization formulae for $D_2$ and groomed jet mass and derive new factorization formulae for the groomed $D_2$ cross section in $e^+e^-$ collisions.  Collinear factorization enables us to use all of the results from $e^+e^-$ collisions to formulate the factorized cross section for the groomed $D_2$ observable for jets produced in $pp$ collisions in \Sec{sec:factpp}.  From the form of the factorization formula, we are immediately able to make all-orders statements regarding properties of the distribution.  We will review those properties that are identical to those of the groomed jet mass and discuss in some detail reduced hadronization corrections for the $D_2$ distribution in \Sec{sec:consfact}.  In \Sec{sec:eepred}, we present numerical results, comparing our NLL predictions in $e^+e^-$ collisions to parton shower Monte Carlo. A comparison of robust qualitative features of the distribution derived from the factorization formula for $pp$ are compared with parton shower Monte Carlo predictions in \Sec{sec:pppred}. We conclude in \Sec{sec:conc}.  Technical details and calculations are presented in appendices.

\section{Observables}\label{sec:obs}

In this section, we review the definitions of the observables and grooming procedures that will be studied in this paper.  As mentioned in the introduction, we will restrict our focus to grooming with mMDT \cite{Dasgupta:2013ihk,Dasgupta:2013via} or equivalently, soft drop with angular exponent $\beta = 0$  \cite{Larkoski:2014wba}.  Definitions of mMDT and energy correlation functions will be presented for jets produced in both $e^+e^-$ and $pp$ collisions.

\subsection{Modified Mass Drop/Soft Drop Grooming}\label{sec:sd}

The modified mass drop/ soft drop groomer proceeds in the following way.  The identified jet is reclustered with the Cambridge/Aachen algorithm \cite{Dokshitzer:1997in,Wobisch:1998wt,Wobisch:2000dk}, which orders emissions in the jet by their relative angle.  Then, starting at the widest angle, the two branches following from each splitting in the jet are required to satisfy an energy fraction constraint.  For branches $i$ and $j$ in a jet produced in $e^+e^-$ collisions, this requirement is
\begin{equation}\label{eq:eesd}
\frac{\min[E_i,E_j]}{E_i+E_j}>\zcut\,,
\end{equation}
where $E_i$ is the energy of branch $i$.
For a jet from $pp$ collisions, the requirement is
\begin{equation}\label{eq:ppsd}
\frac{\min[p_{Ti},p_{Tj}]}{p_{Ti}+p_{Tj}}>\zcut\,,
\end{equation}
where $p_{Ti}$ is the transverse momentum with respect to the proton beam of branch $i$.
If these requirements are not satisfied, the softer (lower energy/transverse momentum) branch is removed from the jet, and the procedure iterates to the next splitting on the harder branch.  The process terminates when the two branches in a splitting in the jet satisfy \Eq{eq:eesd} or \Eq{eq:ppsd}, as appropriate.  The cut parameter $\zcut$ is typically taken to be $\zcut \sim 0.1$.  In this paper, we will formally assume that $\zcut \ll 1$, so that emissions that fail these requirements are necessarily soft.  Once a jet has been groomed, any observable can be measured on its remaining constituents. 

\subsection{Energy Correlation Functions and $D_2$}\label{sec:D2def}

For powerful identification of hadronic decays of electroweak bosons, we use the energy correlation functions \cite{Larkoski:2013eya}.\footnote{Another common approach is to use the $N$-subjettiness ratio observables \cite{Thaler:2010tr,Thaler:2011gf}. However, these are poorly behaved in perturbation theory \cite{Larkoski:2015uaa}.}  The $n$-point energy correlation function is sensitive to radiation about $n-1$ hard cores in the jet.  Therefore, for this application, we use the two- and three-point energy correlation functions, which are sensitive to radiation about one or two hard cores in the jet.  For a set of particles $\{i\}$, in a jet $J$, the energy correlation functions in $e^+e^-$ collisions are defined as
\begin{align}\label{eq:eee2}
\left.\ecf{2}{\alpha}\right|_{e^+e^-}&=\frac{1}{E_J^2}\sum_{i<j\in J} E_i E_j\left(
\frac{2p_i\cdot p_j}{E_i E_j}
\right)^{\alpha/2}\,, \\
\left.\ecf{3}{\alpha}\right|_{e^+e^-}&=\frac{1}{E_J^3}\sum_{i<j<k\in J} E_i E_j E_k\left(
\frac{2p_i\cdot p_j}{E_i E_j}\frac{2p_i\cdot p_k}{E_i E_k}\frac{2p_j\cdot p_k}{E_j E_k}
\right)^{\alpha/2}\,.
\end{align}
For jets produced in $pp$ collisions, the energy correlation functions are simply modified as
\begin{align}\label{eq:ppe2}
\left.\ecf{2}{\alpha}\right|_{pp}&=\frac{1}{p_{TJ}^2}\sum_{i<j\in J} p_{Ti} p_{Tj}R_{ij}^\alpha\,, \\
\left.\ecf{3}{\alpha}\right|_{pp}&=\frac{1}{p_{TJ}^3}\sum_{i<j<k\in J} p_{Ti} p_{Tj}p_{Tk}R_{ij}^\alpha R_{ik}^\alpha R_{jk}^\alpha\,.
\end{align}
Here $R_{ij}$ is the distance between particles $i$ and $j$ in the pseudorapidity-azimuth angle plane.  For jets that are central ($p_{TJ} \sim E_J$) and if all emissions in the jet are collinear, the definitions of the two- and three-point energy correlation functions for jets in $pp$ collisions are equivalent to those for $e^+e^-$ collisions.

The angular exponent $\alpha$ in the definition of the energy correlation functions is a parameter that controls sensitivity to wide-angle emissions.  For jets that consist of massless particles, the two-point energy correlation function in $e^+e^-$ collisions reduces to a function of the jet mass, $m_J$, if $\alpha=2$:
\begin{equation}
\left.\ecf{2}{2}\right|_{e^+e^-} = \frac{m_J^2}{E_J^2}\,.
\end{equation} 
It has been shown that the optimal observable, formed from $\ecf{2}{\alpha}$ and $\ecf{3}{\alpha}$, for discrimination of boosted hadronic decays of electroweak bosons (with a hard two-prong substructure) from QCD jets (typically with a single hard prong) is a particular ratio of the energy correlation functions called $D_2^{(\alpha)}$ \cite{Larkoski:2014gra,Larkoski:2015kga}.  $D_2^{(\alpha)}$ is defined as
\begin{equation}
D_2^{(\alpha)} = \frac{\ecf{3}{\alpha}}{(\ecf{2}{\alpha})^3}\,.
\end{equation}
In this paper, for jets which have been groomed with mMDT, we subsequently measure $D_2^{(\alpha)}$.  For brevity, we will often denote $D_2^{(\alpha)}$ generically as $D_2$.

\subsection{Phase Space and Behavior of Groomed $D_2$}\label{sec:ps_struc}

In this section, we use the power counting techniques of \cite{Larkoski:2014gra} to review the phase space structure of a jet on which the observables $\ecf{2}{\alpha}$ and $\ecf{3}{\alpha}$ are measured. We then discuss how this is modified when the jet is groomed with mMDT or soft drop. In particular, we emphasize the parametric features of the groomed $D_2$ distribution that will be reproduced by our calculation, as well as demonstrating that $D_2$ remains a powerful discriminant even after grooming has been applied.

We begin with a brief review of the structure of the $\ecf{2}{\alpha}$ and $\ecf{3}{\alpha}$ phase space without grooming, as was considered in detail in \cite{Larkoski:2014gra,Larkoski:2015kga}.  With no grooming the dominant emissions in a one-prong jet with $\ecf{2}{\alpha},\ecf{3}{\alpha}\ll 1$ are either soft (low energy) or collinear to the jet core.
These contributions to the value of $\ecf{2}{\alpha}$ and $\ecf{3}{\alpha}$ scale like
\begin{align}
\ecf{2}{\alpha}&\sim \theta_{cc}^\alpha+z_s  \,,  \nn \\
\ecf{3}{\alpha}&\sim \theta_{cc}^{3\alpha}+\theta_{cc}^\alpha z_s+z_s^2 \,,
\end{align}
where $z_s$ is the characteristic energy fraction of soft emissions and $\theta_{cc}$ is the characteristic angle of collinear emissions.  Depending on the assumptions made about the relative scaling of $z_s$ and $\theta_{cc}$, one finds the upper and lower boundaries of the phase space for one-prong jets.
The upper boundary is the so-called soft haze region where
\begin{align}
\ecf{3}{\alpha}\sim (\ecf{2}{\alpha})^2\,,
\end{align}
while the lower boundary corresponds to the scaling
\begin{align}
\ecf{3}{\alpha}\sim (\ecf{2}{\alpha})^3\,.
\end{align}
Therefore, up to order-1 coefficients, one-prong jets have a measured value of $\ecf{3}{\alpha}$ that lies between $(\ecf{2}{\alpha})^3$ and $(\ecf{2}{\alpha})^2$.

\begin{figure}
\begin{center}
\subfloat[]{\label{fig:unsoftdropped}
\includegraphics[width=7cm]{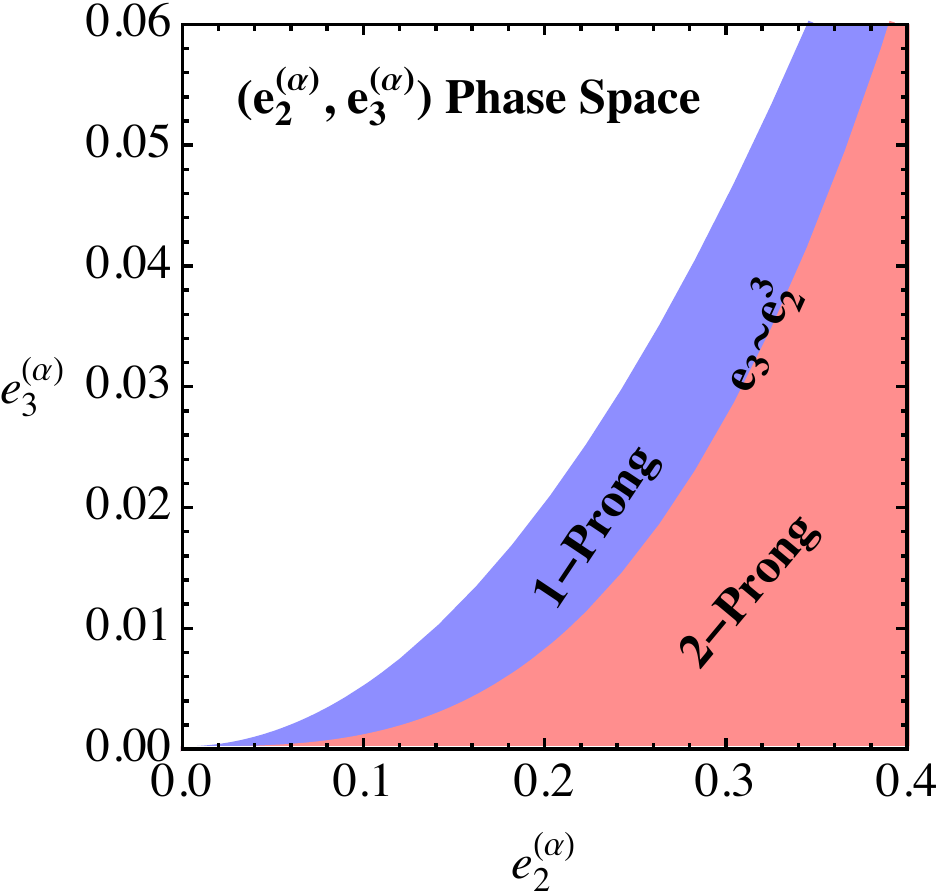}    
}\qquad
\subfloat[]{\label{fig:softdropped}
\includegraphics[width=7cm]{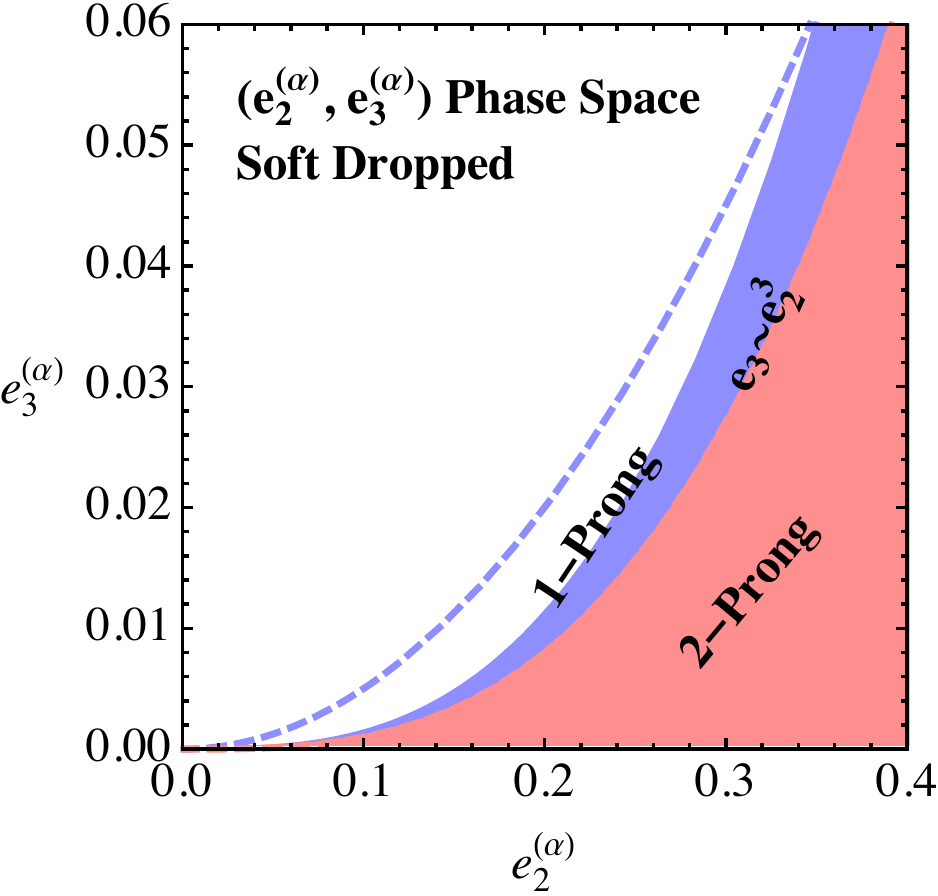}
}
\end{center}
\caption{The phase space for one- and two-prong jets on which the energy correlation functions $\ecf{2}{\alpha}$ and $\ecf{3}{\alpha}$ are measured.  (a) Phase space for ungroomed jets, with the absolute upper boundary scaling like $\ecf{3}{\alpha} \sim (\ecf{2}{\alpha})^2$.  (b) Phase space for mMDT/soft drop groomed jets, where the two-prong region is unchanged, while the absolute upper boundary now scales like $\ecf{3}{\alpha} \sim (\ecf{2}{\alpha})^3/\zcut$.
}
\label{fig:phasespace}
\end{figure}

To determine the region of phase space where a two-prong jet lives, it is sufficient to consider a jet with two hard, collinear prongs, with all other radiation at much lower energy.  In this case, the scaling of the contributions to $\ecf{2}{\alpha}$ and $\ecf{3}{\alpha}$ are
\begin{align}
\ecf{2}{\alpha}&\sim \theta_{ab}^\alpha\,, \nn \\
\ecf{3}{\alpha} &\sim \theta_{cc}^\alpha \theta_{ab}^{2\alpha} + \theta_{ab}^{3\alpha}z_{cs} +\theta_{ab}^{\alpha}z_s \,.
\end{align}
Here, $\theta_{ab}$ is the angle between the hard prongs, $\theta_{cc}$ is the characteristic angular size of each of the hard prongs individually, $z_{cs}$ is the energy fraction of collinear-soft radiation emitted from the dipole of the two hard prongs, and $z_s$ is the energy fraction of soft radiation emitted at large angles.  The requirement that the hard prongs are well-defined restricts the energy fraction of the collinear-soft radiation to be small:
\begin{align}
z_{cs} \sim \frac{\ecf{3}{\alpha}}{(\ecf{2}{\alpha})^3}\ll 1\,.
\end{align}
That is, two-prong jets have measured values of $\ecf{3}{\alpha}$ that are much smaller than $(\ecf{2}{\alpha})^3$.  The one- and two-prong regions of phase space in the $(\ecf{2}{\alpha},\ecf{3}{\alpha})$ plane are illustrated in \Fig{fig:unsoftdropped}.

We now consider how this phase space is modified in the presence of grooming.  First, we consider the parametric scaling of contributions in the one-prong region of phase space. We assume that there exists radiation in the groomed jet whose energy fraction is set by $\zcut$ at a characteristic angle $\theta_{sc}$ from the jet core, in addition to the collinear modes. The soft modes (which had energy fraction $z_s$) have been removed by the mMDT/soft drop procedure.  The contributions to the observables are then
\begin{align}
\ecf{2}{\alpha} &\sim \zcut \theta_{sc}^\alpha +\theta_{cc}^\alpha\,, \nn \\
\ecf{3}{\alpha} &\sim \theta_{cc}^{3\alpha}+\theta_{cc}^\alpha \theta_{sc}^{2\alpha} \zcut+\theta_{sc}^{3\alpha} \zcut^2\,.
\end{align}
From this we can determine the boundaries of the phase space by imposing different relationships between the characteristic angles and energy fractions.  As in the ungroomed case, the lower bound of this phase space region occurs when collinear emissions dominate the value of $\ecf{2}{\alpha}$:
\begin{align}
\ecf{2}{\alpha} \sim \theta_{cc}^\alpha \gg \zcut \theta_{sc}^\alpha\,, \qquad  \ecf{3}{\alpha}\sim (\ecf{2}{\alpha})^3\,.
\end{align}
This is the same lower boundary as with ungroomed jets. This implies in particular, that $D_2$ remains a powerful discriminant, even after grooming has been applied.

The upper boundary of this region is more interesting.  Assuming that the contributions to $\ecf{2}{\alpha}$ are democratic
\begin{align}
\ecf{2}{\alpha} &\sim \zcut \theta_{sc}^\alpha +\theta_{cc}^\alpha\,,
\end{align}
we find the characteristic angular scale of the radiation sensitive to $\zcut$ to be
\begin{align}
\theta_{sc}^\alpha \sim \frac{\ecf{2}{\alpha}}{ \zcut  }\,.
\end{align}
Using this scaling, for $\ecf{3}{\alpha}$ for groomed jets, we then find
\begin{align}
\ecf{3}{\alpha}
&\sim \frac{(\ecf{2}{\alpha})^3}{\zcut}\,.
\end{align}
We refer to the region of phase space near this upper boundary as ``collinear-soft haze''.  Note that, assuming that two-prong jets have two hard prongs, their phase space is unchanged from the ungroomed case.

This is quite interesting. Since we formally assume $\zcut \ll 1$, there is a separation of the collinear-soft haze region from the lower boundary of groomed one-prong jets.  However, both boundaries have the same cubic relationship between
$\ecf{2}{\alpha}$ and $\ecf{3}{\alpha}$. 
The phase space for groomed jets is illustrated in \Fig{fig:softdropped}. 
Grooming the jet removes the region of phase space where $\ecf{3}{\alpha}\sim(\ecf{2}{\alpha})^2$.  From this analysis, it is straightforward to determine the maximal value of $D_2$ with and without grooming.
For the ungroomed jet, the maximal value of $D_2$ is when
\begin{align}
\left.\Dobs{2}{\alpha}\right|_{\max}\sim \left.\frac{\ecf{3}{\alpha}}{(\ecf{2}{\alpha})^3}\right|_{\ecf{3}{\alpha}\sim(\ecf{2}{\alpha})^2} \sim \frac{1}{2\ecf{2}{\alpha}}\,.
\end{align}
With a more careful analysis (discussed in \Ref{Larkoski:2014gra}), one can derive the factor of $1/2$ in the location of the endpoint.  Note that for $\alpha=2$ this maximum value is sensitive to both the energy and mass of the jet:
\begin{align}
\left.\Dobs{2}{2}\right|_{\max}\sim \frac{E_J^2}{2m_J^2}\,.
\end{align}
The endpoint of the $D_2$ distribution formally increases without bound as the energy of the jet increases, for a fixed mass cut.\footnote{Of course, this isn't quite true because there is a characteristic mass scale of QCD.  Even perturbatively this isn't true because the Sudakov factor will exponentially suppress low masses.}
On the other hand, when the jet is groomed, we find
\begin{align}
\left.\Dobs{2}{\alpha}\right|_{\max, \text{ soft drop}}\sim \left.\frac{\ecf{3}{\alpha}}{(\ecf{2}{\alpha})^3}\right|_{\ecf{3}{\alpha}\sim(\ecf{2}{\alpha})^3/\zcut} \sim \frac{1}{2\zcut}\,.
\end{align}
Therefore, when a jet is groomed with mMDT or soft drop, the endpoint of the $D_2$ distribution is independent of both the jet mass and energy.  This property will be one part of the reason why the groomed $D_2$ distribution is incredibly robust to changes in energy and/or mass cuts.

\begin{figure}
\begin{center}
\subfloat[]{\label{fig:ps_MCa}
\includegraphics[width=7cm]{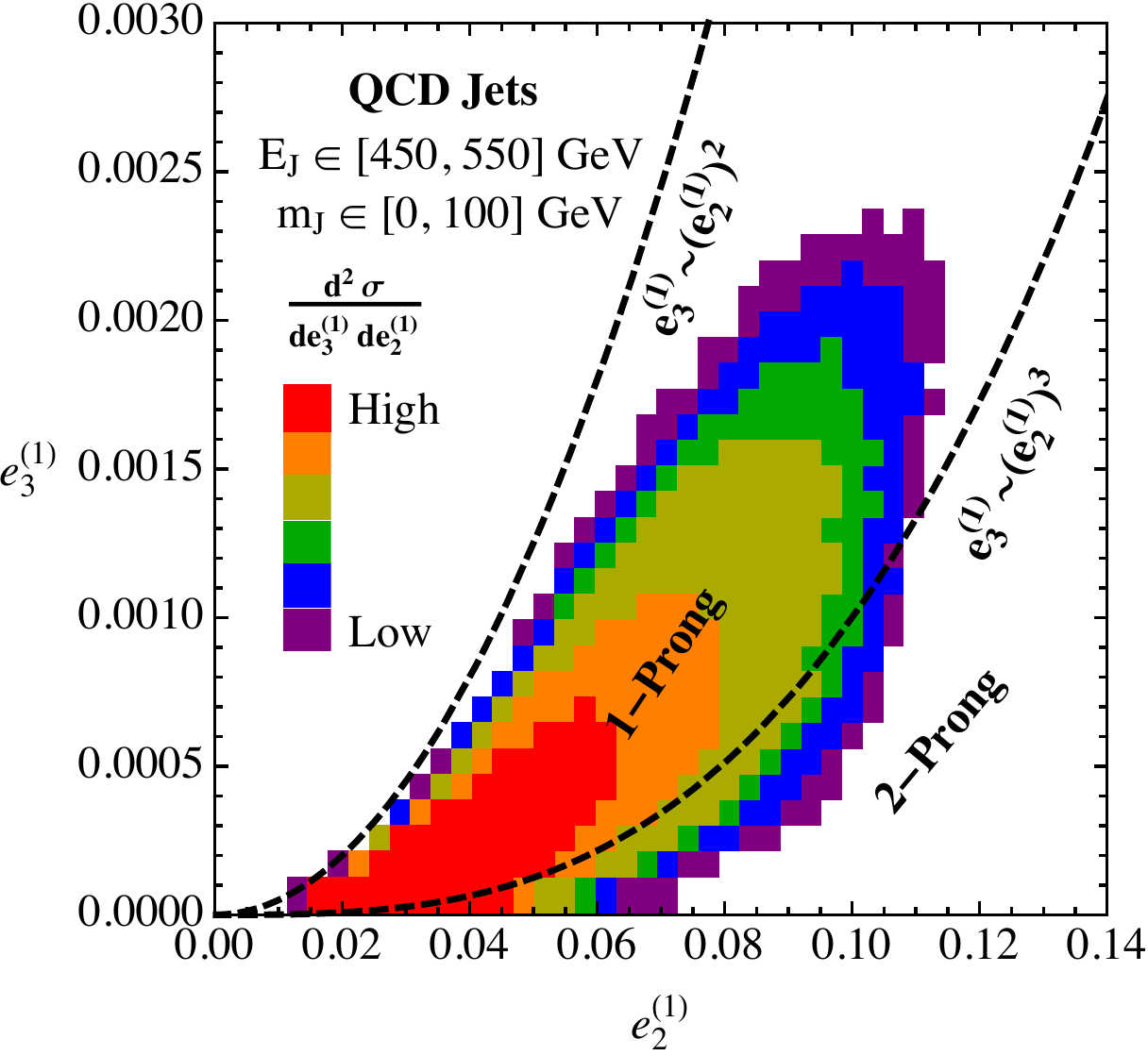}    
}\qquad
\subfloat[]{\label{fig:ps_MCb}
\includegraphics[width=7cm]{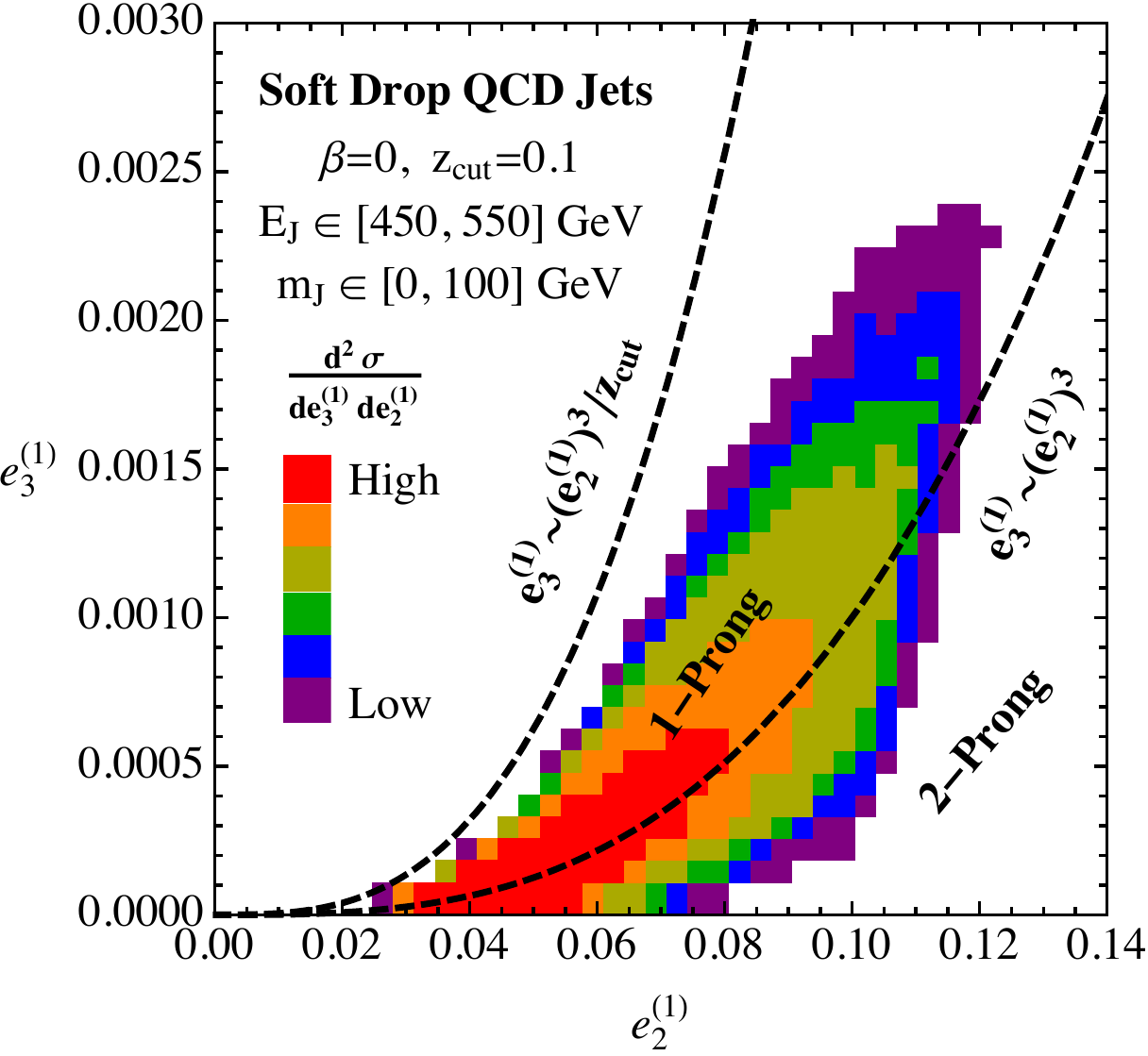}
}
\end{center}
\caption{Illustration of the population of jets from 1 TeV $e^+e^-$ collisions in the $(\ecf{2}{\alpha},\ecf{3}{\alpha})$ phase space plane as simulated in parton shower Monte Carlo.  Here, $\alpha=1$ and the mass of the jet is restricted as $m_J<100$ GeV.  (a) Ungroomed jets, that extend up to $\ecf{3}{\alpha}\sim(\ecf{2}{\alpha})^2$.  (b) mMDT/soft drop groomed jets, with $\zcut=0.1$, that extend up to $\ecf{3}{\alpha}\sim(\ecf{2}{\alpha})^3/\zcut$.
}
\label{fig:phasespace_MC}
\end{figure}

To demonstrate that this scaling is satisfied in simulation, in \Fig{fig:phasespace_MC} we plot the distribution of jets in the $(\ecf{2}{\alpha},\ecf{3}{\alpha})$ phase space plane as simulated in parton shower Monte Carlo.  The details of the Monte Carlo simulation will be described in \Sec{sec:eepred}.  Here we use the angular exponent $\alpha=1$ and impose an upper cut on the mass of $m_J<100$ GeV, which more clearly illustrates the phase space regions.  The same general features are present for other values of $\alpha$.  On these jets we then measure $\ecf{2}{\alpha}$ and $\ecf{3}{\alpha}$, either before grooming or after mMDT/soft drop grooming, with $\zcut = 0.1$.  In \Fig{fig:ps_MCa}, we show the ungroomed phase space, and jets populate up to the curve where $\ecf{3}{\alpha}\sim(\ecf{2}{\alpha})^2$.  Once grooming is applied, however, jets only populate up to the curve $\ecf{3}{\alpha}\sim(\ecf{2}{\alpha})^3/\zcut$, as illustrated in \Fig{fig:ps_MCb}.  This demonstrates that our parametric scaling analysis of the phase space is satisfied by parton shower simulation. More detailed tests will be provided in \Sec{sec:eepred}, when we study the structure of the $D_2$ distribution in our analytic calculation, and in parton shower Monte Carlo.

The location of the endpoint for the groomed $D_2$ distribution also has important consequences for its calculation. In particular, for the ungroomed $D_2$ distribution, the endpoint is at $\sim1/\ecf{2}{\alpha}$. In the limit $\ecf{2}{\alpha}\ll 1$, this is formally large, and can be neglected.  This is what was done in \Ref{Larkoski:2015kga}.  However, for the groomed $D_2$ distribution, the endpoint of the distribution is at $\sim1/\zcut$. Since we assume $\zcut \gg \ecf{2}{\alpha}$, we must compute the matrix element in this region of the phase space, and match it to our resummed calculation, to accurately predict the endpoint of the distribution.

\section{Factorized Cross Section in $e^+e^-$ Collisions}\label{sec:factee}

In this section we present factorization formulae for mMDT/soft drop groomed $D_2$. These allow for a systematically improvable calculation of the $D_2$ distribution, and the resummation of logarithmically enhanced terms in all regions of phase space to be resummed by renormalization group evolution. The factorization formulae are presented in the language of SCET \cite{Bauer:2000yr,Bauer:2001ct,Bauer:2001yt,Bauer:2002nz}, an effective field theory describing soft and collinear radiation in the presence of a hard scattering. For the case of jet substructure observables, where multiple hierarchies are present within the jet, extensions of SCET are required. These have been developed in  \Refs{Bauer:2011uc,Larkoski:2015zka,Larkoski:2015kga,Pietrulewicz:2016nwo}, and were discussed in detail in the context of the $D_2$ observable in \Ref{Larkoski:2015kga}. In this section we will restrict ourselves to giving physical descriptions of the functions appearing in the factorization formulae. Field theoretic definitions, and one-loop calculations, are given in the Appendices. 

Our approach to deriving the factorization formulae will closely follow the techniques used to study the groomed jet mass and the ungroomed $D_2$ observable. In particular, we will begin from the factorization for the groomed jet mass cross section, and then perform a refactorization of the resolved substructure. Because of this, we begin in \Sec{sec:fact_review} with a review of the factorization formulae in both these cases. We then present the factorization for groomed $D_2$ in \Sec{sec:fact_D2}. Throughout this section, we will restrict ourselves to the case of $e^+e^-$ collisions for simplicity. In \Sec{sec:factpp}, we will then show that the grooming algorithm allows us to trivially extend this factorization formula to the case of $pp$ collisions.

\subsection{Review of Known Results}\label{sec:fact_review}

Factorization formulae are known for both the groomed jet mass \cite{Frye:2016okc,Frye:2016aiz} and for the ungroomed $D_2$ observable \cite{Larkoski:2015kga}. Since our factorization for the groomed $D_2$ observable will rely heavily on ingredients from both these analyses, we begin by reviewing the essential ingredients of the factorization formulae for these two cases. The discussion will be brief, and more details can be found in the respective papers.

\subsubsection{Groomed Jet Mass Factorization Formula}\label{sec:fact_review_sdmass}

A factorization formulae was presented in SCET for the soft dropped two-point energy correlation functions $\ecf{2}{\alpha}$, and was used to calculate the distribution to NNLL order \cite{Frye:2016okc,Frye:2016aiz}.  Throughout this section we will always take the soft drop parameter $\beta=0$. The case $\beta >0$ follows an identical logic, and is discussed in detail in \Refs{Frye:2016okc,Frye:2016aiz}.

The factorization formula is valid in the limit $\ecf{2}{\alpha} \ll \zcut\ll 1$. It can be derived through a multi-stage matching procedure from the standard SCET involving a global soft function and jet functions. The first stage of the matching is a soft and collinear factorization, with the soft virtuality set by $Q\zcut$, and the collinear virtuality set by $\ecf{2,R}{\alpha}$:
\begin{align}\label{eq:sd_mother}
\frac{d^2\sigma}{d\ecf{2,L}{\alpha} d\ecf{2,R}{\alpha} }= H(Q^2) S(\zcut) \left[ J(\ecf{2,L}{\alpha},\zcut)  \right]\left[ J(\ecf{2,R}{\alpha}, \zcut)  \right]\,.
\end{align}
The soft drop grooming has isolated the jet dynamics from the rest of the event, due to the angular ordering of the algorithm. However, this factorization still contains large logarithms within the collinear sector. These can be resummed by refactorizing into a collinear-soft function, which allows for the resummation of all logarithms of $\ecf{2,R}{\alpha}$. The final factorization formula for measuring $\ecf{2}{\alpha}$ in each of the groomed hemispheres in $e^+e^-$ collisions is given by
\begin{align}\label{eq:sd_mass}
\frac{d^2\sigma}{d\ecf{2,L}{\alpha} d\ecf{2,R}{\alpha} }= H(Q^2) S(\zcut) \left[ J(\ecf{2,L}{\alpha}) \otimes S_c(\ecf{2,L}{\alpha} \zcut)  \right]\left[ J(\ecf{2,R}{\alpha}) \otimes S_c(\ecf{2,R}{\alpha} \zcut)  \right]\,.
\end{align}
The physical interpretation of the functions entering this factorization formula are as follows (field theoretic definitions can be found in \cite{Frye:2016okc,Frye:2016aiz}):
\begin{itemize}
\item $H(Q^2)$ is the standard hard function, describing in this case the production of two back to back jets in an $e^+e^-$ collision.
\item $S(\zcut)$ is the global soft function. It describes wide angle soft radiation, which is removed by the groomer. It is therefore independent of the observable, and depends just on $\zcut$.
\item $J(\ecf{2}{\alpha})$ is a jet function describing collinear radiation. Since this radiation is energetic, it is not affected by the groomer, so that this function does not depend on $\zcut$.
\item $S(\ecf{2}{\alpha} \zcut^{\alpha-1})$ describes collinear soft radiation, which contributes to the observable, but is sensitive to the groomer. It can be shown that it depends only on the scales $\zcut$ and $\ecf{2}{\alpha}$ through the combination $\ecf{2}{\alpha}\zcut^{\alpha-1}$, as indicated by the argument of the function.
\end{itemize}

\begin{figure}
\begin{center}
\includegraphics[width=8cm]{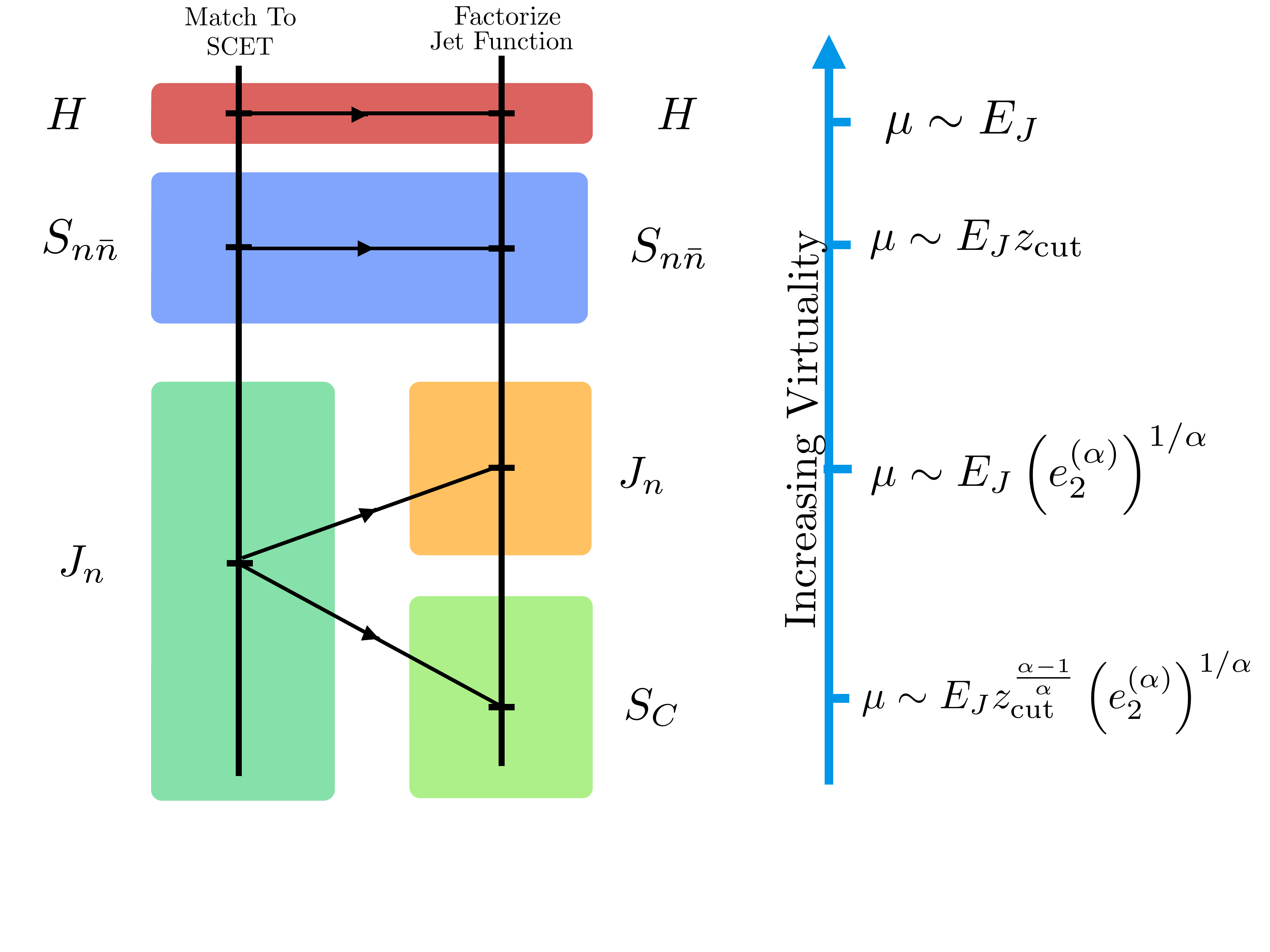}
\end{center}
\caption{An illustration of the multistage matching procedure for the factorization of the soft dropped energy correlation function $\ecf{2}{\alpha}$. For simplicity, we have taken the soft drop angular exponent, $\beta=0$.
}
\label{fig:sd_mass_match}
\end{figure}

The multi-stage matching procedure is shown in \Fig{fig:sd_mass_match}, which also shows the virtualities of the modes contributing to the factorization formula. The results for all functions appearing in the factorization formula of \Eq{eq:sd_mass} allowing for resummation up to NNLL were computed in  \Refs{Frye:2016okc,Frye:2016aiz}

\subsubsection{Ungroomed $D_2$ Factorization Formula}\label{sec:fact_review_ungD2}

In \Ref{Larkoski:2015kga} a factorization formula was presented for the $D_2$ observable. For a two-prong substructure observable such as $D_2$, multiple kinematic regimes with distinct hierarchies exist, each of which contribute to a different region of the multi-dimensional phase space discussed in \Sec{sec:ps_struc}. The approach taken in \Ref{Larkoski:2015kga} was to identify all parametric regions of phase space where hierarchies occur, and to develop distinct effective field theories describing each of these regions. The different effective field theories can then be pieced together to give a complete description of the entire phase space region.

\begin{figure}
\begin{center}
\subfloat[]{\label{fig:soft_haze}
\includegraphics[width=3.95cm, trim =0 -0.5cm 0 0]{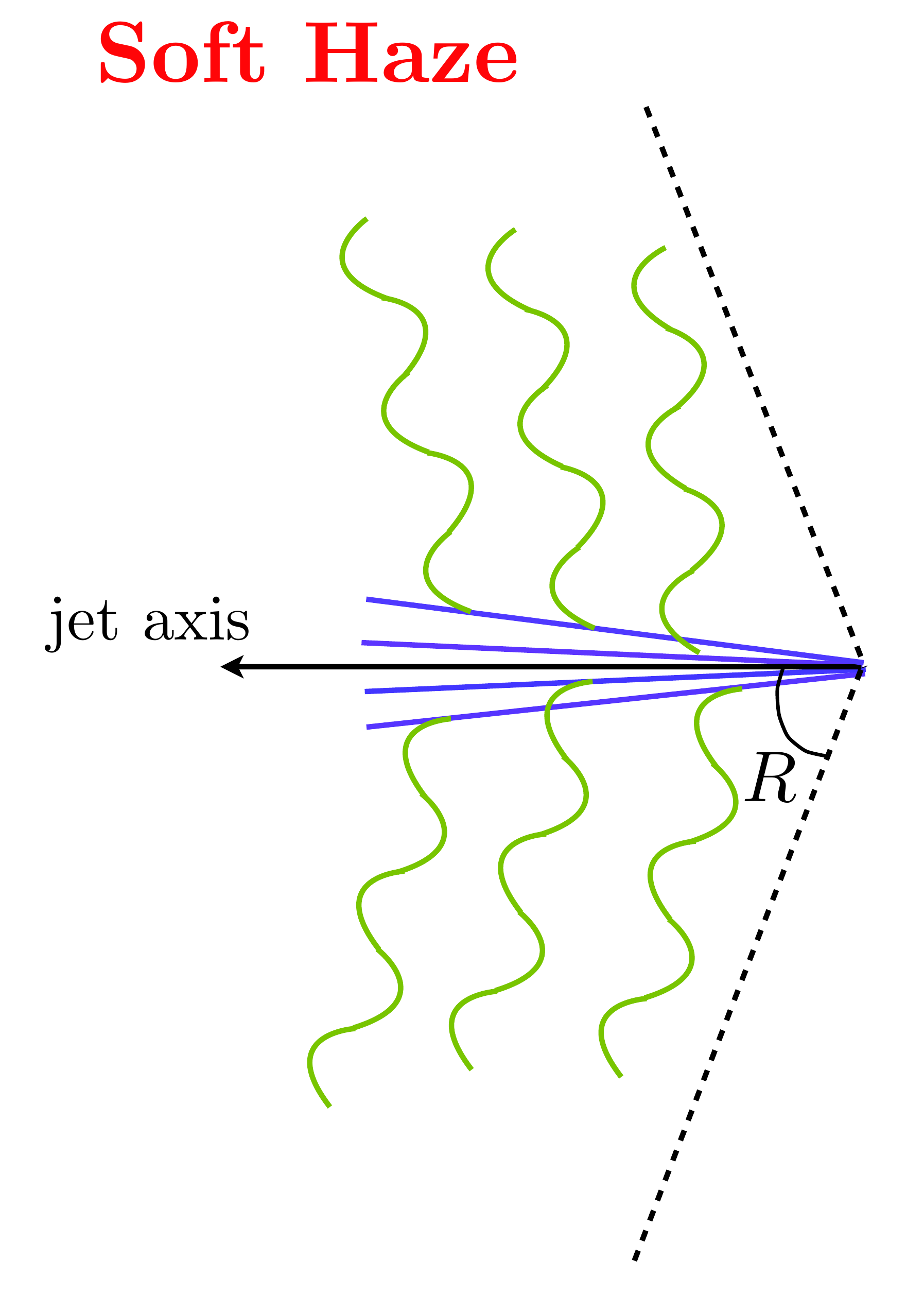}
}\qquad
\subfloat[]{\label{fig:ninja}
\includegraphics[width=4.1cm, trim =0 0 0 0]{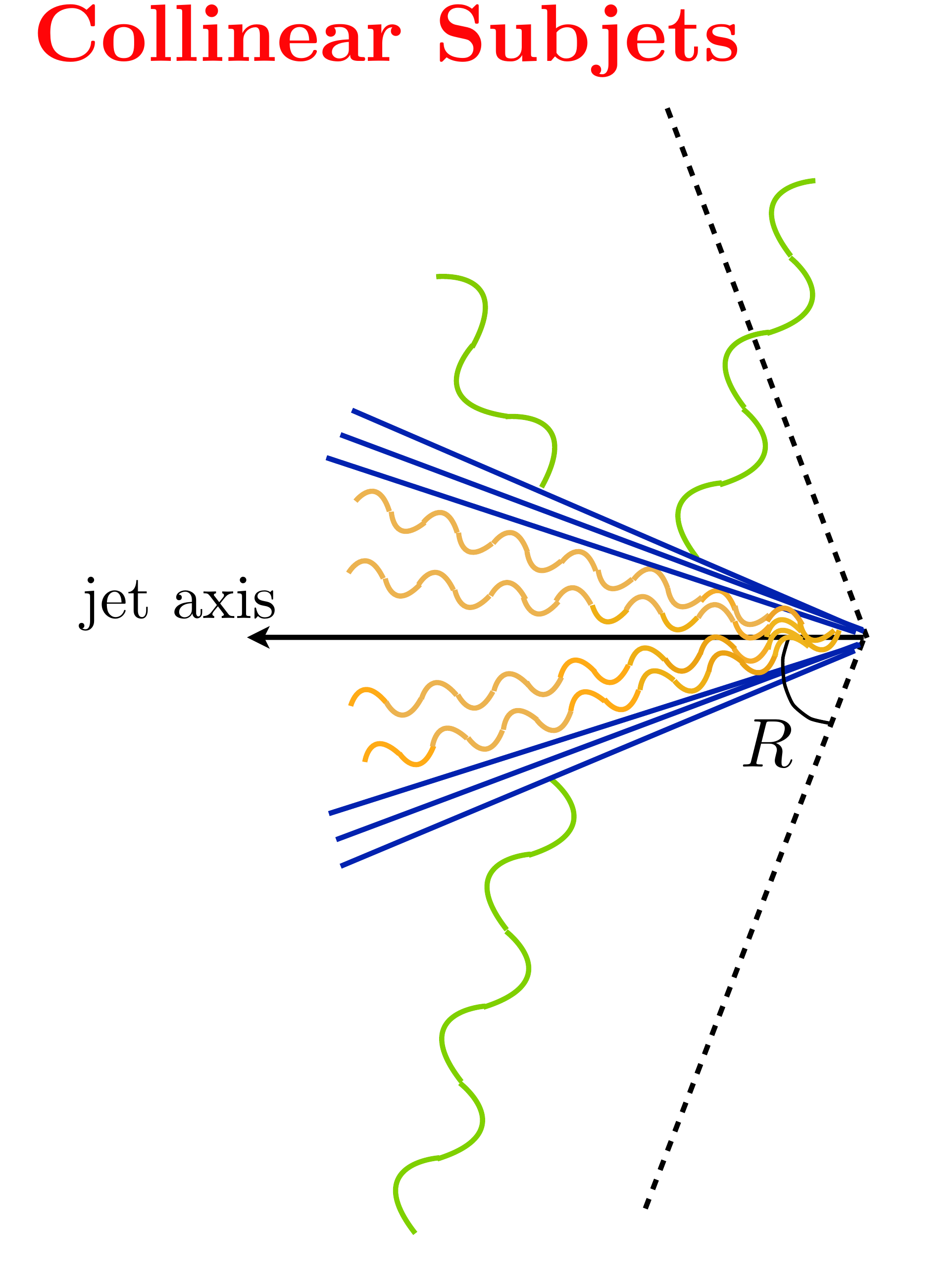}
}\qquad
\subfloat[]{\label{fig:soft_jet}
\includegraphics[width=3.9cm, trim =0 -0.75cm 0 0]{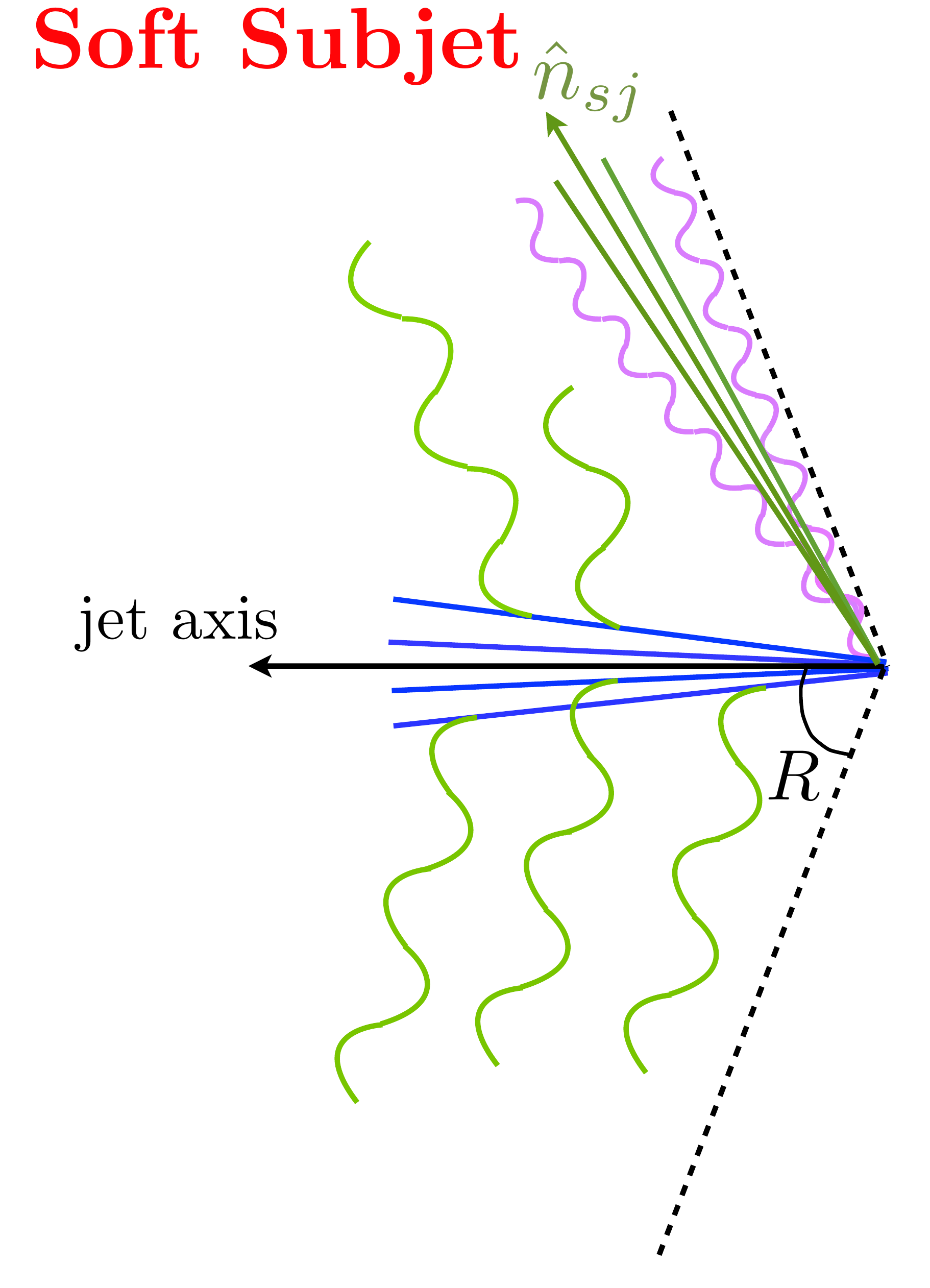}
}
\end{center}
\caption{Regions of interest for studying the two-prong substructure of a jet. (a) Soft haze region in which no subjets are resolved. (b) Collinear subjets with comparable energy and a small opening angle. (c) Soft subjet carrying a small fraction of the total energy, and at a wide angle from the hard subjet.}
\label{fig:diff_jets}
\end{figure}

In \Ref{Larkoski:2015kga}, three phase space regions were required to provide a description of the $D_2$ observable:
\begin{itemize}
\item Soft Haze: The jet does not have a resolved substructure, and is formed from unresolved soft and collinear radiation, as in \Fig{fig:soft_haze}. Here the factorization formula involves multi-differential jet and soft functions as developed in \Refs{Larkoski:2014tva,Procura:2014cba,Lustermans:2016nvk}.

\item Collinear Subjets: The jet is formed of two subjets with small opening angle, and large energies, as shown in \Fig{fig:ninja}. The factorization formula in this region of phase space is formulated in the SCET$_+$ theory of \Ref{Bauer:2011uc}. In addition to the standard soft and collinear modes of SCET, it involves collinear-soft modes emitted from the dipole formed by the two subjets. In this factorization formula, the modes describing the radiation within the subjets are not sensitive to the presence of the jet boundary.

\item Soft Subjet: The jet is formed of a single highly-energetic subjet, and a wide-angle subjet with energy fraction $z_{sj}\ll 1$, as shown in \Fig{fig:soft_jet}. The effective field theory description of this region of phase space was first presented in \Ref{Larkoski:2015zka}. Its complexity arises due to the fact that the soft subjet is sensitive to the presence of the jet boundary.

\end{itemize}
A smooth transition between the collinear subjets and soft subjet regions of phase space was achieved using a zero bin-like procedure to remove any overlap. A similar approach was advocated in \Ref{Pietrulewicz:2016nwo}.

\subsection{Groomed $D_2$ Factorization Formula}\label{sec:fact_D2}

Having reviewed the factorization formula for the soft dropped energy correlation functions, as well as for the $D_2$ observable, we can now combine these two approaches to provide a factorized description for groomed $D_2$. This will be accomplished by refactorizing an analogous parent effective theory to the expression \Eq{eq:sd_mother} in the different parametric regions:
\begin{align}\label{eq:sd_mother_D2}
\frac{d^4\sigma}{d\ecf{2,L}{\alpha}d\ecf{3,L}{\alpha} d\ecf{2,R}{\alpha}d\ecf{3,R}{\alpha} }= H(Q^2) S(\zcut) \left[ J(\ecf{2,L}{\alpha},\ecf{3,L}{\alpha},\zcut)  \right]\left[ J(\ecf{2,R}{\alpha},\ecf{3,R}{\alpha},\zcut)  \right]\,.
\end{align}

 The merging and region analysis will be similar to that performed in \Ref{Larkoski:2015kga} for $D_2$ without soft drop. Before giving a detailed discussion of each of the factorized expressions in the different phase space regions, we give a brief overview of the different regions of phase space that can contribute and the dynamics occurring in each region, as well as comparing them to the three phase space regions which contributed to ungroomed $D_2$, as shown in \Fig{fig:diff_jets}.

To describe the $D_2$ distribution of a jet on which the soft drop grooming algorithm has been applied, we will similarly need three regions of phase space. Note, however, that since all the factorizations will appear as refactorizations of the \Eq{eq:sd_mother_D2}, all components of the factorized expression which contribute to $D_2$ will be collinear in nature. This will significantly simplify the analysis. In particular, the wide-angle soft subjet region of phase space is completely removed from contributing to the observables by the soft drop algorithm. In the soft subjet region of phase space, we would have $\ecf{2}{\alpha}\sim z_{sj}$. However, by assumption, we take $\ecf{2}{\alpha}\ll \zcut$, and therefore, the wide angle soft subjet is removed by the soft drop algorithm. This region of phase space will instead be replaced by a collinear-soft subjet which has characteristic energy fraction $z_{cs}\sim \zcut$. The effective field theory description for this hierarchy is new, and will be described in \Sec{sec:resolved_soft}.

\begin{figure}
\begin{center}
\subfloat[]{\label{fig:collinear_subjets_c}
\includegraphics[width = 5cm]{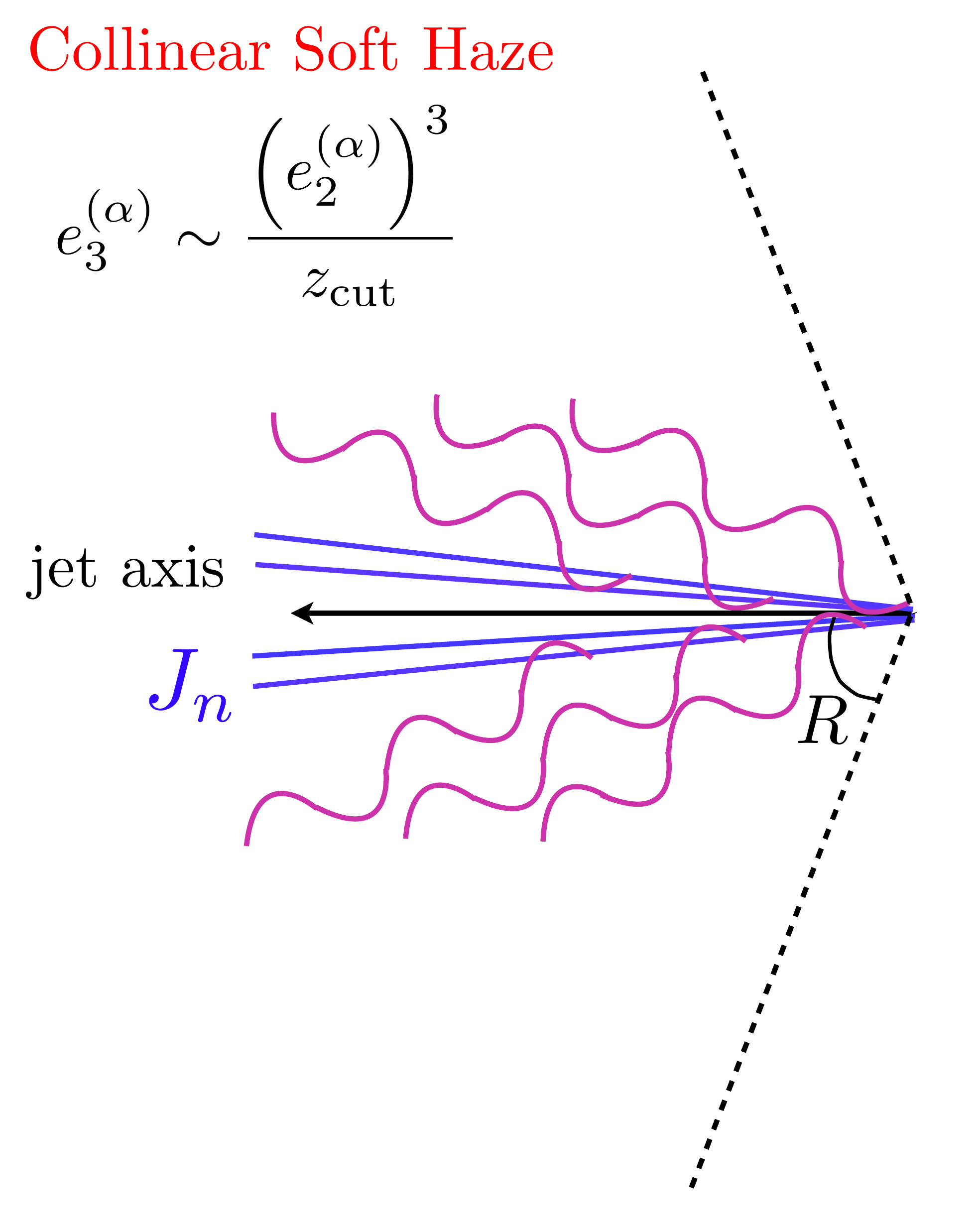}
}
\subfloat[]{\label{fig:collinear_subjets_a}
\includegraphics[width=5cm]{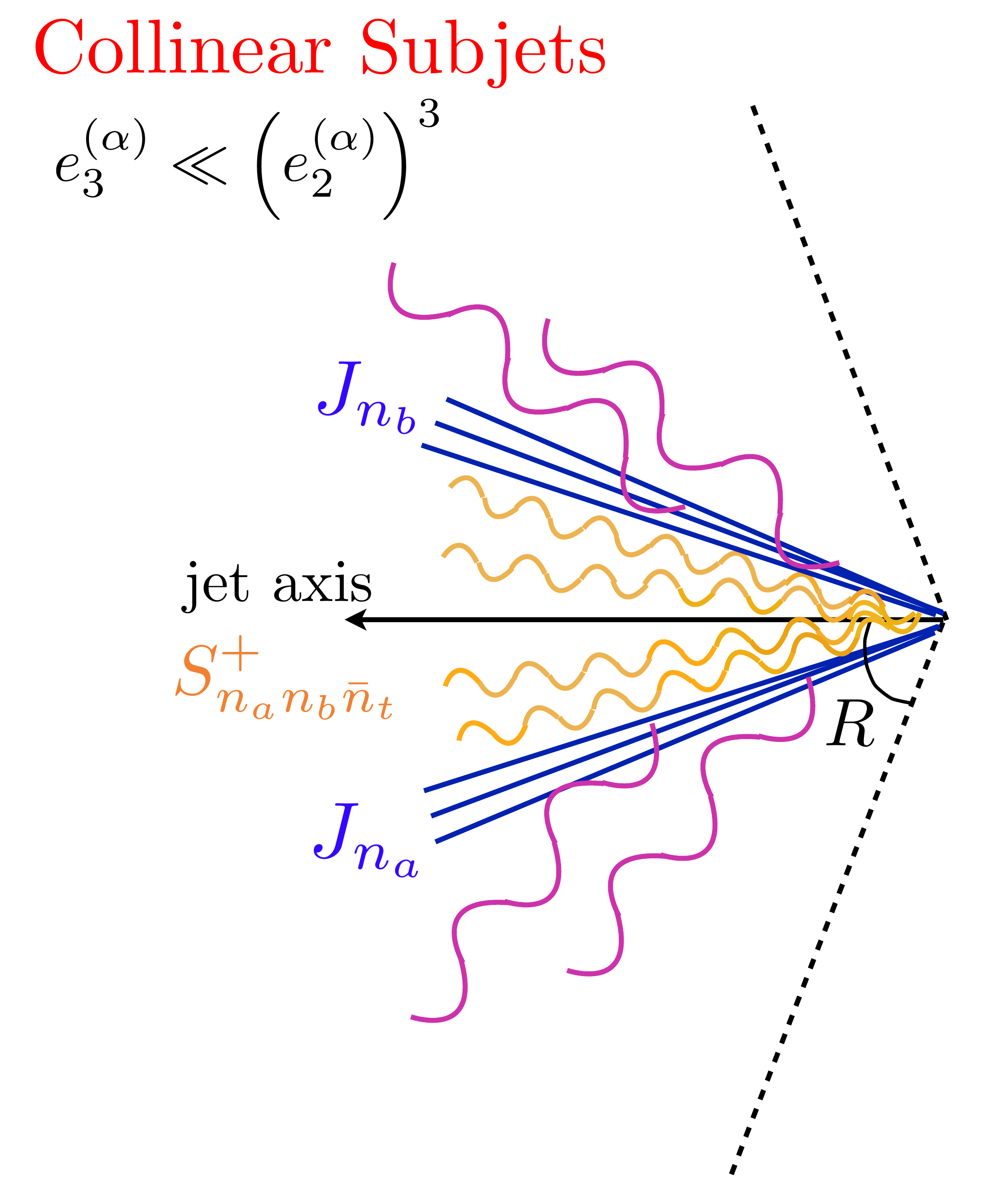}
}
\subfloat[]{\label{fig:collinear_subjets_b}
\includegraphics[width = 5cm]{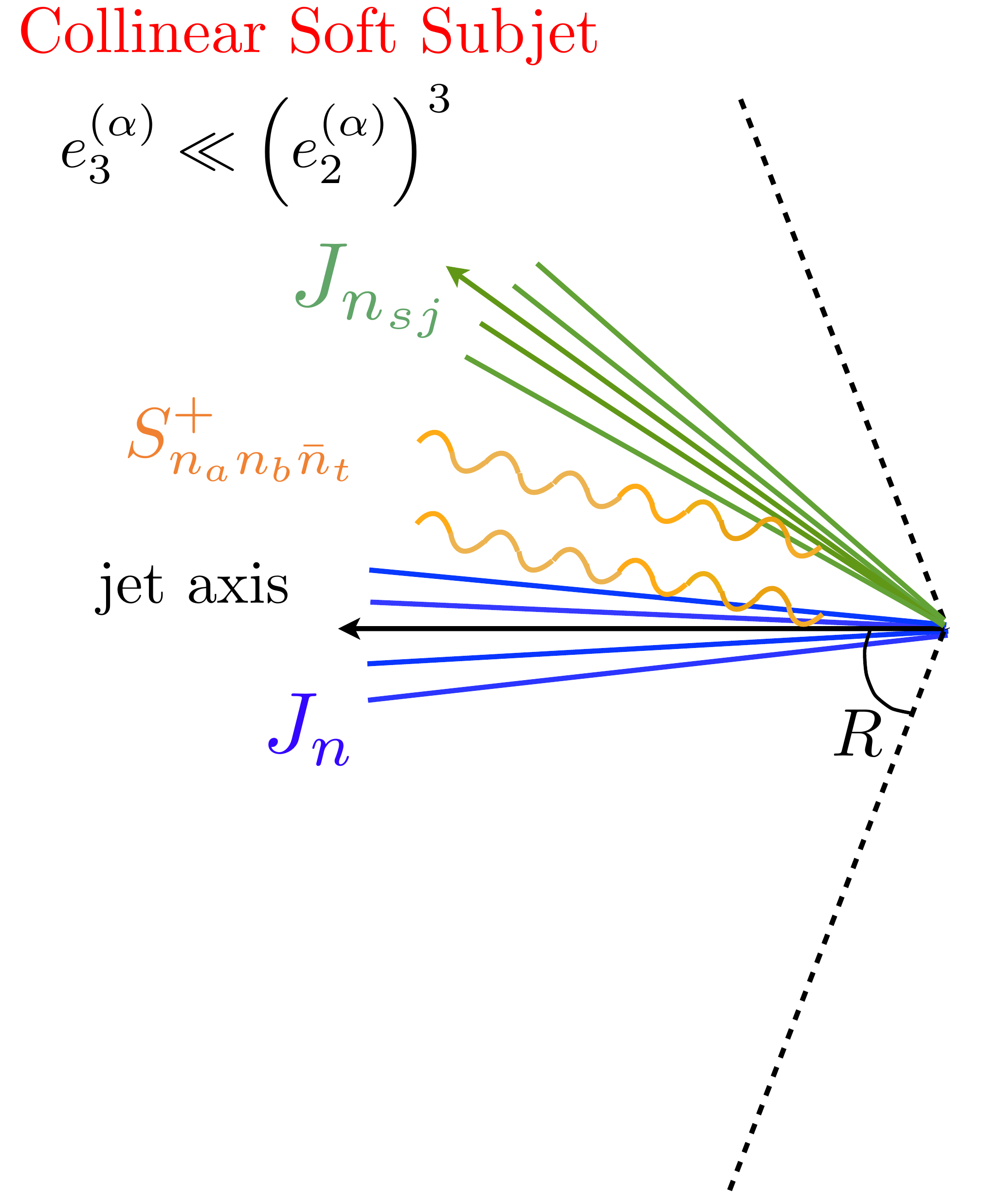}
}
\end{center}
\caption{Regions of interest for studying the two-prong substructure of a jet on which the soft drop grooming algorithm has been applied. (a) Collinear-soft haze region in which no subjets are resolved. (b) Collinear subjets with comparable energy and a small opening angle. (c) Collinear-soft subjet carrying a small fraction of the total energy, with $z_{cs} \sim \zcut$. 
}
\label{fig:subjets_summary}
\end{figure}

The three phase space regions that will contribute to the $D_2$ observable as measured on a soft dropped jet are shown schematically in \Fig{fig:subjets_summary}. A brief description of each of the different phase space regions is as follows:
\begin{itemize}

\item Collinear-Soft Haze: The jet does not have a resolved substructure. It is formed entirely from unresolved collinear-soft radiation. This is shown schematically in \Fig{fig:collinear_subjets_c}.

\item Collinear Subjets: As shown in \Fig{fig:collinear_subjets_a}, in the collinear subjets region of phase space, the jet consists of two subjets of approximately equal energies, and a small opening angle, surrounded by collinear-soft radiation.

\item Collinear-Soft Subjets: The jet is formed of two subjets, of parametrically different energies, with the softer jet energy set by $\zcut$, but where the opening angle between the jets is still assumed to be small. Unlike the previous two phase space regions, because $\zcut$ sets the energy of the soft jet, there is no additional collinear-soft radiation at a wider angle than the soft subjet. This is shown schematically in \Fig{fig:collinear_subjets_b}.

\end{itemize}

It is interesting to contrast the different phase space regions for the $D_2$ observable with and without the soft drop grooming algorithm applied, as shown in \Figs{fig:diff_jets}{fig:subjets_summary}. These configurations are similar, with the exception that the wide angle radiation is removed by the soft drop algorithm, so that only collinear-soft radiation remains. Importantly, this radiation is boosted along the direction of the jet. It is therefore not sensitive to the directions of other jets in the event, all of which appear boosted in the opposite direction, and it is also not sensitive to the radius of the jet. This will lead to a large degree of universality for the soft dropped $D_2$ distributions, and simplify their calculation in the presence of additional jets.

We now discuss each of the phase space regions in \Fig{fig:subjets_summary} in detail, and present factorization formulae describing the radiation in these different regions of phase space. These factorization formulae will allow for the radiation at each hierarchical scale to be described by a different function, allowing for large logarithms in the perturbative calculation to be resummed. A complete description of the groomed $D_2$ distribution can then be obtained by merging these different factorization formulae. We will discuss how this is done in \Secs{sec:fact_merge}{sec:fact_merge_res_unres}.

\subsubsection{Unresolved Substructure: Collinear-Soft Haze}\label{sec:fact_un}

\begin{figure}
\begin{center}
\includegraphics[width = 10cm]{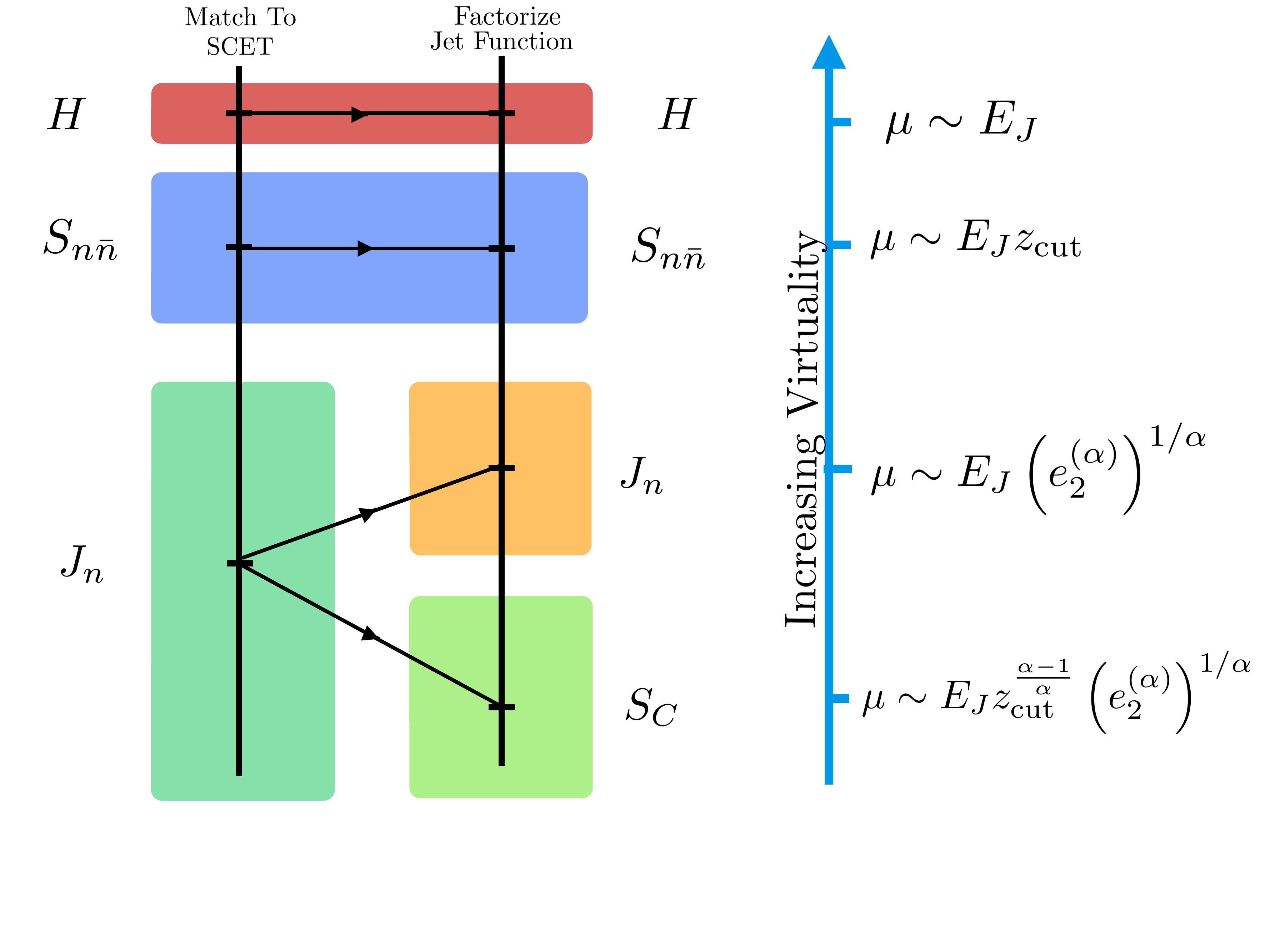}
\end{center}
\caption{An illustration of the multistage matching procedure and relevant scales for the collinear-soft haze region of phase space.
}
\label{fig:csoft_haze_setup}
\end{figure}

We begin by discussing the factorization in the region of phase space where the jet has no resolved subjets. The factorization formula in this region of phase space will follow almost identically the soft-haze factorization formula of \Ref{Larkoski:2015kga}, except that the soft function will be replaced by a boosted collinear-soft function due to the implementation of the soft drop algorithm. Recall that in this region of phase space we have collinear modes and collinear-soft modes, and the power counting for the observables is
\begin{align}
\ecf{2}{\alpha} &\sim \zcut \theta_{sc}^\alpha +\theta_{cc}^\alpha\,, \nn \\
\ecf{3}{\alpha} &\sim \theta_{cc}^\alpha \theta_{sc}^{2\alpha} \zcut+\theta_{sc}^{3\alpha} \zcut^2\,.
\end{align}
In this region of phase space, $\ecf{3}{\alpha}$ can be expressed as the sum of two contributions.  One is just the three-point energy correlation function $\ecf{3}{\alpha}$ measured in the collinear-soft function and the other is a product of two-point energy correlation functions with different exponents as measured in the jet function and the collinear-soft function.  The factorization formula in this region of phase space is then given by
\begin{align}\label{eq:fact_haze}
\hspace{-0.25cm}
\frac{d^2\sigma}{d\ecf{2}{\alpha} d\ecf{3}{\alpha}}&= H S(\zcut) \int de_2^c \, de_2^{sc}\, de_2^{(2\alpha) sc}\, de_3^{sc}\, \delta(\ecf{3}{\alpha}-e_3^{sc}-e_2^c\cdot  e_2^{(2\alpha) sc}  )  \delta(\ecf{2}{\alpha}-e_2^{sc}-e_2^c) \nonumber\\
& 
\hspace{7cm}
\times J(e_2^c) S_{sc}(e_2^{sc},e_2^{(2\alpha) sc}, e_3^{sc}, \zcut)\,. 
\end{align}
For brevity, we only write the $\ecf{2}{\alpha}$ and $\ecf{3}{\alpha}$ dependence of a single hemisphere; including both hemispheres is trivial.  Since the $\ecf{3}{\alpha}$ observable is first non-zero with two emissions, this factorization formula first gives a non-trivial contribution at NNLL$'$ order, i.e., it requires the two-loop matrix elements (and the product of two one-loop matrix elements).

A brief description of the different functions entering the factorization formula in the collinear soft haze region is as follows
\begin{itemize}
\item $H(Q^2)$  is the hard function describing the underlying hard process, namely $e^+e^- \to q\bar q$.

\item $S(\zcut)$ is the global soft function, describing radiation which has been removed by the soft drop procedure. It depends only on $\zcut$, and not on the observables $\ecf{2}{\alpha}$, or $\ecf{3}{\alpha}$.

\item $J(e_2^c)$ describes the collinear dynamics within the jet. It is independent of the soft drop algorithm. It contributes to the $\ecf{3}{\alpha}$ observable only through the product form entering \Eq{eq:fact_haze}.

\item $S_{sc}(e_2^{sc},e_2^{(2\alpha) sc}, e_3^{sc}, \zcut)$ describes the soft collinear radiation within the jet. It is sensitive both to the soft drop criterion as well as contributing to the $\ecf{2}{\alpha}$, $\ecf{2}{2\alpha}$, and $\ecf{3}{\alpha}$ observables.  Here, $\ecf{2}{2\alpha}$ means that the angular exponent is $2\alpha$.

\end{itemize}

A similar factorization formula was proposed in \Ref{Larkoski:2015kga} for describing the unresolved region of phase space for the $D_2$ observable without the soft drop algorithm. This region also first contributed to the observable at NNLL$'$ order, and was therefore not considered. This was because the endpoint of the ungroomed distribution is $1/\ecf{2}{\alpha}\gg1$, and therefore the distribution has a smooth long tail, which can be well-approximated by simply extending the factorization formulae from the two-prong region of phase space. However, in the case that the soft drop algorithm is applied, it was shown in \Sec{sec:ps_struc} that the $D_2$ distribution has an upper boundary at $D_2^{\max} =1/(2\zcut)$. This endpoint feature is not described by the factorization formulae in the two-prong region of phase space, as it is expanded away. Matrix elements in the collinear-soft haze region of phase space are required to describe this kinematic feature. We will therefore compute the fixed-order matrix elements at $\mathcal{O}(\alpha_s^2)$ and match within the effective theory.

The most convienent way to calculate the $D_2$ distribution in the soft haze region to $\mathcal{O}(\alpha_s^2)$ is to integrate the appropriate $1\to3$ splitting functions, as described in \App{app:1to3}. This is equivalent to calculating the $D_2$ distribution in the parent theory of \Eq{eq:sd_mother_D2}. One can explicitly check that one reproduces the matrix elements of the collinear-soft haze factorization when two of the emissions in the $1\to3$ splitting functions are taken to be soft, and when two are taken to be collinear and one is soft. These contributions reproduce the two-loop soft function, and the convolution between the one-loop jet and one-loop soft functions within the factorization formula of \Eq{eq:fact_haze}.

A critical feature of the  $1\to3$ splitting functions, as shown in \App{app:1to3}, and by extension, also the collinear-soft haze factorization formula, is that at $\mathcal{O}(\alpha_s^2)$ all the $\ecf{2}{\alpha}$ dependence explicitly scales out of the matrix element when we scale $\ecf{3}{\alpha}$ to the $D_2$ ratio. Thus the $\ecf{2}{\alpha}$ dependence merely becomes a multiplicative factor to the shape of the $D_2$ distribution in the collinear-soft haze region. This implies to N$^3$LL logarithmic counting in the $\ecf{2}{\alpha}$ logarithms, that the $\ecf{2}{\alpha}$ spectrum is simply \emph{multiplicative} to the normalized $D_2$ distribution. This is consistent with the arguments given in \Sec{sec:ps_struc} about the endpoints of the groomed and ungroomed $D_2$ distributions. The endpoint of the ungroomed $D_2$ distribution is set by the value of $\ecf{2}{\alpha}$ at fixed order, so the functional dependence of the ungroomed $D_2$ spectrum is highly nontrivial. One would have to convolve the Sudakov resummation of the $\ecf{2}{\alpha}$ spectrum with the ungroomed $D_2$ distribution as a function of $\ecf{2}{\alpha}$ in order to accurately describe even the \emph{normalized} endpoint of the the ungroomed $D_2$ distribution. The grooming procedure decouples the shape of the endpoint from the value of $\ecf{2}{\alpha}$, significantly simplifying the calculation of the $D_2$ distribution at large values of $D_2$. We explain in more detail the importance of these observations when considering the matching between resolved and unresolved limits in \Sec{sec:fact_merge_res_unres}.

As a check of the splitting function integration, we also compute the $D_2$ distribution with \eventtwo~\cite{Catani:1996vz} and then match to the factorization formulae for the two-prong phase space regions. 

\subsubsection{Resolved Substructure: Collinear Limit}\label{sec:resolved}

Here, we will determine the factorization formula in the limit when the jet has two relatively hard collinear subjets.  To derive this factorization formula, we must return to the parent theory of \Eq{eq:sd_mother_D2} (for brevity, we just focus on one hemisphere):
\begin{align}
\frac{d\sigma}{d\ecf{2}{\alpha}d\ecf{3}{\alpha} }= H(Q^2) S(\zcut)  J\big(\ecf{2}{\alpha},\ecf{3}{\alpha},\zcut\big) \,.
\end{align}
Now, on this soft dropped jet on which we have measured $\ecf{2}{\alpha}$, we additionally measure $\ecf{3}{\alpha}$, with the assumption that $\ecf{3}{\alpha}\ll (\ecf{2}{\alpha})^3$.  In this limit, and using the mode decomposition outlined in \Ref{Larkoski:2015kga}, we can factorize the jet function into a hard, collinear splitting:
\begin{equation}
J\big(\ecf{2}{\alpha},\ecf{3}{\alpha},\zcut\big) \to H_2(z,\ecf{2}{\alpha})J_1(\ecf{3}{\alpha})\otimes J_2(\ecf{3}{\alpha})\otimes C_s(\ecf{3}{\alpha},\zcut)\,.
\end{equation}
Here, $z$ is the momentum fraction of one of the subjets, and $H_2(z,\ecf{2}{\alpha})$ is a function that depends on $\ecf{2}{\alpha}$ that describes the hard, collinear splitting.\footnote{While the momentum fraction of the subjets is not well-defined in the unresolved region, we may use a combination of energy correlation functions with different angular exponents to give a definition to $z$ outside the two prong region; see \Ref{Larkoski:2015kga}.}  $J_1(\ecf{3}{\alpha})$ and $J_2(\ecf{3}{\alpha})$ are the jet functions that describe the collinear radiation off of the two hard prongs in the splitting.  $C_s(\ecf{3}{\alpha},\zcut)$ is the collinear-soft function that describes relatively soft radiation emitted off of the dipole formed by the two hard prongs. In contrast to the ungroomed $D_2$ distribution, there is no global soft contribution (and thus for $e^+e^-$, no two-eikonal line soft function depending on $\ecf{3}{\alpha}$), as the jet has already been isolated by the grooming procedure. The factorization formula in the two-prong collinear limit is then:
\begin{align}
\frac{d^3\sigma}{dz\, d\ecf{2}{\alpha}\, d\ecf{3}{\alpha} }= H(Q^2) S(\zcut) H_2(z,\ecf{2}{\alpha})J_1(\ecf{3}{\alpha})\otimes J_2(\ecf{3}{\alpha})\otimes C_s(\ecf{3}{\alpha},\zcut) \,.
\end{align}
We could stop with this factorization, and begin calculating the resummation of $D_2$; however, it is worthwhile to further analyze the structure of the collinear-soft function.

\begin{figure}
\begin{center}
\includegraphics[width = 10cm]{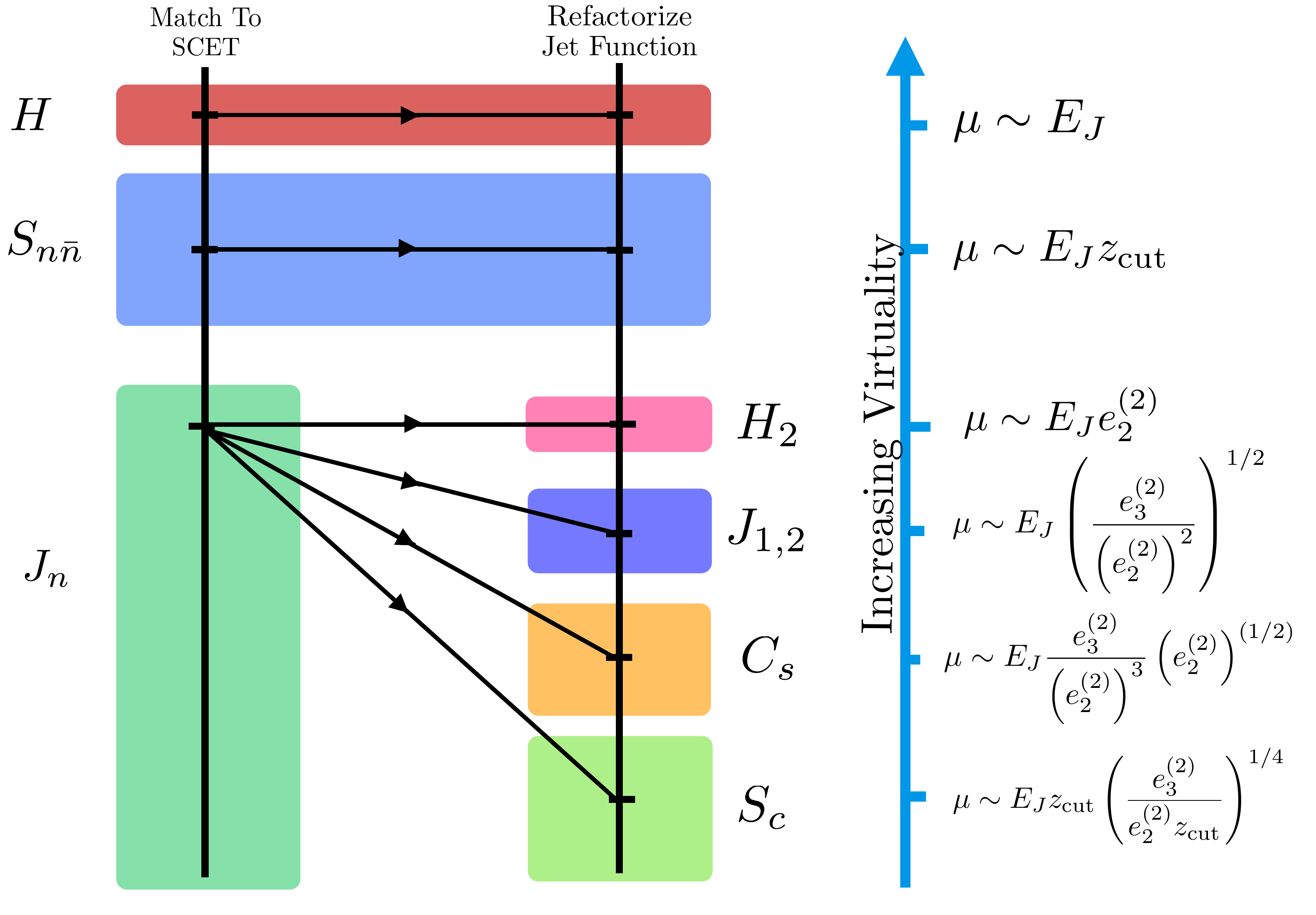}
\end{center}
\caption{A schematic of the multistage matching procedure in the collinear limit of resolved substructure. The function incorporating non-global collinear effects is not shown, but is discussed in the text.
}
\label{fig:collinear_subjets_setup}
\end{figure}

As $\ecf{3}{\alpha}\rightarrow 0$, this forces the energy of the soft modes within $C_s(\ecf{3}{\alpha},\zcut)$ to zero. All emissions generated off of the eikonal lines within the collinear-soft function can only contribute to the observable by being clustered with one of the legs of the hard prongs, before the legs themselves are clustered together. Otherwise, the emission will be at too low an energy scale and too wide of an angle to be included in the groomed jet.  Correspondingly, emissions that do contribute to $\ecf{3}{\alpha}$ from $C_s(\ecf{3}{\alpha})$ cannot be emitted at too wide of an angle.  If these emissions are not first clustered with one of the two hard prongs, then they are necessarily groomed away.  Therefore, we seperate out two angular regions of the collinear soft function and write:
\begin{equation}\label{eq:collinear_soft_mother_f}
C_s(\ecf{3}{\alpha},\zcut) \to C_s(\ecf{3}{\alpha},\theta<\theta_{ab})S_{c}(\zcut,\theta>\theta_{ab})\otimes C_s^{\text{NG}}(\ecf{3}{\alpha},\zcut) \,.
\end{equation}
Again, we emphasize that the constraint $\theta<\theta_{ab}$ or $\theta>\theta_{ab}$ is schematic; the precise constraint will depend on the detailed clustering history, however it is purely geometrical in its implementation. The last function is independent of the renormalization group, and encodes \emph{local-to-the-jet} non-global correlations. The presence of the hard splitting with an opening angle set by $\ecf{2}{\alpha}$ implies that an effective jet area is created within the two prong region. This will lead to non-global correlations between emissions that are groomed away, but emit into this opening angle, and the emissions which come off of the primary hard legs.

\begin{figure}
\begin{center}
\includegraphics[width = 4cm]{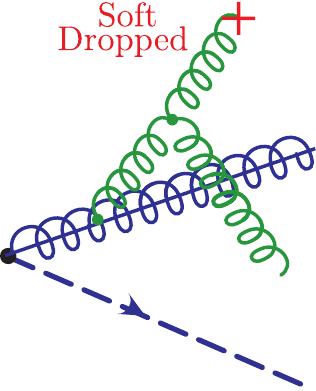}
\end{center}
\caption{
Schematic illustration of a configuration of emissions that contributes to collinear non-global logarithms.  The soft-collinear gluon emission off of the hard-collinear gluon is groomed away, but not before it re-emitted into the ungroomed region between the two hard prongs.
}
\label{fig:collinear_NGLs}
\end{figure}

An illustration of the origin of these non-global logarithms (NGLs) is illustrated in \Fig{fig:collinear_NGLs}.  In this figure, a hard collinear quark and collinear gluon (denoted by the curly curve with a line through it) sets the mass of the jet, and then their dipole emits a soft-collinear gluon.  This soft-collinear gluon has sufficiently low energy and fails soft drop, but re-emits another soft-collinear gluon that is clustered into the hard collinear particles.  Such a re-emission is non-global in origin, as it is simultaneously sensitive to the infrared scales $\zcut$ and $\ecf{3}{\alpha}$.  However, all particles in this picture are collinear, as the jet was already isolated from the rest of the event in the first stage of matching. Therefore the resulting NGLs only depend on the fact that the jet was initiated by a hard quark (in general, they depend on the flavor structure of the splitting).  Because of this universality, these non-global logarithms are significantly less worrying than more standard NGLs that occur in (ungroomed) jet mass distributions. Techniques have been developed for systematic calculation of NGLs \cite{Caron-Huot:2015bja,Larkoski:2015zka,Neill:2015nya,Becher:2015hka,Becher:2016mmh,Larkoski:2016zzc,Neill:2016stq,Becher:2016omr,Becher:2017nof}, and these NGLs associated with the soft drop procedure have interesting features not previously encountered due to the clustering history.  We have performed some preliminary estimations of these NGLs, and find their numerical effect is small, well within our uncertainties for the purely global (Sudakov) contributions. At leading logarithmic accuracy in the large-$N_c$ limit, they can be computed using an extension of the Monte Carlo algorithm of Dasgupta and Salam \cite{Dasgupta:2001sh}, which is described in \App{sec:NGL_alg}. 

With these replacements, the factorization formula now becomes
{\small\begin{align}\label{eq:ninjafact}
\frac{d^3\sigma}{dz\, d\ecf{2}{\alpha}\, d\ecf{3}{\alpha} }= H(Q^2) S(\zcut) &S_c(\zcut,\theta>\theta_{ab}) H_2(\ecf{2}{\alpha})\nonumber\\
&\qquad J_1(\ecf{3}{\alpha})\otimes J_2(\ecf{3}{\alpha})\otimes C_s(\ecf{3}{\alpha},\theta<\theta_{ab}) \otimes C_s^{\text{NG}}(\ecf{3}{\alpha},\zcut)\,,
\end{align}}
where the functions are as follows
 \begin{itemize}
 \item $H(Q^2)$ is the hard function, in this case for $e^+e^-\to$ dijets.
 \item $S(\zcut)$ is the soft function describing wide angle soft radiation which has been soft dropped.
 \item $S_c(\zcut,\theta>\theta_{ab})$ describes collinear soft radiation at $\theta>\theta_{ab}$.
 \item $H_2(\ecf{2}{\alpha})$ is a hard function describing the production of the two collinear subjets.
 \item $J_{1,2}(\ecf{3}{\alpha})$ are the jet functions for the collinear subjets.
 \item $C_s(\ecf{3}{\alpha},\theta<\theta_{ab})$ describes collinear soft radiation emitted from the dipole formed from the subjets at $\theta<\theta_{ab}$.
 \item $C_s^{\text{NG}}(\ecf{3}{\alpha},\zcut)$ describes the entanglement between the groomed soft-collinear emissions and the two-prong region.
 \end{itemize}
The calculations of the functions in this factorization formula to one-loop accuracy are presented in \App{app:collinear_sub}.  There, we also demonstrate the consistency of this factorization formula by showing that the sum of anomalous dimensions is indeed 0.

\subsubsection{Resolved Substructure: Soft Limit}\label{sec:resolved_soft}

When the jet has two subjets whose energies are hierarchically separated we can determine the form of the appropriate factorization formula in the same manner as in the previous section. As in that case, we start with the parent jet function from \Eq{eq:sd_mother_D2}:
\begin{align}
J\big(\ecf{2}{\alpha},d\ecf{3}{\alpha},\zcut\big) \to HC_{s2}(z,\ecf{2}{\alpha},\ecf{3}{\alpha},\zcut)\otimes J_{sc}(\ecf{3}{\alpha})\otimes J_2(\ecf{3}{\alpha})\,.
\end{align}
This preliminary factorization has removed the collinear subjet contributions, but has not distentangled all the soft scales. This requires a matching procedure that cannot be implemented at the level of the amplitude, but must be performed at the amplitude-squared.\footnote{This is similar in spirit to \Ref{Neill:2015nya}.} The matching procedure is complicated by essentially the same physics that determines the NGLs encountered in the collinear-subjets of the resolved region, but now we must take the energy scale of one of the legs to be just above $\zcut$, and hence not parametrically separated from the soft-collinear emissions which are groomed away just below $\zcut$. Thus we write
\begin{align}\label{eq:soft_jet_factorization}
 HC_{s2}(z,\ecf{2}{\alpha},\ecf{3}{\alpha},\zcut)\to H_{2}^{sj}(z,\ecf{2}{\alpha},\zcut)C_s(\ecf{3}{\alpha},\theta<\theta_{ab})\otimes C_s^{sj-\text{NG}}(\ecf{3}{\alpha},\zcut)\,.
\end{align}
Where $C_s^{sj-\text{NG}}(\ecf{3}{\alpha},\zcut)$ denotes hard matching contributions where additional Wilson lines and jet functions are introduced to capture the non-global correlations. The function $C_s(\ecf{3}{\alpha},\theta<\theta_{ab})$ is the same as found in \Eq{eq:ninjafact}. As it stands, this factorization is sufficient to resum all large global (Sudakov) logarithms, and to leading logarithmic accuracy in the NGLs, the function $C_s^{sj-\text{NG}}(\ecf{3}{\alpha},\zcut)$ is identical to that found in \Eq{eq:ninjafact}\footnote{At higher orders we would have to keep track of the color correlations between multiple directions at the soft subjet scale being integrated out and groomed away, and the soft emissions into the opening angle of the $1\rightarrow 2$ splitting.}.

\begin{figure}
\begin{center}
\subfloat[]{\label{fig:soft_subjet_match_abelian}
\includegraphics[width = 5.2cm]{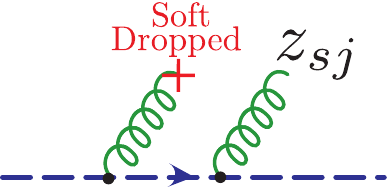}
}
\subfloat[]{\label{fig:soft_subjet_match_nonabelian}
\includegraphics[width=5cm]{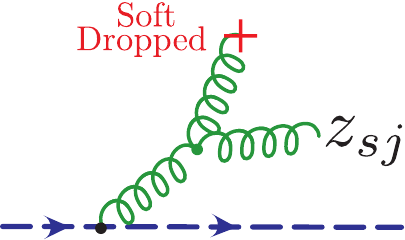}
}
\subfloat[]{\label{fig:soft_subjet_match_quarks}
\includegraphics[width = 5cm]{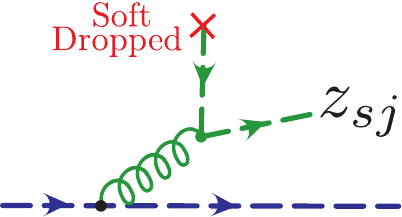}
}
\end{center}
\caption{
Structure of emissions at one-loop order that contribute to the hard matching function in the soft subjet factorization formula for groomed jets.  The emission at a wider angle will be groomed away, while the other emission sets the mass of the jet.  (a) corresponds to the Abelian emission of gluons, and will contribute proportional to $C_F$; (b) corresponds to non-Abelian gluon emissions, and will contribute proportional to $C_A$; and (c) corresponds to a gluon splitting to quarks, which contributes proportional to $n_F$.
}
\label{fig:soft_subjet_matching}
\end{figure}

There are a few things to note about this factorization formula.  First, there is a jet function $J_{sc}(\ecf{3}{\alpha})$ that describes collinear radiation off of the soft subjet in the larger jet.  To leading power, this subjet is always a gluon and is identical to the $z\to 0$ limit of the corresponding gluon jet function in the factorization formula in the case of hard, collinear subjets.  Additionally, there is the identical collinear-soft function as in the hard collinear subjets factorization formula.  Because we assume that $\Dobs{2}{\alpha}\ll \zcut$, emissions that set the value of $\Dobs{2}{\alpha}$ must be at parametrically lower energies than either of the subjets.  Therefore, the soft drop constraint on these emissions is just a geometric constraint that enforces the emissions to first cluster with one of the hard subjets.  This geometrical constraint is necessarily independent of the energy of the subjets of the larger jet, and therefore this collinear-soft function is identical to that which appears in \Eq{eq:ninjafact}.

The novel part of this factorization formula is the hard matching function, $H_2^{sj}(z,\ecf{2}{\alpha},\zcut)$ that describes the production of the soft subjet.  This function has now two contributions relevant for a NLL resummation, or one-loop calculation.  First, there are the standard virtual contributions, which just correspond to the $z\to 0$ limit of the corresponding matching coefficient in \Eq{eq:ninjafact}.  There is, however, a new contribution to the matching function in this factorization formula.  Because we apply soft drop, it is possible that there is an initial emission in the jet that fails soft drop, and so does not seed the production of a soft subjet.  However, a secondary emission could then pass soft drop, and produce the soft subjet. These different configurations are shown in \Fig{fig:soft_subjet_matching}.

The calculation of this two emission contribution to the hard matching function $H_2^{sj}(z,\ecf{2}{\alpha},\zcut)$ is presented in \App{app:hardfunc_sd}, but we will describe its features here.   We must consider all possible pairs of soft emissions which are reclustered in such a way that the first angular-ordered emission fails soft drop, while the second passes.  This is the reason why we explicitly show the $\zcut$ dependence in this function.  Note that the constraint that one emission fails soft drop while the other passes eliminates the collinear singularity when the emissions become close in angle.  If the two emissions are sufficiently close in angle compared to their collective angle to the hard jet core, then they will be clustered together first, which is forbidden by assumption.  This implies that the contribution to this hard function from the emission of a soft quark--anti-quark pair does not contribute to NLL order.  The emission of soft gluons will contribute at NLL order. \Fig{fig:soft_subjet_matching} shows a schematic picture of these two-emission contributions to the hard matching function.

We therefore find that the complete factorization formula for a soft dropped groomed jet with a soft subjet is
{\small\begin{align}\label{eq:sc_sj_fact}
\frac{d^3\sigma}{dz d\ecf{2}{\alpha} \, d\ecf{3}{\alpha}}= H(Q^2) S(\zcut) H_2^{sj}(z,\ecf{2}{\alpha},\zcut)  C_s(\ecf{3}{\alpha},\theta<\theta_{ab}) \otimes J_{sc}(\ecf{3}{\alpha})\otimes J(\ecf{3}{\alpha})\otimes C_s^{sj,\text{NG}}(\ecf{3}{\alpha},\zcut) \,.
\end{align}}
\begin{figure}
\begin{center}
\includegraphics[width = 10cm]{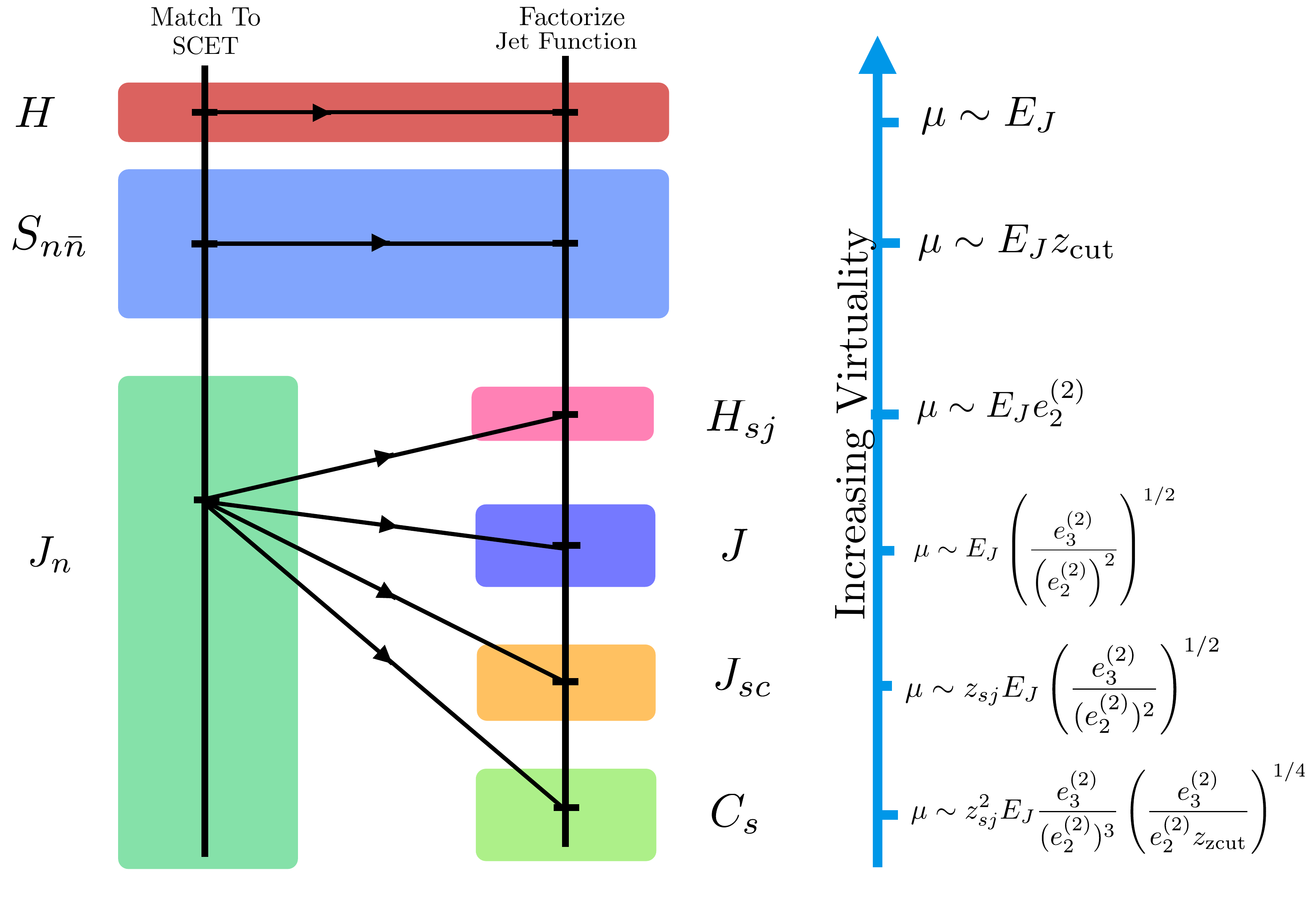}
\end{center}
\caption{A schematic of the multistage matching procedure in the soft-collinear subjet region of phase space.  The function incorporating non-global collinear effects is not shown, but is discussed in the text.
}
\label{fig:csoft_subjet_setup}
\end{figure}
A brief review of the different functions entering the factorization is as follows
 \begin{itemize}
 \item $H(Q^2)$ is the hard function, in this case for $e^+e^-\to$ dijets.
 \item $H_2^{sj}(z,\ecf{2}{\alpha},\zcut) $ is a hard function describing the production of the soft subjet.
 \item  $J(\ecf{3}{\alpha})$ is the jet function for the energetic jet.
 \item $J_{sc}(\ecf{3}{\alpha})$ is the jet function for the soft jet.
 \item $C_s(\ecf{3}{\alpha},\theta<\theta_{ab})$ is the collinear-soft function describing the radiation entering the dipole off of the primary eikonal lines.
 \item $C_s^{sj,\text{NG}}(\ecf{3}{\alpha},\zcut)$ describes the non-global correlations arising from groomed soft-collinear emissions.
 \item $S(\zcut)$ is the soft function describing wide angle soft radiation which has been soft dropped.
 \end{itemize}
For future use, we record the virtualities of the different modes, which are also shown in \Fig{fig:csoft_subjet_setup}:
\begin{align}
\mu_{sj} \sim z_{sj} E_J   \left(  \frac{ \ecf{3}{\alpha}  }{  ( \ecf{2}{\alpha} )^2}  \right)^{1/\alpha}\,, \qquad \mu_{cs} \sim z_{sj}^2 E_J \frac{\ecf{3}{\alpha}  }{  (\ecf{2}{\alpha} )^3   }   \left( \frac{\ecf{3}{\alpha}}{\ecf{2}{\alpha}  \zcut}    \right)^{1/(2\alpha)}\,.
\end{align}
The scalings of these modes will play an important role when studying the behavior of non-perturbative power corrections.

\subsubsection{Merging Collinear and Soft Resolved Limits}\label{sec:fact_merge}

To perform a complete calculation, we must merge our description of the different resolved regions.
We merge between the soft-collinear subjet and collinear subjet region by subtracting their overlap. This gives:
\begin{align}
\frac{d^3\sigma}{dz\,d\ecf{2}{\alpha} \, d\ecf{3}{\alpha}}&= H(Q^2) S(\zcut) \Bigg[S_c(\zcut,\theta>\theta_{ab})C_s^{\text{NG}}(\ecf{3}{\alpha},\zcut)\Big[ H_2(z,\ecf{2}{\alpha})- H_2(\ecf{2}{\alpha},z\to \zcut)\Big]\nonumber\\
&
\hspace{-1.2cm}
+ H_2^{sj}(z,\ecf{2}{\alpha},\zcut)C_s^{sj,\text{NG}}(\ecf{3}{\alpha},\zcut)  \Bigg]\otimes C_s(\ecf{3}{\alpha},\theta<\theta_{ab}) \otimes J_{sc}(\ecf{3}{\alpha})\otimes J(\ecf{3}{\alpha}) \,.
\end{align}
However, to NLL accuracy, including NGLs, we can show that the collinear factorization suffices to capture all large logarithms with the appropriate scale setting. First we note that the tree-level results of the subtracted hard matching of the collinear factorization and the soft-collinear subjet agree. Then all one must check is that the resummation in the collinear sector arising from running $S_c(\zcut,\theta>\theta_{ab})$ naturally merges with the resummation in the soft-collinear sector of the function $H_2^{sj}(z,\ecf{2}{\alpha},\zcut)$. For this to happen, the natural renormalization scales of product:
\begin{align}
S_c(\zcut,\theta>\theta_{ab}) H_2(z,\ecf{2}{\alpha})\,,
\end{align} 
must merge to the natural renormalization scale found in $H_2^{sj}(z,\ecf{2}{\alpha},\zcut)$ when $z\rightarrow \zcut$. This is accomplished so long as we use the transverse momentum of the collinear splitting as the renormalization scale for the collinear hard splitting function. We then compare the scales take from \App{app:scales_resummation} (for simplicity, we take $\alpha =2$):
\begin{align}
\mu_{H_2}^2&=\frac{z(1-z)\ecf{2}{2}Q^2}{4}\,,\\
\mu_{S_c}^2&=\frac{z_{\text{cut}}^2\ecf{2}{2}Q^2}{4z(1-z)}\,.
\end{align}
In the limit $z\rightarrow z_{\text{cut}}$, we have:
\begin{align}
\mu_{H_2}^2\rightarrow \mu_{S_c}^2\,,
\end{align}
showing that the two scales merge.

Finally, we note that the sum of the anomalous dimensions in \Eq{eq:collinear_subjets_hard_matching_anom_dim} gives \Eq{eq:collinear_soft_subjet_hard_matching_anom_dim} in the limit $1-z_q\rightarrow z_{\text{cut}}$, that is, the collinear subjets approach the soft-collinear region. That this must be the case stems from the purely geometrical character of the soft drop constraint. Regardless of the relative energy scales between the emissions that sets the $1\rightarrow 2$ splitting, and the emission which fails soft drop, once all additional emissions are required to fail on their own, whether or not it can contribute to $D_2$ depends on whether it is clustered into the hard splitting, that is, the angular structure of the emissions. Indeed, we exploit this fact to simplifiy the calculation of the collinear-soft subjet matching presented in \App{app:csoft_sub}. Thus to NLL accuracy, the merging of the collinear and soft resolved limits is accomplished by simply running the splitting scale to the transverse momentum of the collinear splitting that sets $\ecf{2}{2}$, and only using the collinear factorization formula for the resummation. Note that the analogous simplification could not be made in the case of the ungroomed $D_2$ distribution, mainly due to the presence of boundary soft modes in the soft-subjet factorization.

\subsubsection{Matching Resolved and Unresolved Limits}\label{sec:fact_merge_res_unres}

In this section we discuss how the factorization formulae in the resolved and unresolved limits can be merged to provide a complete description of the entire $D_2$ distribution. We consider two distinct merging schemes: one using profile functions \cite{Ligeti:2008ac,Abbate:2010xh} to turn off the resummation at the endpoint of the distribution, and a second using only canonical scales for the resummation at all values of $D_2$, never turning off the resummation. We scale set at the level of the cumulative distribution, and take the derivative for the differential cross section. When using profiles, we retain all the logarithms of $\mu$ over the natural scale of the function in the matching, jet, and soft-collinear/collinear-soft functions to order $\mathcal{O}(\alpha_s)$.\footnote{If we had also retained the constants, this would be equivalent to NLL$^{\prime}$.} Thus when we turn off the resummation by taking all scales to be the factorization scale, we are left with the singular terms of the fixed-order $D_2$ distribution. When resummation is fully turned on by the profile function, this trivializes the contribution from the factorized functions. The specific profile used and the canonical scale choices are summarized in \App{app:scales_resummation}.

 For all distributions (quark, gluon, and signal), the use of canonical scales in the resummation gives a resummed distribution that completely over-shoots the singular terms of the fixed order result throughout the range of $D_2$, leading to an unphysical endpoint of the distribution much greater than $1/(2z_{\text{cut}})$. If one were to additively match and normalize, the resulting curve would be equivalent to just the normalized resummed canonical prediction, with completely unphysical behavior in the large $D_2$ region, and with a peak much too low due to the broad tail. Thus we adopt a strategy of multiplicative matching for the distribution
\begin{align}\label{eq:matching_to_fo}
\left.\frac{d\sigma}{d\ecf{2}{2}d\ecf{3}{2}}\right|_{\text{matched}}&=\left.\frac{d\sigma}{d\ecf{2}{2}d\ecf{3}{2}}\right|_{\text{resum}}\left(\frac{\frac{d\sigma}{d\ecf{2}{2}d\ecf{3}{2}}\Big|_{\text{fo}}}{\frac{d\sigma}{d\ecf{2}{2}d\ecf{3}{2}}\Big|_{\text{resum fo}}}\right)\,.
\end{align}
Here ``fo" stands for the fixed-order distribution, which is determined from the $1\rightarrow 3$ splitting functions (as discussed in \App{app:1to3}) or from {\tt EVENT2}~\cite{Catani:1996vz}. Thus, regardless of whether we use canonical scale choices in the resummed distribution or profiles to turn off the resummation, the distribution will always terminate at the physical value $1/(2z_{\text{cut}})$. 

The only subtlety is if the singular distribution has a zero in the physical range of $D_2$. This occurs in some cases, and we are then forced to only use distributions where the resummation is turned off via profiles before this zero is reached. We find this to be the case generically for the signal distributions if $z_{\text{cut}}<0.2$, and for the quark distribution if $z_{\text{cut}}\leq 0.05$. 

It is worthwhile to understand how the merging interplays with the resummation of the $\ecf{2}{\alpha}$ spectrum, and the counting of logs of $D_2$ versus logs of $\ecf{2}{\alpha}$. As can be directly seen from \App{app:1to3}, the $D_2$ spectrum in the large $D_2$ region, which is controlled by the factorization in the soft-haze region of \Sec{sec:fact_un}, is \emph{independent} of the value of $\ecf{2}{\alpha}$. Since the $D_2$ spectrum at leading order is set by the two-loop matrix elements in the soft-haze region, we may write
\begin{align}\label{eq:fixed_order_d2_with_ecf2_resum}
\frac{d\sigma}{d\ecf{2}{\alpha}\,dD_2}&=\frac{d\sigma}{d\ecf{2}{\alpha}}\Bigg|_{\text{N}^3\text{LL}}\times\big[F(D_2,\zcut)\big]_++O\Big(\alpha_s^4~\log~\ecf{2}{\alpha}\Big)\,,
\end{align}
where $F(D_2,\zcut)$ reproduces the fixed order spectrum in $D_2$. The subscript N$^3$LL indicates that this expression is valid up to N$^3$LL order. Although the fixed order distribution diverges at $D_2=0$, the plus distribution ensures that the singularity at $D_2=0$ is formally cancelled by the appropriate virtual corrections, so that we have
\begin{align}
\int_0^{\infty}dD_2\big[F(D_2,\zcut)\big]_+=1\,.
\end{align}
The factor multiplying the fixed order $D_2$ spectrum is simply the groomed $\ecf{2}{\alpha}$ spectrum to N$^3$LL accuracy. Once we match the resummed $D_2$ spectrum to the fixed-order $D_2$ spectrum, we replace
\begin{align}
\big[F(D_2,\zcut)\big]_+\rightarrow F^{\text{matched}}(D_2,\zcut,Q\ecf{2}{\alpha})\,.
\end{align}
The matched function satisfies the properties:
\begin{align}
\int_0^{\infty}dD_2F^{\text{matched}}(D_2,\zcut,Q\ecf{2}{\alpha})&=1\,,\\
F^{\text{matched}}(D_2,\zcut,Q\ecf{2}{\alpha})&\rightarrow F(D_2,\zcut)\text{ as } D_2 \rightarrow \frac{1}{2\zcut}\,.
\end{align}
The resummation ensures the integrability of the matched distribution, and the matching ensures that in the region of validity of the soft-haze factorization, we reproduce the soft-haze spectrum. Thus we may simply replace the plus-distribution for the fixed order result within \Eq{eq:fixed_order_d2_with_ecf2_resum}
\begin{align}\label{eq:fixed_order_d2_with_ecf2_resum_II}
\frac{d\sigma}{d\ecf{2}{\alpha}\,dD_2}&=\frac{d\sigma}{d\ecf{2}{\alpha}}\Bigg|_{\text{N}^3\text{LL}}\times F^{\text{matched}}(D_2,\zcut,Q\ecf{2}{\alpha})+O\Big(\alpha_s^4~\log~\ecf{2}{\alpha}\Big)\,.
\end{align}
This result is still valid to the same logarithmic accuracy in the log counting for the $\ecf{2}{\alpha}$ spectrum, while maintaining the correct sum rules on the $D_2$ variable, and gives the correct shape of the end-point of the distribution where the soft-haze factorization applies. Since the resummation of the groomed $\ecf{2}{\alpha}$ spectrum is multiplictative to the $D_2$ spectrum, we can correctly predict the shape of the distribution for both large and small $D_2$ without resumming any logs of $\ecf{2}{\alpha}$ to at least N$^4$LL accuracy, which is far beyond the practically achievable accuracy.

We stress that such a simple matching procedure does not work when considering the ungroomed $D_2$ distribution. For the ungroomed distribution in the soft haze region, we are forced to write
\begin{align}\label{eq:fixed_order_d2_with_ecf2_resum_no_grooming}
\frac{d\sigma}{d\ecf{2}{\alpha}\,dD_2}&=\frac{d\sigma}{d\ecf{2}{\alpha}}\Bigg|_{\text{N}^3\text{LL}}\otimes\big[F(D_2,\ecf{2}{\alpha})\big]_++O\Big(\alpha_s^4~\log~\ecf{2}{\alpha}\Big)\,,
\end{align}
where we now have a \emph{convolution} in the $\ecf{2}{\alpha}$ variable, denoted by the $\otimes$. We may still replace the fixed order distribution in $D_2$ with the matched distribution, both normalized to obey the correct $D_2$ sum rule, but now we must perform a convolution in $\ecf{2}{\alpha}$! Given that the endpoint of the ungroomed distribution behaves as the inverse of $\ecf{2}{\alpha}$, performing such a convolution would be daunting and computationally expensive, since at each $\ecf{2}{\alpha}$ value, one would need to calculate the full matched and normalized $D_2$ distribution.

\subsubsection{Signal Jets}\label{sec:fact_sig}

\begin{figure}
\begin{center}
\subfloat[]{\label{fig:signal_a}
\includegraphics[width=7cm]{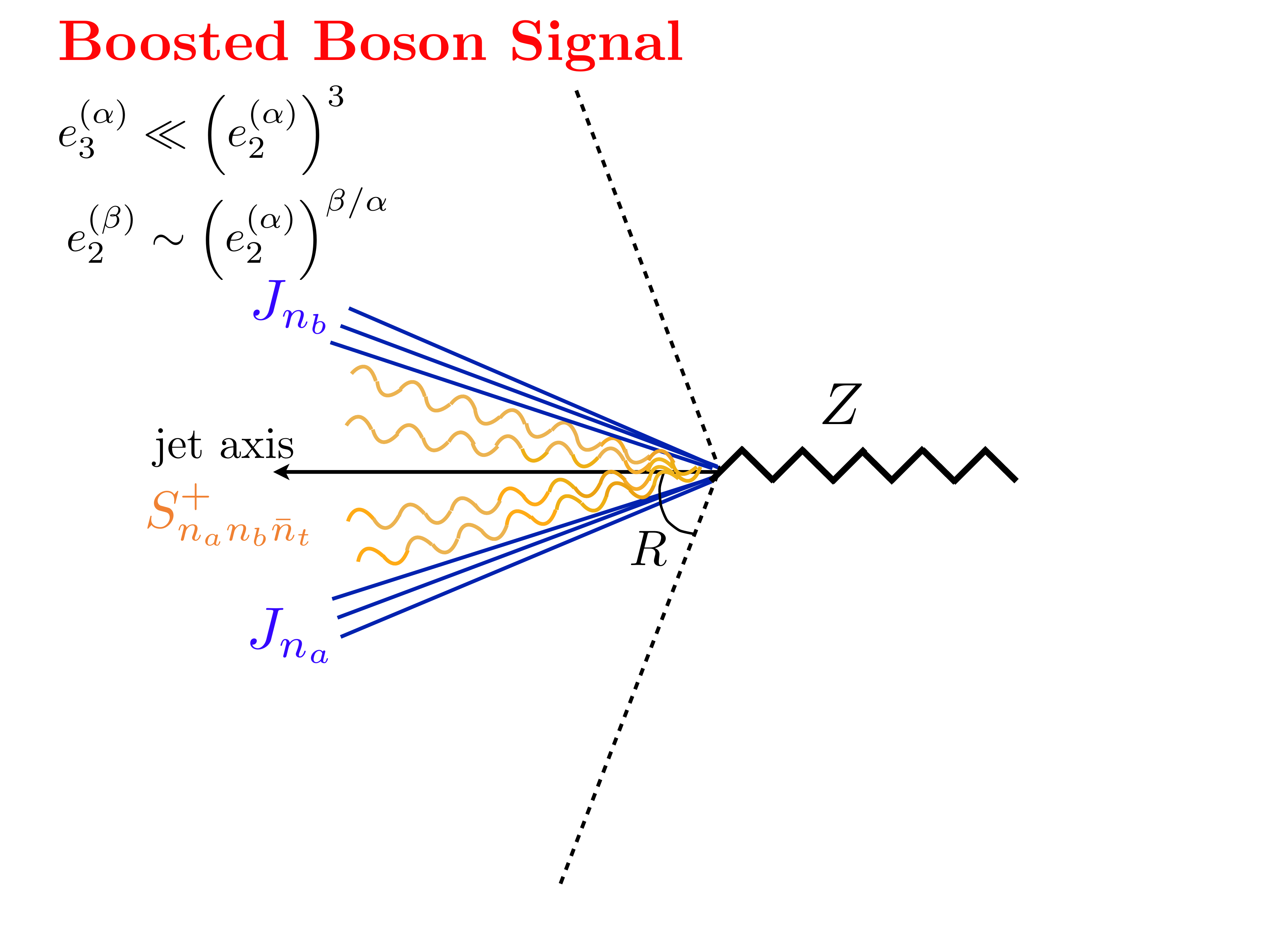}
}\ \ 
\subfloat[]{\label{fig:signal_b}
\includegraphics[width = 8cm]{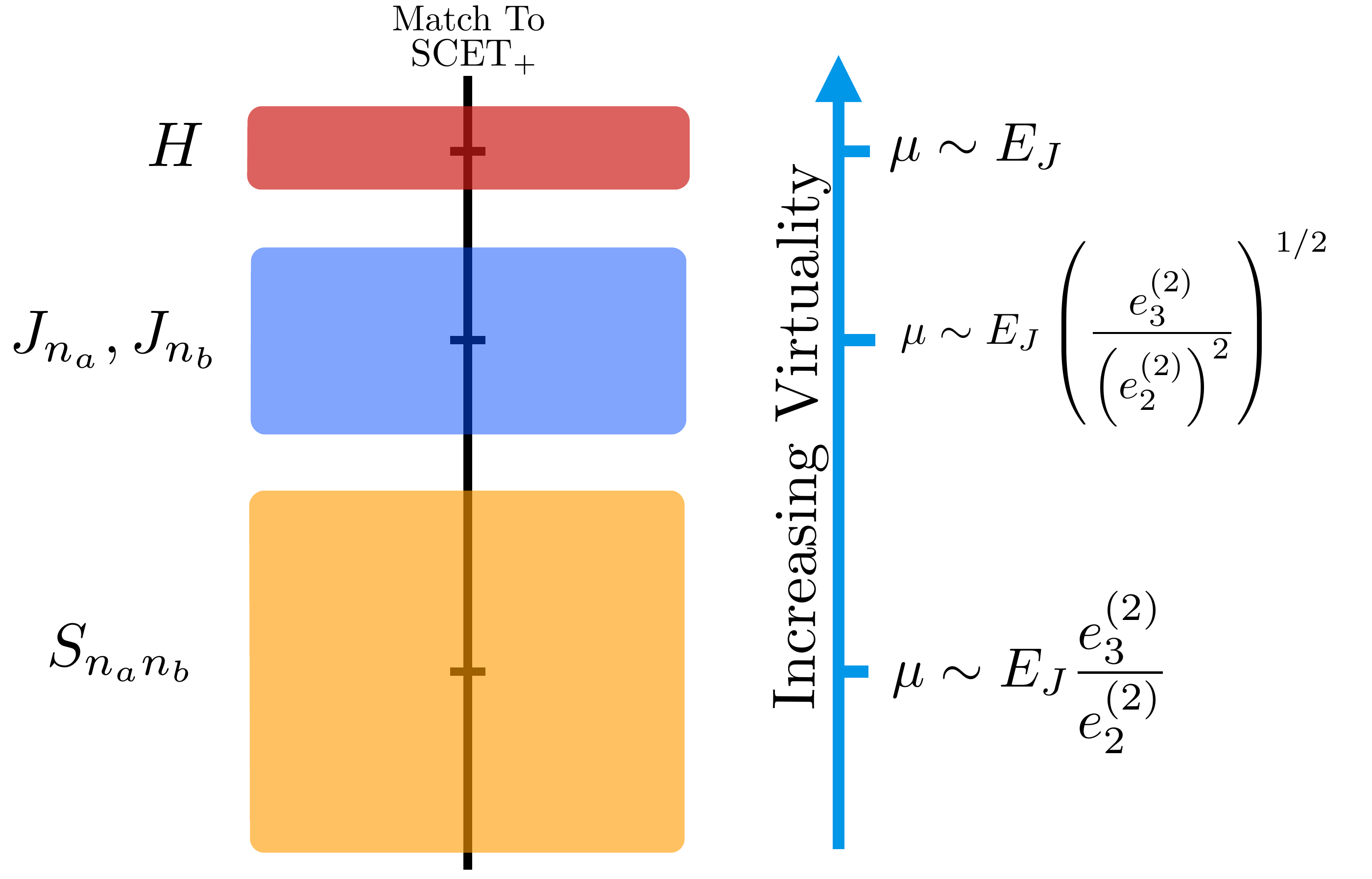}
}
\end{center}
\caption{The collinear subjets configuration for a boosted color singlet decay is shown in (a).  The structure of factorization formula is shown in (b). Figure from \cite{Larkoski:2015kga}.
}
\label{fig:signal_setup}
\end{figure}

In this section, we give the effective field theory description for a groomed hadronically-decaying color singlet, which we take for concreteness to be a $Z$ boson:
\begin{align}\label{eq:signal_fact}
\frac{d^3\sigma}{dz\, d\ecf{2}{\alpha}d\ecf{3}{\alpha}}= H(Q^2)H_2^{Z\rightarrow q \bar q}(\ecf{2}{\alpha},m_Z^2) J_a(\ecf{3}{\alpha})\otimes J_b(\ecf{3}{\alpha})\otimes S(\ecf{3}{\alpha},\zcut) \,. 
\end{align}
A brief description of the functions appearing in \Eq{eq:signal_fact} is as follows:
 \begin{itemize}
 \item $H(Q^2)$ is the hard function describing the production of the on-shell $Z$ boson.
 \item $H_2^{Z\rightarrow q \bar q}(z,\ecf{2}{\alpha},m_Z^2)$ is a hard function describing the decay of the $Z$ boson into a $q\bar q$ pair.  
 \item  $J_a(\ecf{3}{\alpha})$, $J_b(\ecf{3}{\alpha})$ are the jet functions describing the two collinear subjets.
 \item  $S(\ecf{3}{\alpha},\zcut)$ is the collinear-soft function describing the radiation emitted from the $q\bar q $ dipole.
 \end{itemize}
The factorization formula, and the region of phase space it describes is shown schematically in \Fig{fig:signal_setup}. One-loop calculations are given in \App{app:signal}.

The factorization formula of \Eq{eq:signal_fact} is valid when $\ecf{3}{\alpha} \ll  ( \ecf{2}{\alpha} )^{3}$. The structuring of the radiation within $S(\ecf{3}{\alpha},\zcut)$ is similar to the collinear-soft function in \Eq{eq:collinear_soft_mother_f}, and can also be refactorized similarily, except that there is no third Wilson line corresponding to the hard recoil direction of the jet. When $\ecf{3}{\alpha} \gtrsim  ( \ecf{2}{\alpha} )^{3}$, one must match to the full $Z\to q\bar q g$ matrix element.

\section{Factorized Cross Section in $pp$ Collisions}\label{sec:factpp}

In this section we will discuss the extension of the $e^+e^-$ factorization formulae of \Sec{sec:fact_D2} to $pp$ collisions. In particular, we show that for phenomenologically relevant parameters for the jet mass and $p_T$, the assumptions of the factorization formula hold, and no new ingredients are required to extend the factorization formula to $pp$. The only process dependence is carried by the quark, anti-quark and gluon fractions of the process. This will follow straightforwardly from the universality of collinear factorization and the fact that all the factorization formulae of \Sec{sec:fact_D2} were obtained through a refactorization of the jet or collinear-soft functions.  For concreteness, in this section we will consider the factorization for the groomed $D_2$ observable in $pp\to Z+j$.  We identify the highest $p_T$ jet satisfying $|\eta_J|< \eta_{\text{max}}$, groom it with the mMDT/soft drop algorithm, and then measure $D_2$ on the groomed jet. It is important to emphasize that here we are completely inclusive over additional hadronic activity throughout the event. We do not need to apply any form of veto on out-of-jet radiation, as is sometimes imposed to study ungroomed jet mass (for example, see \Ref{Jouttenus:2013hs}).

Since our factorization formula for the groomed $D_2$ observable is obtained as a refactorization of the cross section for mMDT/soft drop groomed $\ecf{2}{\alpha}$ observable, we begin by summarizing its factorization in $pp$ collisions.
In \Ref{Frye:2016aiz} it was shown that in the region where the factorization formula applies, namely $\ecf{2}{\alpha} \ll \zcut \ll 1$, the cross section can be written as
\begin{align}\label{eq:fac_pp_e2}
\frac{d\sigma^{pp}}{d\ecf{2}{\alpha}}=\sum\limits_{k=q,\bar q, g}D_k(p_T^{\text{min}}, y_\text{max}, \zcut, R) S_{C,k}(\zcut, \ecf{2}{\alpha})\otimes J_k (\ecf{2}{\alpha})\,.
\end{align}
Here, it is important to emphasize that since we are inclusive over hadronic activity in the event, a strict factorization into jet and soft functions does not apply. Indeed, it is clear that this must be the case, since the number of jets in the event is not fixed. Nevertheless, \Eq{eq:fac_pp_e2} shows that all dependence on the rest of the event can be absorbed into a process dependent normalization factor $D_k$, which does not depend on the $\ecf{2}{\alpha}$ observable.  In general, $D_k$ depends on the minimum $p_T$ cut, the jet radius $R$, rapidity cuts,  parton distributions, $\zcut$, etc.  The $\ecf{2}{\alpha}$ observable is set by universal collinear physics described by the convolution between the collinear-soft function and the jet function.  Since these are collinear matrix elements, they depend only on the collinear dynamics of the particular jet in question, and are independent of other jets in the event. In particular, global color correlations are absent.

The $D_k$ functions depend on the parton flavor, which must be summed over, an added complication of jets in $pp$ collisions. While parton flavor is not in general an IRC safe quantity, due to the fact that soft partons can radiate flavor into or out of the jet, it was shown in \Ref{Frye:2016aiz}  that the parton flavor can be defined on soft dropped jets in the limit $\ecf{2}{\beta} \ll \zcut \ll 1$, where the factorization formula applies. On a soft dropped jet, we can define the flavor of the jet as
\begin{align}
f_J=\sum\limits_{i\in J_{\text{SD}}} f_i\,,
\end{align}
where $f_q=1$, $f_{\bar q}=-1$, $f_g=0$, and $J_{\text{SD}}$ indicates the constituents of the jet after the soft drop algorithm has been applied. If $f_J=\pm 1$, then the jet is defined as quark type, while if $f_J=0$, the jet is defined as gluon type. In the normalized distribution,  the $D_k$ can therefore  be interpreted as quark, anti-quark, and gluon jet fractions in the event sample under consideration, and can easily be extracted from fixed-order Monte Carlo codes, such as MCFM \cite{Campbell:1999ah,Campbell:2010ff,Campbell:2011bn}.

The factorization of $D_2$ in $pp$ collisions now follows trivially from combining the factorization formula of \Eq{eq:fac_pp_e2} for soft dropped $\ecf{2}{\alpha}$ with $\ecf{2}{\alpha} \ll \zcut \ll 1$ with the factorization formulae derived for $e^+e^-$ in \Sec{sec:fact_D2}.  To proceed, starting from \Eq{eq:fac_pp_e2}, we refactorize the jet and collinear-soft functions, as appropriate.
This also implies that the same process dependent functions $D_k$ also appear in the expression of the cross section of $D_2$.  We can then write
\begin{align}\label{eq:fac_pp_coll}
\frac{d^2\sigma^{pp,\text{coll}}}{d\ecf{2}{\alpha} d\ecf{3}{\alpha}}=\sum\limits_{k=q,\bar q, g}D_k  \left[  H_{2}(\ecf{2}{\alpha})C_s(\ecf{3}{\alpha})\otimes J_1(\ecf{3}{\alpha})\otimes J_2(\ecf{3}{\alpha}) \otimes S_c(\ecf{3}{\alpha}, \zcut)  \right]\,,
\end{align}
in the collinear subjets region of phase space,  
\begin{align}\label{eq:fac_pp_cs}
\frac{d^2\sigma^{pp,\text{c-soft}}}{d\ecf{2}{\alpha} d\ecf{3}{\alpha}}=\sum\limits_{k=q,\bar q, g}D_k  \left[  H_{2}^{sj}(\ecf{2}{\alpha},\zcut )C_s(\ecf{3}{\alpha})\otimes J_{sc}(\ecf{3}{\alpha})\otimes J(\ecf{3}{\alpha})  \right]\,,
\end{align}
in the soft subjet region of phase space, and
\begin{align}\label{eq:fac_pp_cshaze}
\frac{d^3\sigma^{pp,\text{cs haze}}}{d\ecf{2}{\alpha} d\ecf{2}{2\alpha} d\ecf{3}{\alpha}}=\sum\limits_{k=q,\bar q, g}D_k  \left[  J(\ecf{2}{\alpha}) \otimes S_{sc}(\ecf{2}{\alpha},\ecf{2}{2\alpha}, \ecf{3}{\alpha}, \zcut)  \right]\,,
\end{align}
in the collinear-soft haze region of phase space.

Importantly, since the same $D_k$ factor appears in each of the factorization formulae in the different regions of phase space,  we can then perform the marginalization separately over the different factorization formulae. We can therefore write, for the normalized distribution when summed over the factorization formulae:
\begin{align}
\frac{d\sigma^{pp, \text{norm}}}{d\Dobs{2}{\alpha}}=\sum\limits_{k=q,\bar q, g} \kappa_k   \frac{d\sigma_k^{pp, \text{norm}}}{d\Dobs{2}{\alpha}}\,,
\end{align}
where the $\kappa_k$ can be interpreted as the fraction of jets in the sample with flavor $k$.

\section{Consequences of Factorization Formulae}\label{sec:consfact}

Given the factorization formulae developed in the previous sections, there are several fascinating consequences that immediately follow.  Several of these have been noted before (see \Ref{Frye:2016aiz}), and are consequences of the fact that mMDT or soft drop removes soft, wide angle radiation in a jet from contributing to the observables of interest.  Here, we will briefly mention these general properties of mMDT and soft drop grooming, and discuss in some detail features that are new to measuring $D_2$ on these groomed jets.

The absence of soft, wide angle radiation in the jet eliminates event-wide color correlations and NGLs of the groomed jet observables to all orders in $\alpha_s$.  With the relative scaling that we have assumed between the two-point energy correlation function and $\zcut$, $\ecf{2}{\alpha}\ll \zcut \ll 1$, all radiation that remains in the jet after grooming must be collinear.  Assuming collinear factorization, this then implies that the shape of the mass distribution is independent of the process that created that jet, up to the relative fraction of quark and gluon jets in the sample.  The quark and gluon groomed jet fractions are well-defined to leading power in $\ecf{2}{\alpha}$ and $\zcut$, and can be determined from fixed-order codes.  Because the measurement of $D_2$ is more differential than just the groomed jet mass, all of these properties continue to hold in that case.

Additionally, the mMDT groomed $D_2$ distribution enjoys other properties that actually make its perturbative distribution more well-defined and robust than the jet mass.  Because the cut on the groomed jet mass can be tuned to satisfy $\ecf{2}{2} E_J^2 = m_J^2 \ll \zcut E_J^2\ll E_J^2$, perturbative power corrections to the $D_2$ distribution can formally be made arbitrarily small.  Additionally, because soft, wide angle emissions do not contribute to the groomed observables, non-perturbative corrections are suppressed by powers of the ratio of $\Lambda_\text{QCD}$ to the groomed jet mass.  These make $D_2$ a good candidate for QCD studies at the LHC, and therefore we will discuss these points in some detail.

\subsection{Universality of the Shape of the $D_2$ Distribution}

Typically, resummation is only important in a restricted region of the distribution of a particular observable.  For example, soft and collinear emissions dominate the hadronic final state of $e^+e^-$ collisions when an appropriately chosen event shape, such as thrust \cite{Farhi:1977sg}, is small.  In the case of soft drop groomed jet mass, radiation in the jet is constrained to be collinear if $ m_J^2 \ll \zcut E_J^2\ll E_J^2$; however, this is not the whole allowed phase space.  There are regions where $ m_J^2 \gtrsim \zcut E_J^2$, which are vital to describe correctly to claim a precision description of the distribution.

The entire mMDT/soft drop groomed $D_2$ distribution, however, enjoys a universality.  First, requiring the groomed jet mass to satisfy $ m_J^2 \ll \zcut E_J^2\ll E_J^2$, all radiation that remains in the jet is collinear.  At this stage, both $\zcut$ and $m_J$ are fixed.  Then, with this configuration, we measure $D_2$ on the remaining constituents of the jet.  All remaining emissions in the jet are necessarily collinear, and so any measured value of $D_2$ of these groomed jets is well-described just by resummation.  Perturbative power corrections beyond the resummation (non-singular contributions) are small, and can be made arbitrarily small in perturbation theory by going further into the regime where $ m_J^2 \ll \zcut E_J^2\ll E_J^2$.  Note that this property requires that we restrict the jet mass appropriately and then measure $D_2$, an observable which resolves further substructure of the jet.

For applications to the LHC, it is interesting to briefly consider the values of the jet mass and $p_T$ for which our factorization formula, and therefore this universality, holds. Observables such as $D_2$ are used at the LHC both to identify hadronically decaying $W/Z/H$ bosons, as well as to search for new light particles with $m\lesssim m_Z$ which decaying hadronically  \cite{CMS-PAS-EXO-17-001}. For many of these searches, the bulk of the data is for $p_T>500$ GeV, and extends up to approximately $p_T\sim 1000$ GeV. Using the condition $\ecf{2}{2} \ll \zcut \ll 1$, with $\zcut=0.1$, we expect that our factorization will begin to break down around $p_T=500$ GeV for $m_J \sim m_Z$, if the value of $\zcut=0.1$ is fixed. For lower values of $p_T$, one will become sensitive again to global color correlations from emissions with energy fraction greater than $\zcut$, which do not fail the soft drop criteria, and can contribute to the observable.  Taking as a concrete example a bin from $p_T=600-800$ GeV in which the $D_2$ observable has been measured by ATLAS  \cite{collaboration:2015aa}, for a jet mass of $m_J \sim m_Z$, this has $\ecf{2}{2}\lesssim0.02$. For $\zcut=0.1$ the assumptions of our factorization formula safely hold. For lighter particles the $p_T$ range can be extended, or alternatively, the expansion parameter is smaller. We therefore find that our factorization applies for most of the $p_T$ range of phenomenological interest, and therefore so do our conclusions regarding the universality of the distribution. We believe that this understanding of universality derived from the factorization formula is one of the most important outcomes of our analysis.

\subsection{Hadronization Corrections Suppressed by Perturbative Jet Mass}\label{sec:hadr_suppress}

The dominant non-perturbative corrections to a factorization formula arise from modes whose virtualities approach $\Lambda_\text{QCD}$.  A simple estimate of the size or importance of these non-perturbative effects follows from determining the value of the observable at which the lowest virtuality mode becomes comparable to $\Lambda_\text{QCD}$.  The mode with the lowest virtuality often corresponds to soft, wide angle emissions.  So, by grooming them away with mMDT or soft drop, we can significantly reduce the effect of non-perturbative corrections and render the perturbative distribution more robust.

To see this for $D_2^{(\alpha)}$ on groomed jets, we first review the size of non-perturbative corrections in the ungroomed case.  For concreteness, we will focus on the non-perturbative corrections to the collinear subjets factorization formula.  In the ungroomed case the lowest virtuality mode is that of soft, wide-angle radiation; see \Sec{sec:fact_review_ungD2}.  Its virtuality was identified in \Ref{Larkoski:2015kga} and is
\begin{equation}
\mu_S \simeq \frac{\ecf{3}{\alpha}}{\ecf{2}{\alpha}} E_J \simeq (\ecf{2}{\alpha})^2 D_2^{(\alpha)}E_J\,,
\end{equation}
where we have expressed $\ecf{3}{\alpha}$ in terms of $\ecf{2}{\alpha}$ and $D_2^{(\alpha)}$.  Setting $\mu_S=\Lambda_\text{QCD}$, we find that non-perturbative effects dominate when
\begin{equation}
\left. D_2^{(\alpha)}\right|_\text{np} \simeq \frac{\Lambda_\text{QCD}}{(\ecf{2}{\alpha})^2 E_J}\,.
\end{equation}
If we take $\alpha = 2$ for concreteness, this can be rewritten in terms of the jet mass and energy as
\begin{equation}
\left. D_2^{(2)}\right|_\text{np} \simeq \frac{\Lambda_\text{QCD}E_J^3}{m_J^4}\,.
\end{equation}
Therefore, perhaps surprisingly, as the jet energy increases for a fixed jet mass, non-perturbative corrections increase significantly.  If we assume that $E_J = 500 $ GeV, $m_J = 100$ GeV, and take $\Lambda_\text{QCD} = 1$ GeV, then $\left. D_2^{(2)}\right|_\text{np} \simeq 1$.  That is, we expect non-perturbative physics to dominate right at the boundary between where one- and two-prong jets live in the $D_2^{(\alpha)}$ distribution.

Now, let's do the same analysis but for the mMDT/soft drop $D_2$ cross section.  The lowest virtuality mode that appears in any factorization formula is the collinear-soft radiation of the collinear-soft subjet factorization formula; see \Sec{sec:resolved_soft}.  The virtuality of this mode is
\begin{equation}
\mu_{cs} \sim E_J   \frac{   \zcut^2 \ecf{3}{\alpha} }{  (\ecf{2}{\alpha})^3  }   \left( \frac{    \ecf{2}{\alpha} }{  \zcut  }  \right)^{1/\alpha} =  E_J   \zcut^2 \Dobs{2}{\alpha}     \left( \frac{    \ecf{2}{\alpha} }{  \zcut  }  \right)^{1/\alpha}\,.
\end{equation}
Setting $\mu_{cs}=\Lambda_\text{QCD}$, non-perturbative effects dominate this mode when
\begin{equation}
\left. D_2^{(\alpha)}\right|_{\text{np},cs} \simeq \frac{1}{(\ecf{2}{\alpha})^{1/\alpha}  \zcut^{2-1/\alpha}  }\left(
\frac{\Lambda_\text{QCD}}{ E_J}
\right)\,.
\end{equation}
As before, taking $\alpha = 2$ for concreteness, we find that non-perturbative effects dominate when
\begin{align}
\left. D_2^{(2)}\right|_{\text{np},cs} &\simeq 
\frac{\Lambda_\text{QCD}}{\zcut^{3/2} m_J }\,.
\end{align}

This result is quite remarkable.  Without grooming, non-perturbative effects for $D_2^{(2)}$ become larger, for a fixed jet mass cut, as the energy of the jet is increased.  However, by grooming the jet with mMDT or soft drop, non-perturbative corrections are {\it independent of the jet energy!} Physically this arises since after grooming the jet behaves loosely like a boosted event shape, and it is the jet mass that sets the scale.  As long as the mass cut on the jet is perturbative, hadronization corrections are highly suppressed.  
Importantly, the distribution is perturbative well below $D_2^{(2)}\sim 1$, into the region where two-prong jets live. Taking the numerical values of $\zcut=0.1$, $\Lambda_{\text{QCD}}=1$ GeV, and $m_J=m_Z$, we find  that dominant non-perturbative correction arises from the soft dropped soft subjet region of phase space, and we can estimate that non-pertubative effects becomes important at $\Dobs{2}{2}\sim 0.35$. Non-perturbative corrections for the other regions of phase space in the factorization formulae are further suppressed, and so are ignored.  A more detailed study of non-perturbative effects for the $D_2$ distribution is performed in a companion paper \cite{Larkoski:2017iuy}.

Combined with the fact that the distribution terminates at (as discussed in \Sec{sec:ps_struc})
\begin{equation}
\left.D_2^{(\alpha)}\right|_\text{max} \sim \frac{1}{2\zcut}\,,
\end{equation}
this implies that, for a fixed mass cut, the full distribution, including non-perturbative effects, of the mMDT/soft drop groomed $D_2^{(\alpha)}$ is largely independent of the jet energy!  Unlike the ungroomed $D_2^{(\alpha)}$ distribution, which had both an upper endpoint and location of non-perturbative corrections that depended on the jet energy, the groomed $D_2^{(\alpha)}$ distribution has endpoints and non-perturbative corrections that are independent of the jet energy.  We will demonstrate in \Secs{sec:eepred}{sec:pppred} that both the NLL calculation of the distribution as well as the Monte Carlo simulation respect this prediction.

\subsection{Grooming Efficiency for Signal Jets}\label{sec:eff_signal}

While we have focused on the properties of the mMDT/soft drop $D_2$ distribution for background (QCD) jets, jet grooming can have an effect on the signal distribution as well.  For an unpolarized boosted $Z$ boson that decays to a $q\bar q$ pair, the distribution of the energy fraction $z$ of the quark, say, is approximately flat:
\begin{equation}\label{eq:no_dla_sup}
\frac{d\sigma}{dz}\simeq \Theta(1-z)\Theta(z)\,.
\end{equation}
This implies that when the boosted $Z$ jet is groomed, a fraction $2\zcut$ of the jets will have one prong removed by grooming.  For these jets that lose one prong, they will also typically fail the mass cut, as well as no longer have a clear two-prong structure.  Of course, for $\zcut\ll 1$, this is formally a small effect, but practically, if $\zcut\simeq 0.1$, then about 20\% of the $Z$ jets could have a prong removed.  This effect could have a large effect on the signal $D_2$ distribution.

While at leading-order the distribution of the energy fraction $z$ is approximately flat, when all-orders effects are included the regions with $z\to 0$ and $z\to 1$ are suppressed by a Sudakov factor.  When $z\to 0$, for example, there is of course no divergence in the leading-order $Z$ decay matrix element.  However, a gluon emitted off of the soft decay product will itself necessarily be soft, and result in a divergence at fixed-order.  When all-orders effects are included, these soft gluon divergences arrange themselves into a Sudakov factor that suppresses the probability for a decay product to only carry a small fraction of the energy of the $Z$.  At double logarithmic accuracy (DLA), this Sudakov factor is
\begin{equation}\label{eq:dla_sup}
\frac{d\sigma^\text{DLA}}{dz}\simeq \Theta(1-z)\Theta(z)\exp\left[-\frac{\alpha_s}{2\pi}C_F\left(
\log^2z+\log^2(1-z)
\right)\right]\,,
\end{equation}
which can be derived from the $Z$ boson decay matrix element at next-to-leading order.  This Sudakov factor pushes decay products of the $Z$ to have more equal energies, and reduces the fraction of $Z$ jets that have a subjet that is removed by the jet groomer.  That is, due to all-orders effects, hadronic decays of $Z$ bosons can look more two-prong-like than their fixed-order description would suggest.

In our analytic calculations for the prediction of the $D_2$ distribution on groomed signal jets, we include this resummation to NLL accuracy.  The suppression of the $z\to 0$ and $z\to 1$ regions will be much larger than that suggested by the simple Sudakov factor that exists at DLA accuracy.  Nevertheless, even at DLA accuracy, this suppression is non-trivial.  With $\zcut=0.1$ and using the distribution of \Eq{eq:dla_sup}, only about 15\% of $Z$ jets fail soft drop, as compared to 20\% using \Eq{eq:no_dla_sup}.

\section{NLL Predictions in $e^+e^-$ Collisions}\label{sec:eepred}

In this section we use our factorization formulae to provide numerical results for the $D_2$ distribution in $e^+e^-$ collisions. In \Sec{sec:eepred_singular} we compare our result, expanded to fixed order, with the fixed order code \eventtwo~to ensure that we reproduce the singular behavior of the $D_2$ distribution. In \Sec{sec:eepred_shower} we compare our resummed results, matched to fixed order, with parton shower Monte Carlo.

\subsection{Singular Results and Comparison with \eventtwo}\label{sec:eepred_singular}

\begin{figure}
\begin{center}
\subfloat[]{\label{fig:eesing}
\includegraphics[width=7.5cm]{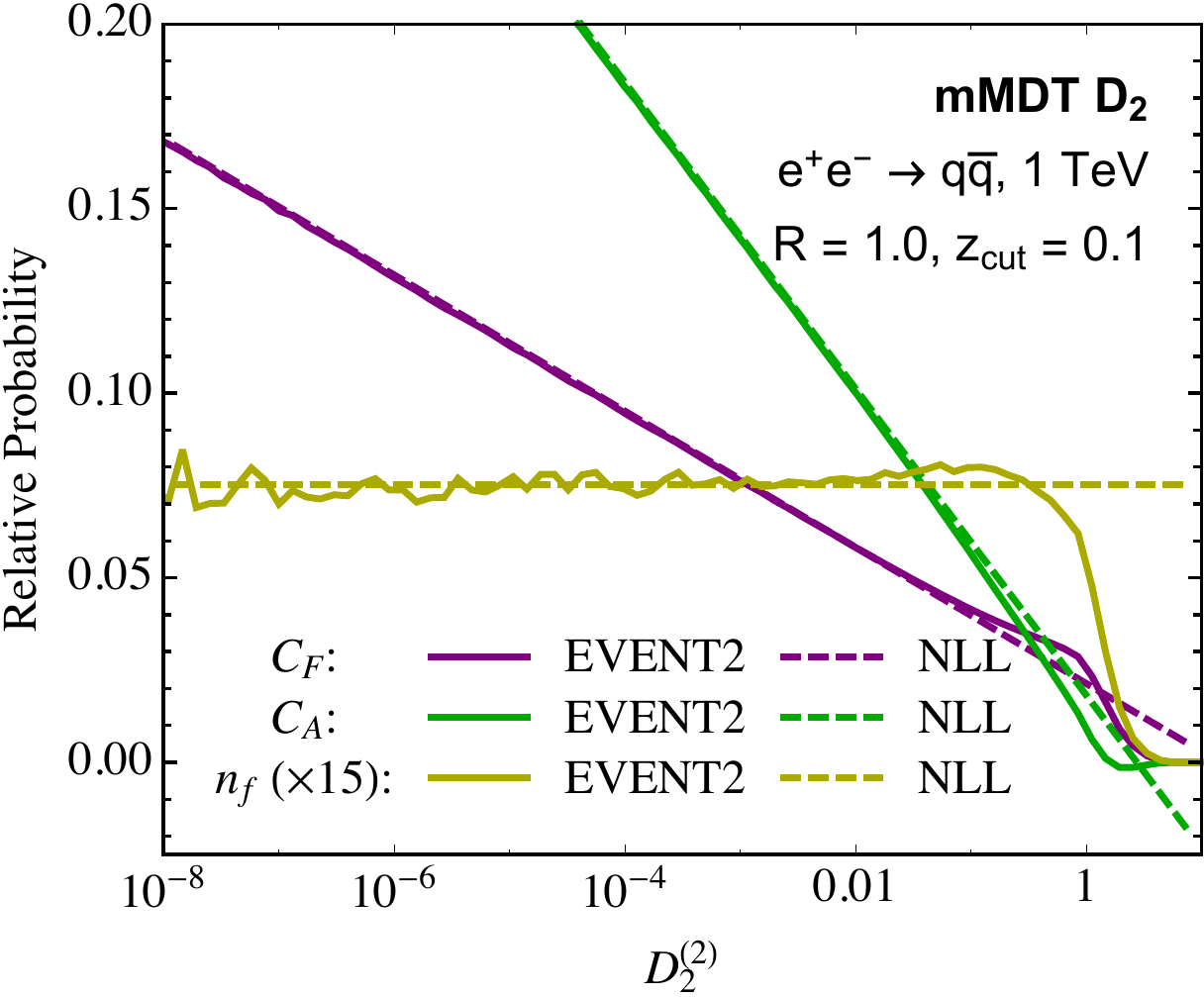}
}\ \ 
\subfloat[]{\label{fig:eesplit}
\includegraphics[width = 7.5cm]{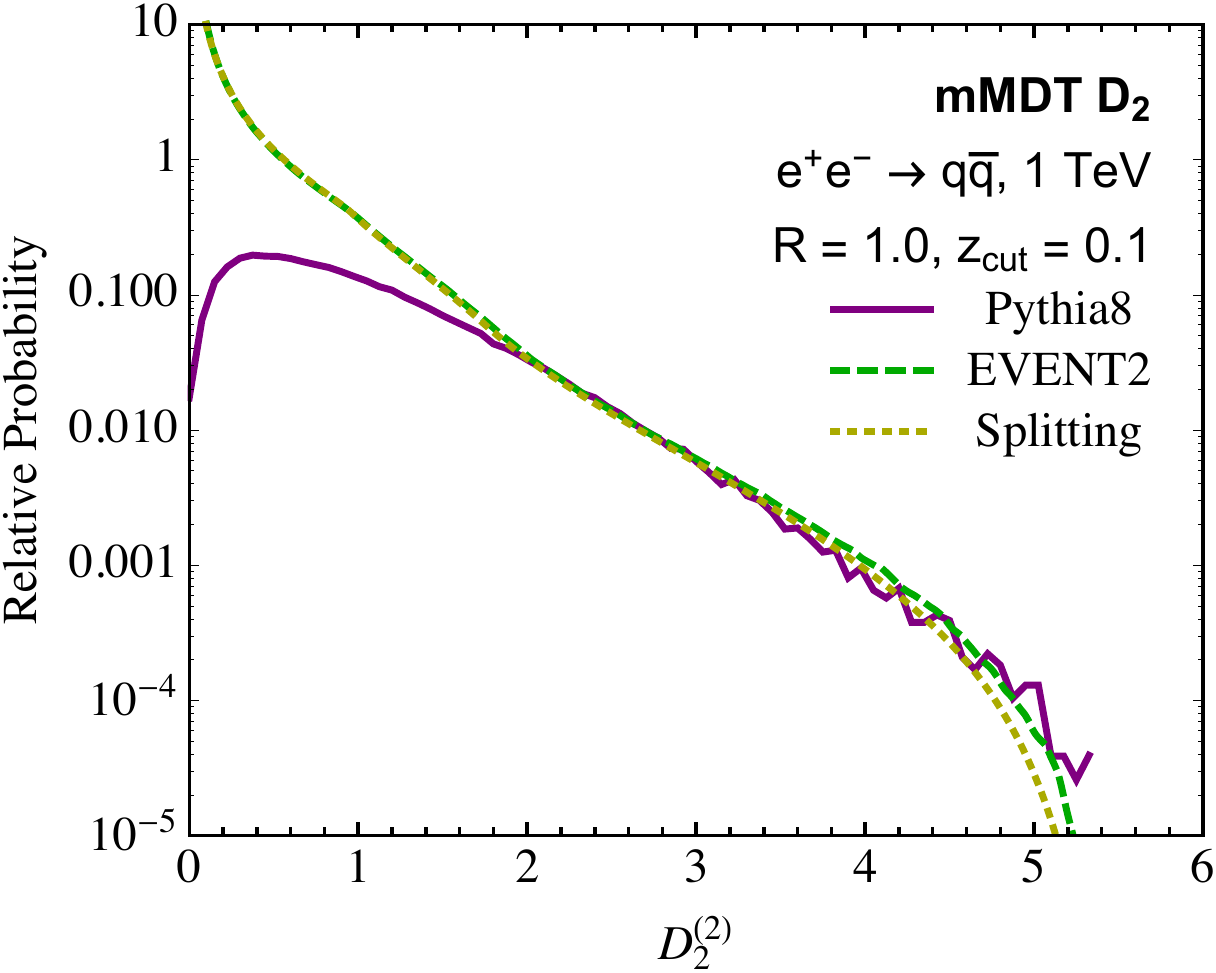}
}
\end{center}
\caption{(a) A comparison of the NLL prediction expanded to fixed order with \eventtwo. The singular behavior of the $D_2$ distribution is reproduced by our factorization. (b) Comparison of the end-point value found by fixed-order codes and Monte Carlo event generators. Note the robustness of the endpoint set by $\zcut$ even in the prescence of multiple emissions.
}
\label{fig:eec_event2}
\end{figure}

To verify that our factorization reproduces the singular behavior of the $D_2$ distribution as $D_2\to 0$, we can compare the results of our factorization formula, expanded to $\alpha_s^2$, with the fixed order generator \eventtwo~\cite{Catani:1996vz}. In \Fig{fig:eesing} we show the result of \eventtwo~in each of the color channels, compared with the expansion of our NLL formula.  We see that at small values of $D_2$, our NLL formula captures the singular structure of the \eventtwo~distribution, as is required. Here we consider $e^+e^-\to$ dijets at $1$ TeV with $\zcut =0.1$, but we have found similar agreement for other values of the $\zcut$ parameter, while verifying the independence on the center-of-mass energy and jet mass bin. This verifies the consistency of our factorization to $\mathcal{O}(\alpha_s^2)$. Due to the complexity of our factorization, this is a highly non-trivial check, and gives us confidence that have correctly incorporated all modes in the effective theory.

In \Fig{fig:eesplit}  we show a linear plot of the $D_2$ distribution, comparing \eventtwo~\cite{Catani:1996vz},  \pythia{8.226} \cite{Sjostrand:2006za,Sjostrand:2014zea}, and a calculation using the $1\to 3$ splitting functions that is discussed in detail in \App{app:1to3}. The details of the \pythia{8.226} result will be discussed in \Sec{sec:eepred_shower}. This figure illustrates two important points. First, as described in \Sec{sec:factee}, our factorization formula for the $D_2$ observable isolates the collinear physics. If we did not want to resum the small $D_2$ behavior, then this shows that the fixed order result can be computed, up to power corrections, using the $1\to 3$ splitting function. This is seen by the excellent agreement between the result of  \eventtwo~and the result computed using the $1\to3$ splitting functions, shown in \Fig{fig:eesplit}. Second, our factorization formulae describe both the small $D_2$ region, where there is a resolved substructure, as well as the large $D_2$ region, where the substructure is unresolved. A correct description of the unresolved region of phase space, with the collinear-soft haze factorization of \Sec{sec:fact_un}, is required to describe the correct endpoint of the distribution, which occurs at $1/(2\zcut)$. In the collinear-soft haze factorization, we do not need to resum logarithms of $D_2$, and therefore we can simply compute to fixed order, which is equivalent to a fixed order calculation using the $1\to 3$ splitting function. In \Fig{fig:eesplit}, we see first of all that all three curves reproduce well the expected $1/(2\zcut)$ endpoint, and second, that the calculation based on the $1\to 3$ splitting function describes relatively well the distribution at large values of $D_2$, and in particular, the approach to the endpoint.\footnote{The precise behavior of the $1\rightarrow 3$ splitting function calculation and \eventtwo$\,$  at the endpoint becomes sensitive to the binning used in this region, since the distribution is rapidly vanishing. One must trade accuracy of reproducing the endpoint for numerical stability of the bins. We have found that using smaller binning always improves the agreement between \eventtwo~and the $1\to 3$ splitting functions, at the expense of having to run longer to achieve adequate accuracy and precision.} This is important, since it illustrates that already at LO one can have a reasonable description of the endpoint of the distribution, and that the phase space of the observable is already reasonably well filled out. In \Sec{sec:eepred_shower} we will further study this in the matched distributions for different values of $\zcut$.

\subsection{Comparison with Parton Shower Monte Carlo}\label{sec:eepred_shower}

Having shown that we reproduce the singular structure of the $D_2$ distribution, in this section we compare our NLL resummed predictions multiplicatively matched to the leading order (LO) {\tt EVENT2} or $1\to 3$ splitting functions with parton shower Monte Carlo. For QCD jets, we consider both $e^+e^-\to q\bar q$, as well as $e^+e^-\to gg$, generated through an off shell Higgs, while for signal, we consider $e^+e^-\to ZZ$ events with both $Z$s decaying hadronically. The events were generated with \madgraph{2.5.5} \cite{Alwall:2014hca}, and showered with \pythia{8.226} \cite{Sjostrand:2006za,Sjostrand:2014zea}. We also verified that similar results are obtained with \vincia{} \cite{Giele:2007di,Giele:2011cb,GehrmannDeRidder:2011dm,Ritzmann:2012ca,Hartgring:2013jma}, although for simplicity we do not show distributions from \vincia{}. Throughout this section we use \fastjet{3.1.2} \cite{Cacciari:2011ma} and the \texttt{EnergyCorrelator} \fastjet{contrib} \cite{Cacciari:2011ma,fjcontrib} for jet clustering and analysis. All jets are clustered using the $e^+e^-$ anti-$k_T$ metric \cite{Cacciari:2008gp,Cacciari:2011ma} using the WTA recombination scheme \cite{Larkoski:2014uqa,Larkoski:2014bia}, with an energy metric.

\begin{figure}
\begin{center}
\subfloat[]{\label{fig:eeanal05}
\includegraphics[width=7.5cm]{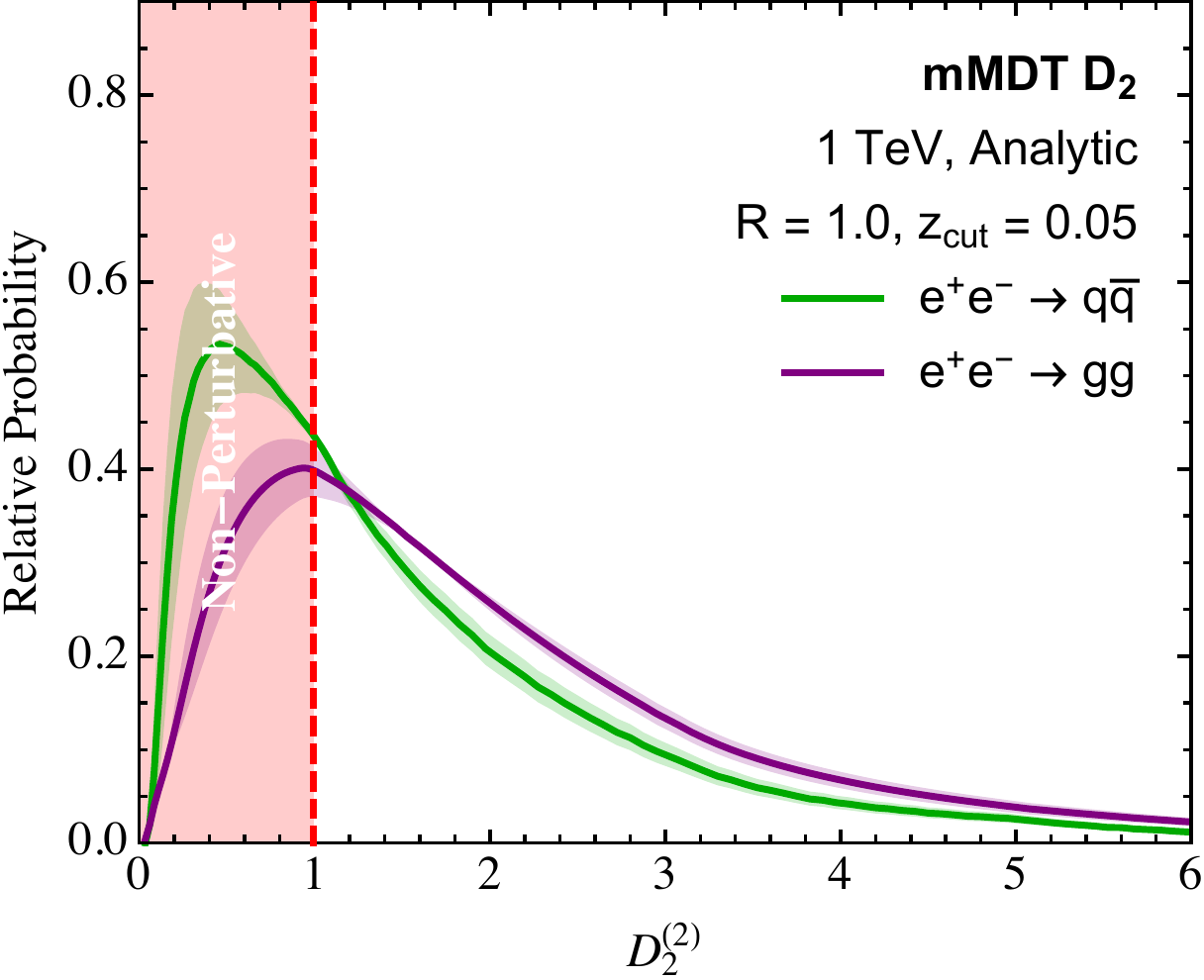}
}\ \ 
\subfloat[]{\label{fig:eepythia05}
\includegraphics[width = 7.5cm]{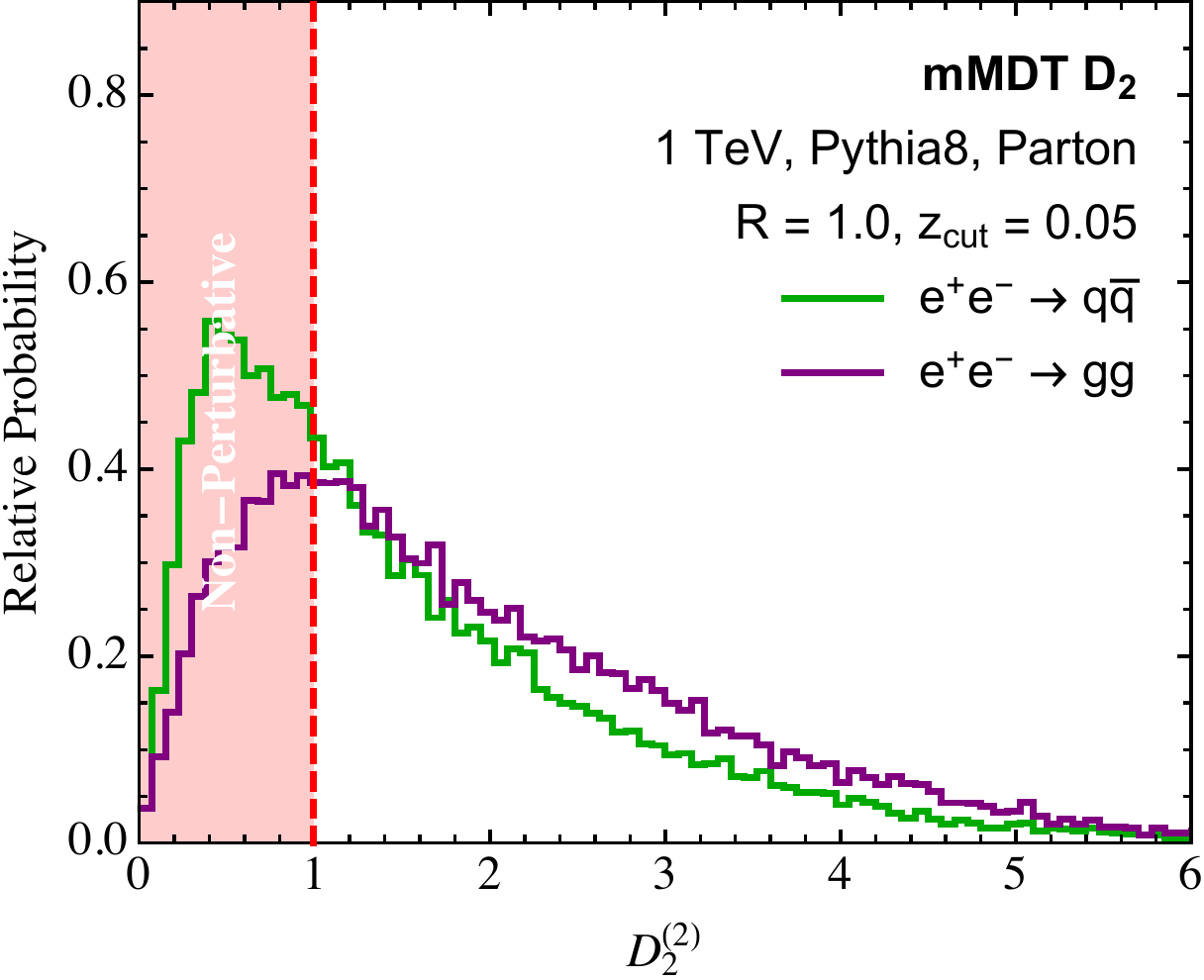}
}\\
\subfloat[]{\label{fig:eeanal}
\includegraphics[width=7.5cm]{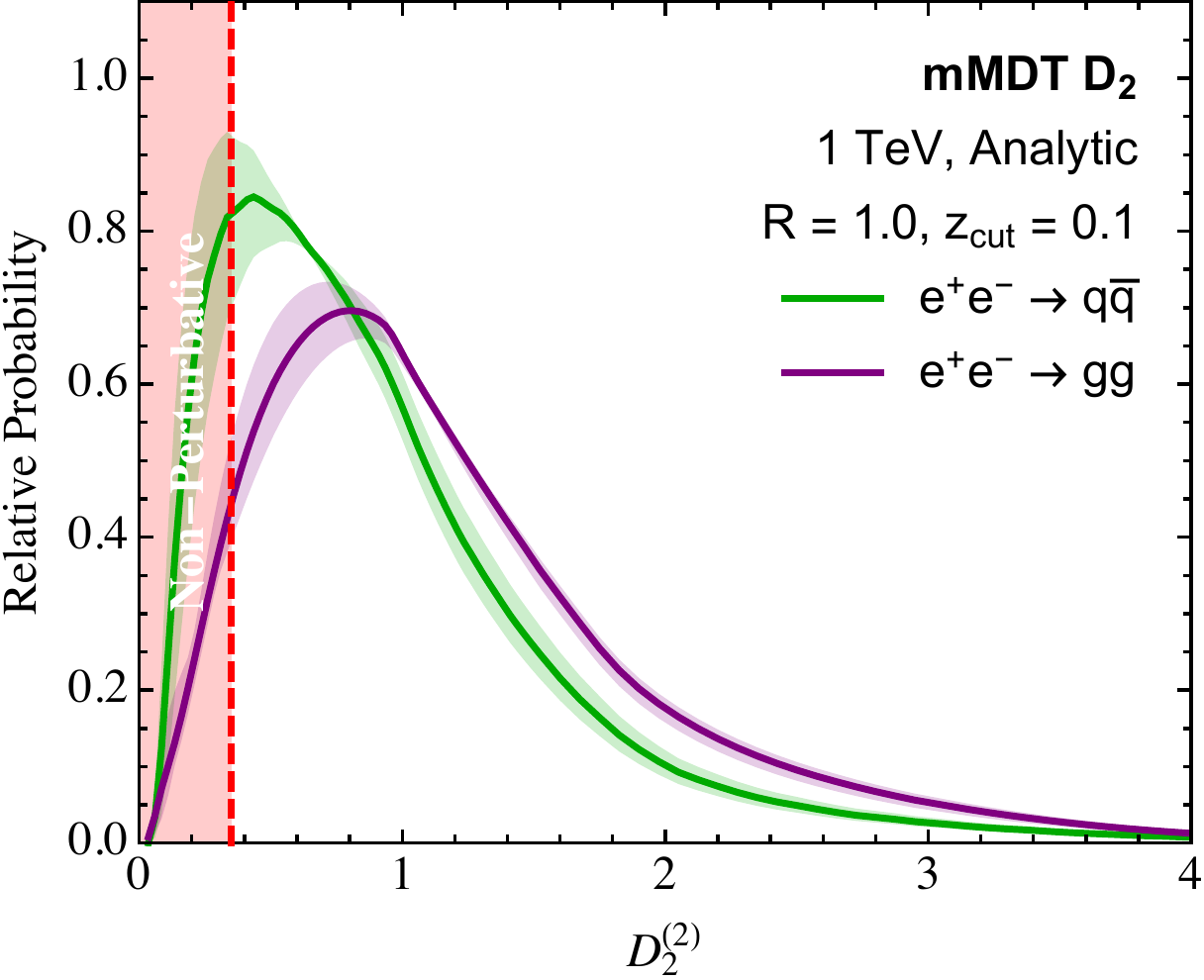}
}\ \ 
\subfloat[]{\label{fig:eepythia}
\includegraphics[width = 7.5cm]{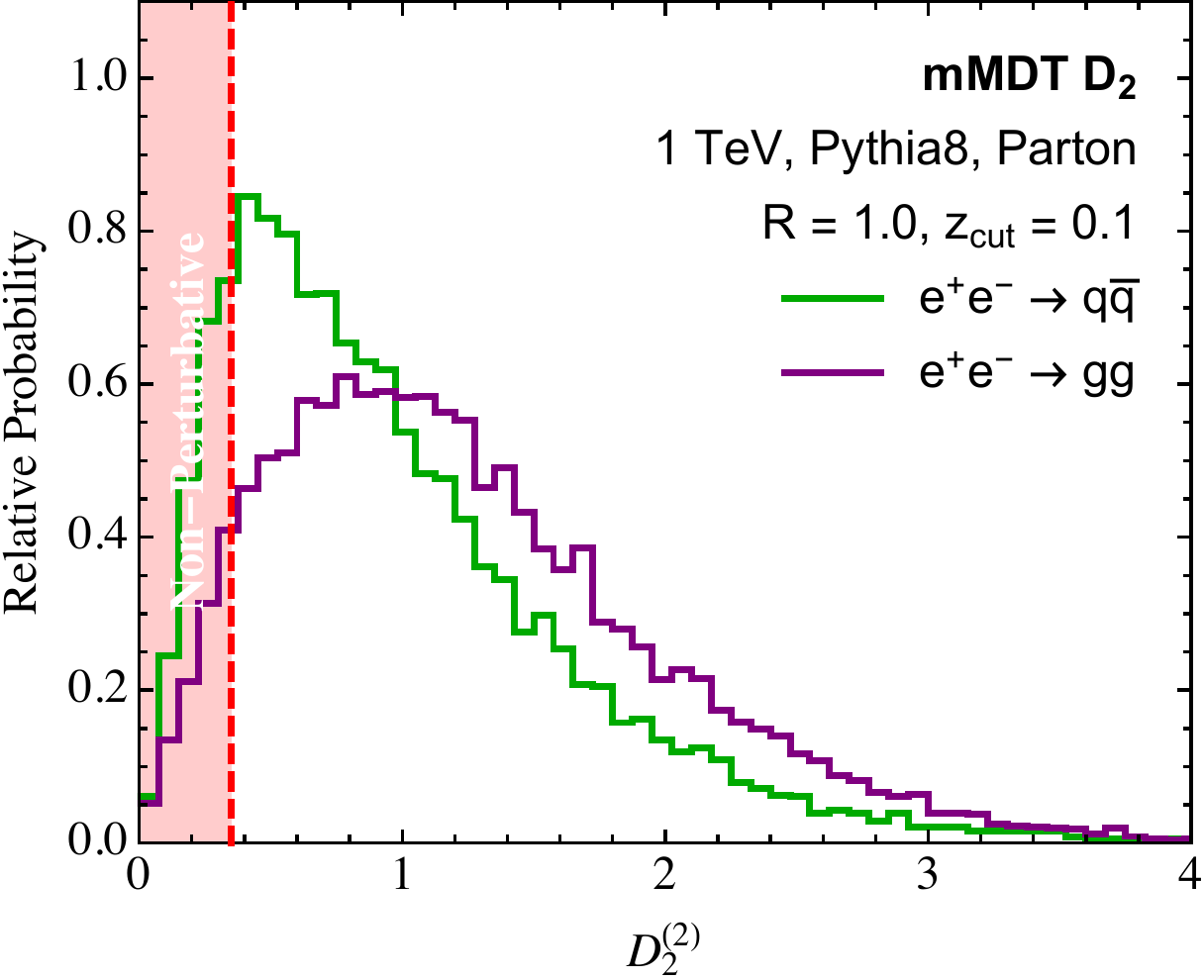}
}\\
\subfloat[]{\label{fig:eeanal2}
\includegraphics[width=7.5cm]{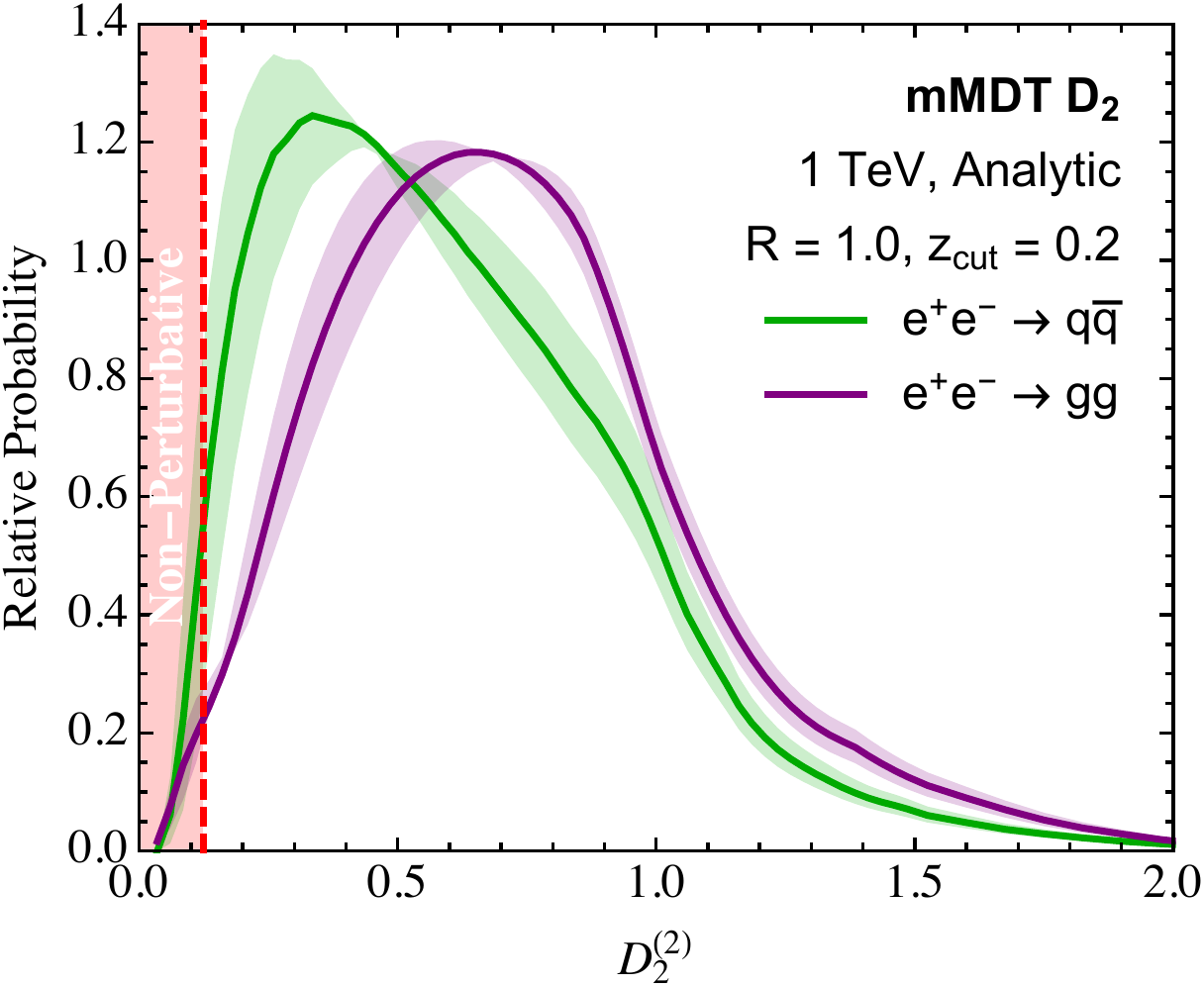}
}\ \ 
\subfloat[]{\label{fig:eepythia2}
\includegraphics[width = 7.5cm]{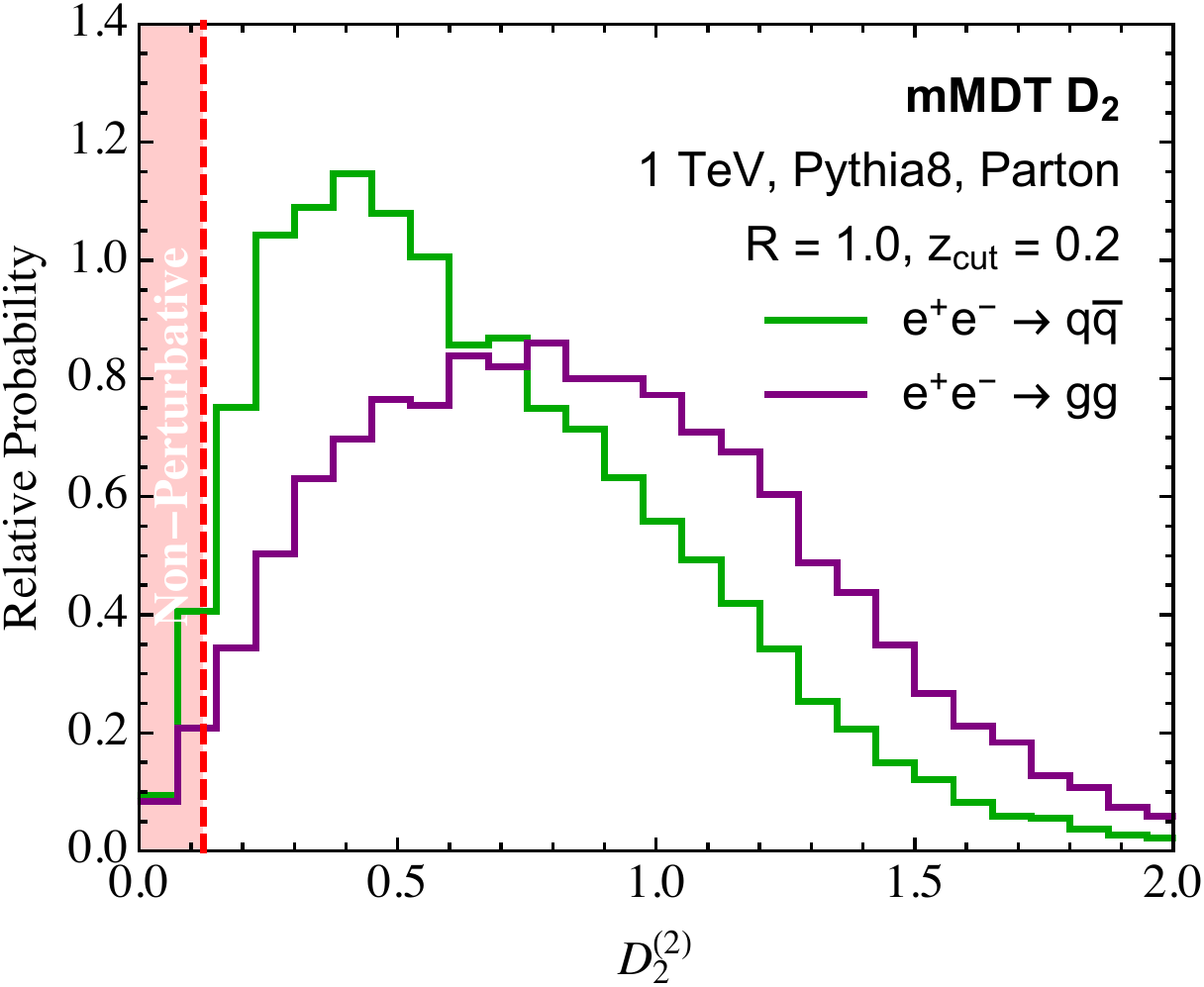}
}
\end{center}
\caption{A comparison of the analytic $D_2$ distributions for both quark and gluons (left column) with parton shower Monte Carlo at parton level (right column), for different values of the $\zcut$ parameter.  A mass cut of $m_J \in [80,100]$ GeV has been applied. Good agreement in the shape is observed, particularly for smaller values of $\zcut$.
}
\label{fig:eecompplots}
\end{figure}

In \Fig{fig:eecompplots} we show comparisons of our analytic predictions (on the left) with parton shower Monte Carlo results at parton level (on the right). Results are shown for both quark and gluon jets. We also highlight the region where non-perturbative effects from hadronization will have a significant impact on the distribution, as will be discussed shortly. The distributions are shown for three different values of the $\zcut$ parameter, namely $\zcut=0.05,0.1,0.2$. Overall, good agreement between the analytic calculation and the parton shower Monte Carlo is observed, and differences between quarks and gluons, as well as the behavior as a function of $\zcut$ are well reproduced. In particular, due to the inclusion of the fixed order corrections, the correct endpoint of the distribution is obtained in the analytic calculation. This is crucial for obtaining agreement of the distributions.

\begin{figure}
\begin{center}
\subfloat[]{\label{fig:eeanalz}
\includegraphics[width=7.5cm]{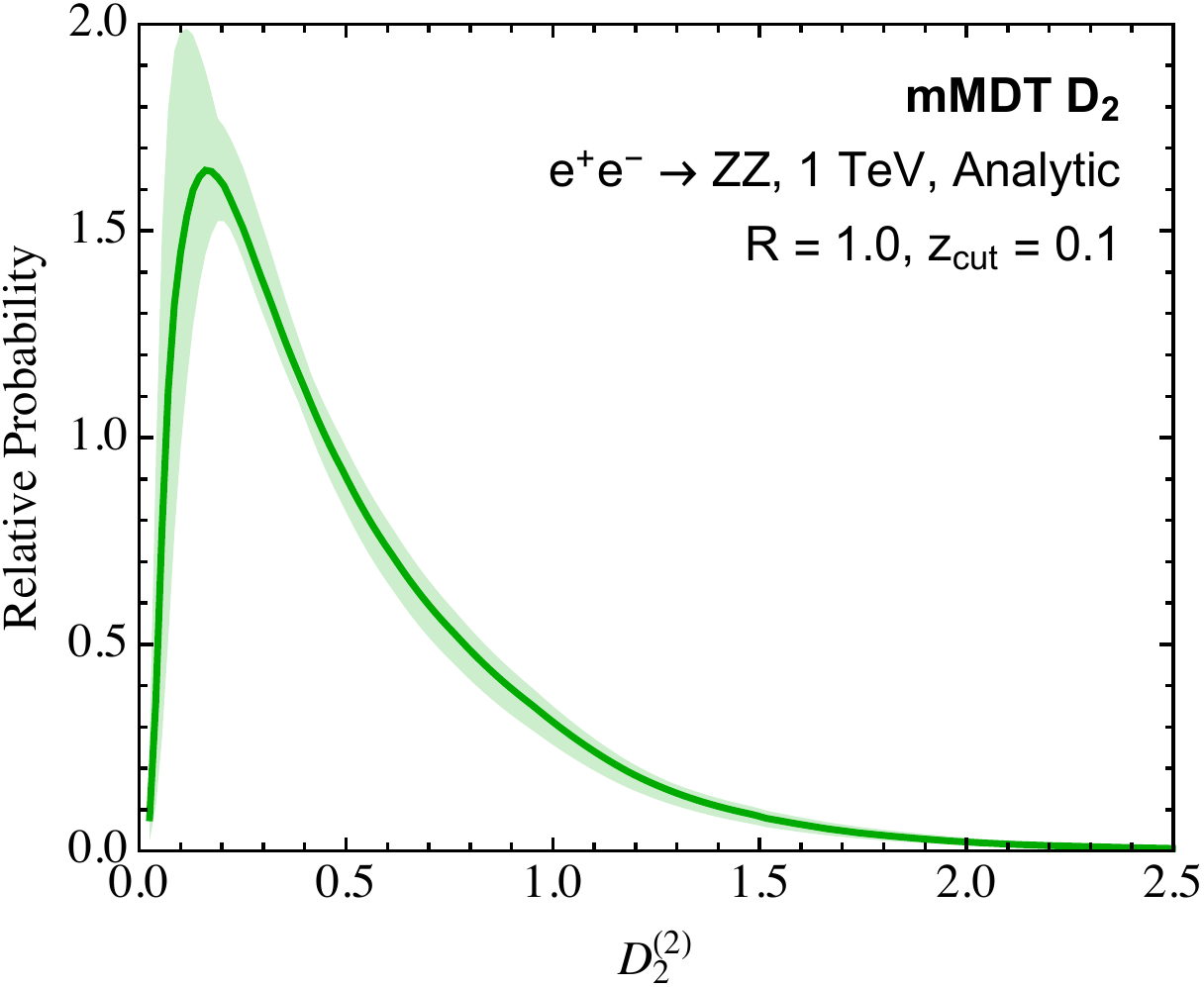}
}\ \ 
\subfloat[]{\label{fig:eepythiaz}
\includegraphics[width = 7.5cm]{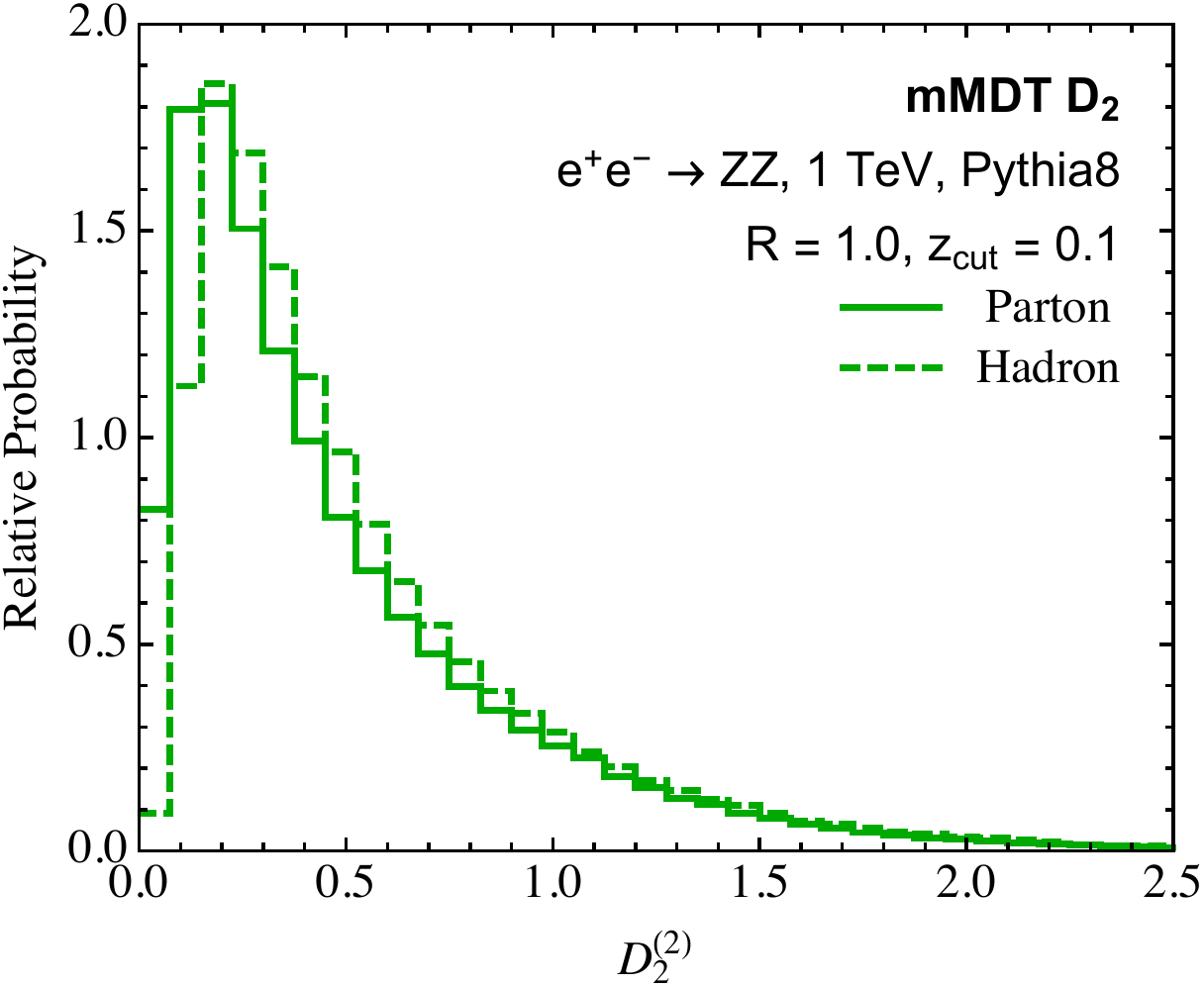}
}
\end{center}
\caption{A comparison of the analytic $D_2$ distribution for signal (Z) jets (left) and the parton shower Monte Carlo distribution (right). A mass cut of $m_J \in [80,100]$ GeV has been applied. Good agreement is observed.
}
\label{fig:eezcompplots}
\end{figure}

\begin{figure}
\begin{center}
\subfloat[]{\label{fig:eeqcomp05}
\includegraphics[width=7.5cm]{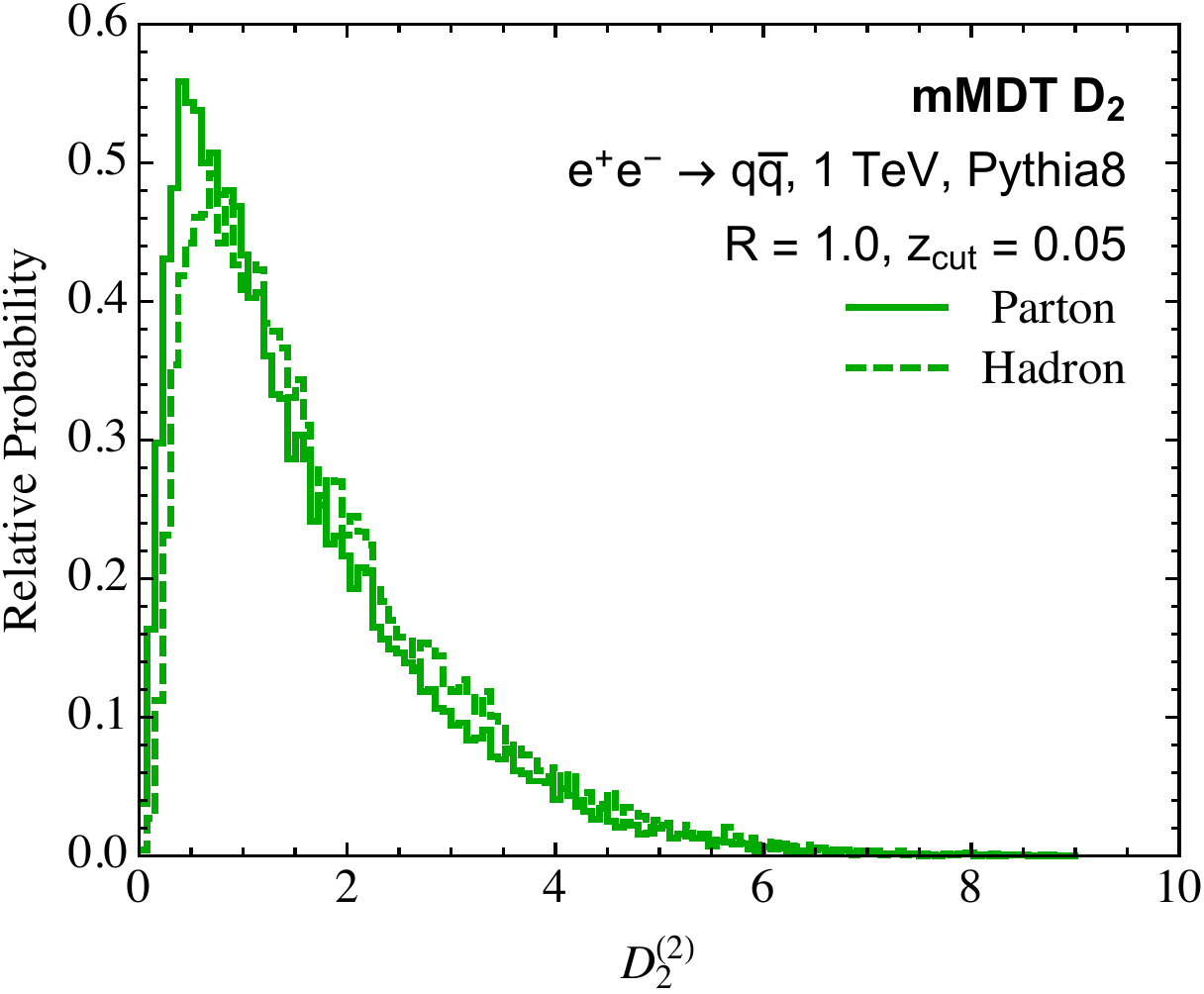}
}\ \ 
\subfloat[]{\label{fig:eegcomp05}
\includegraphics[width = 7.5cm]{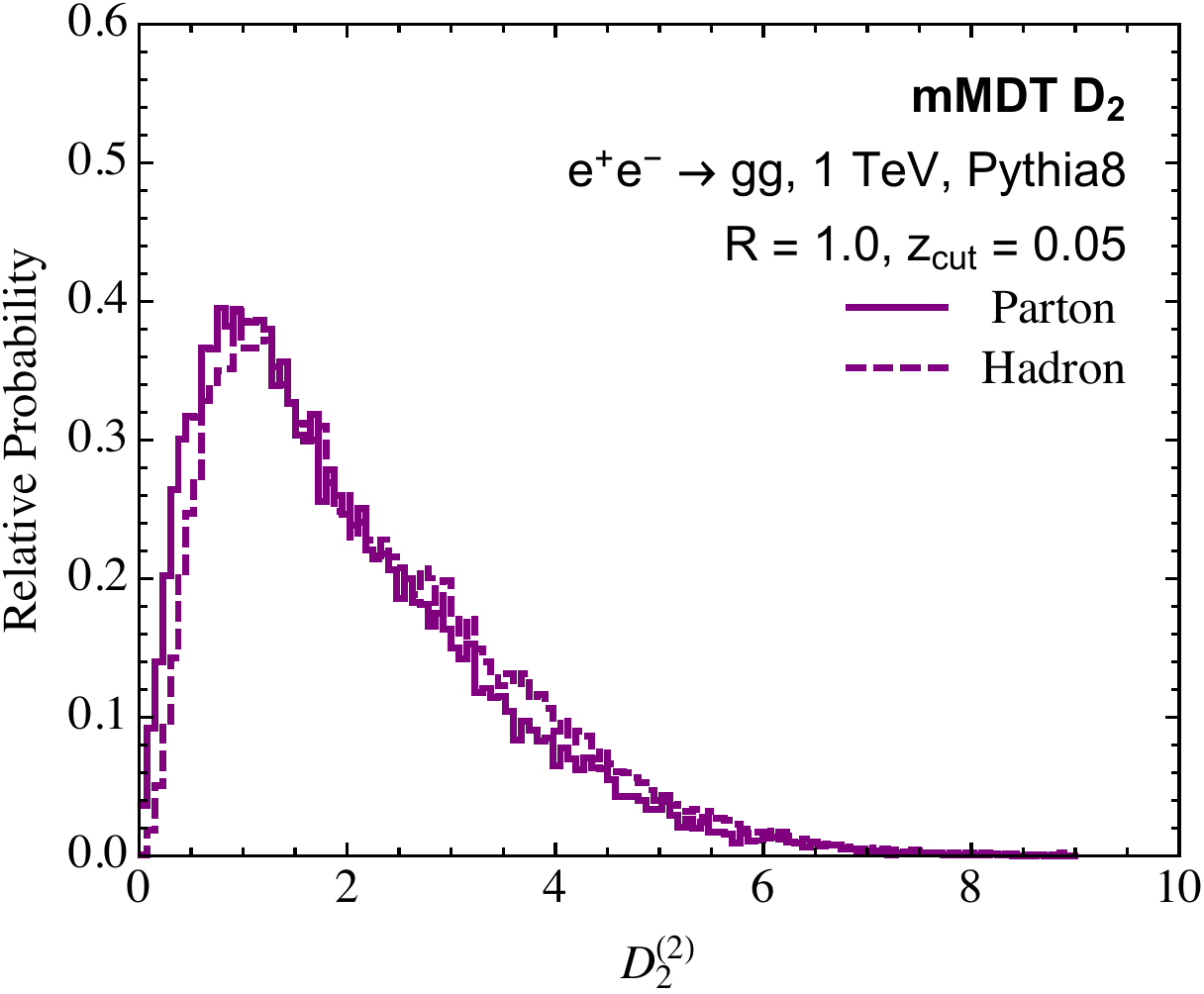}
}\\
\subfloat[]{\label{fig:eeqcomp}
\includegraphics[width=7.5cm]{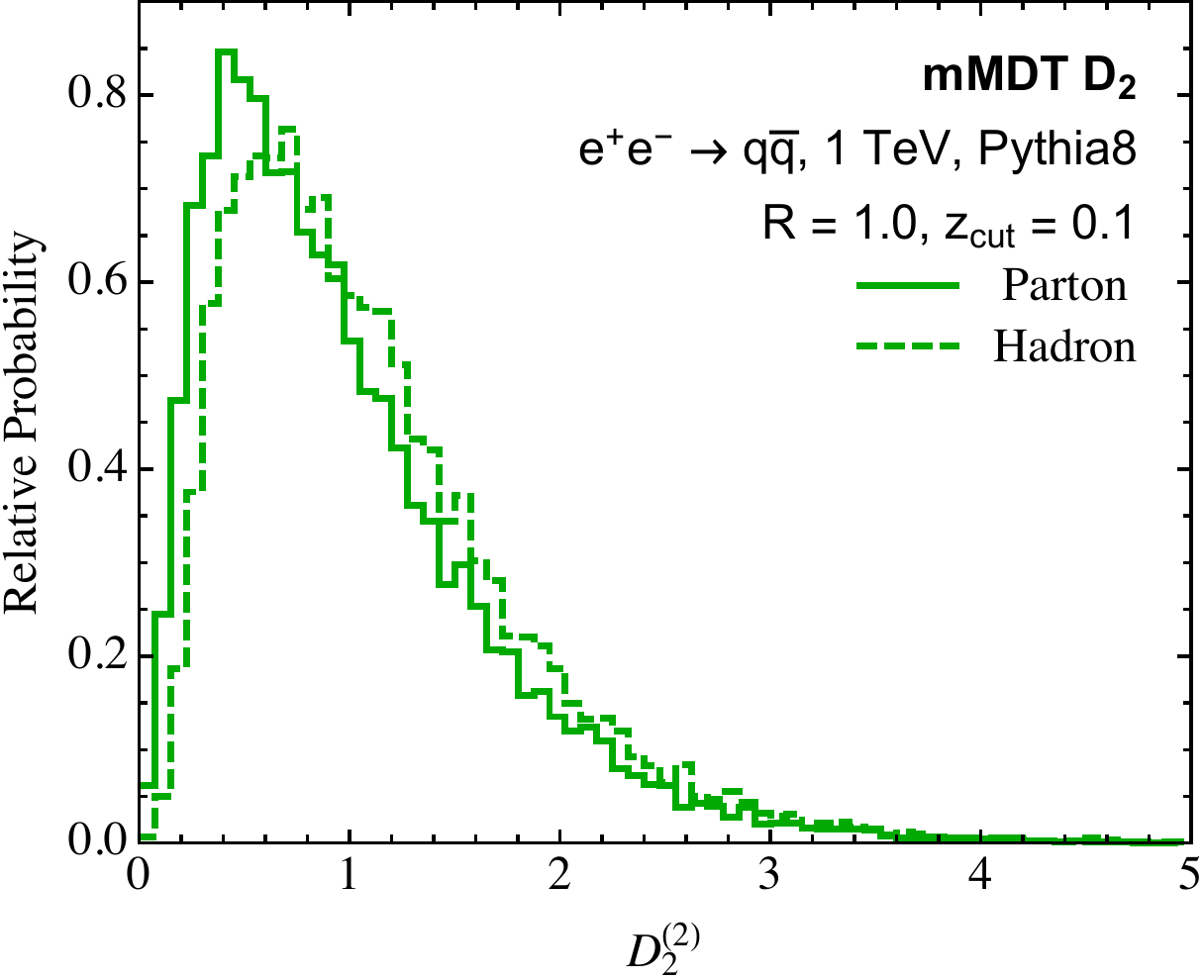}
}\ \ 
\subfloat[]{\label{fig:eegcomp}
\includegraphics[width = 7.5cm]{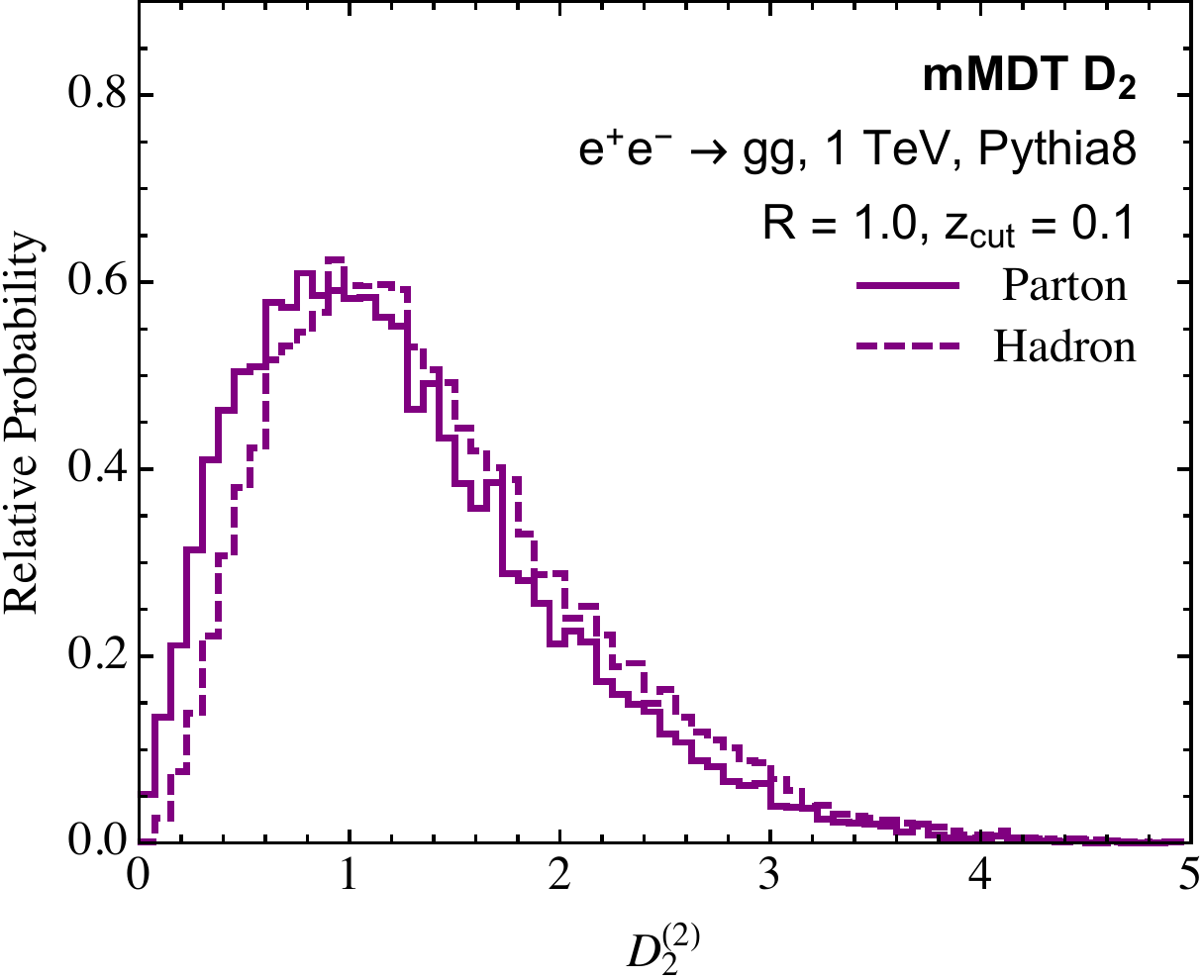}
}\\
\subfloat[]{\label{fig:eeqcomp2}
\includegraphics[width=7.5cm]{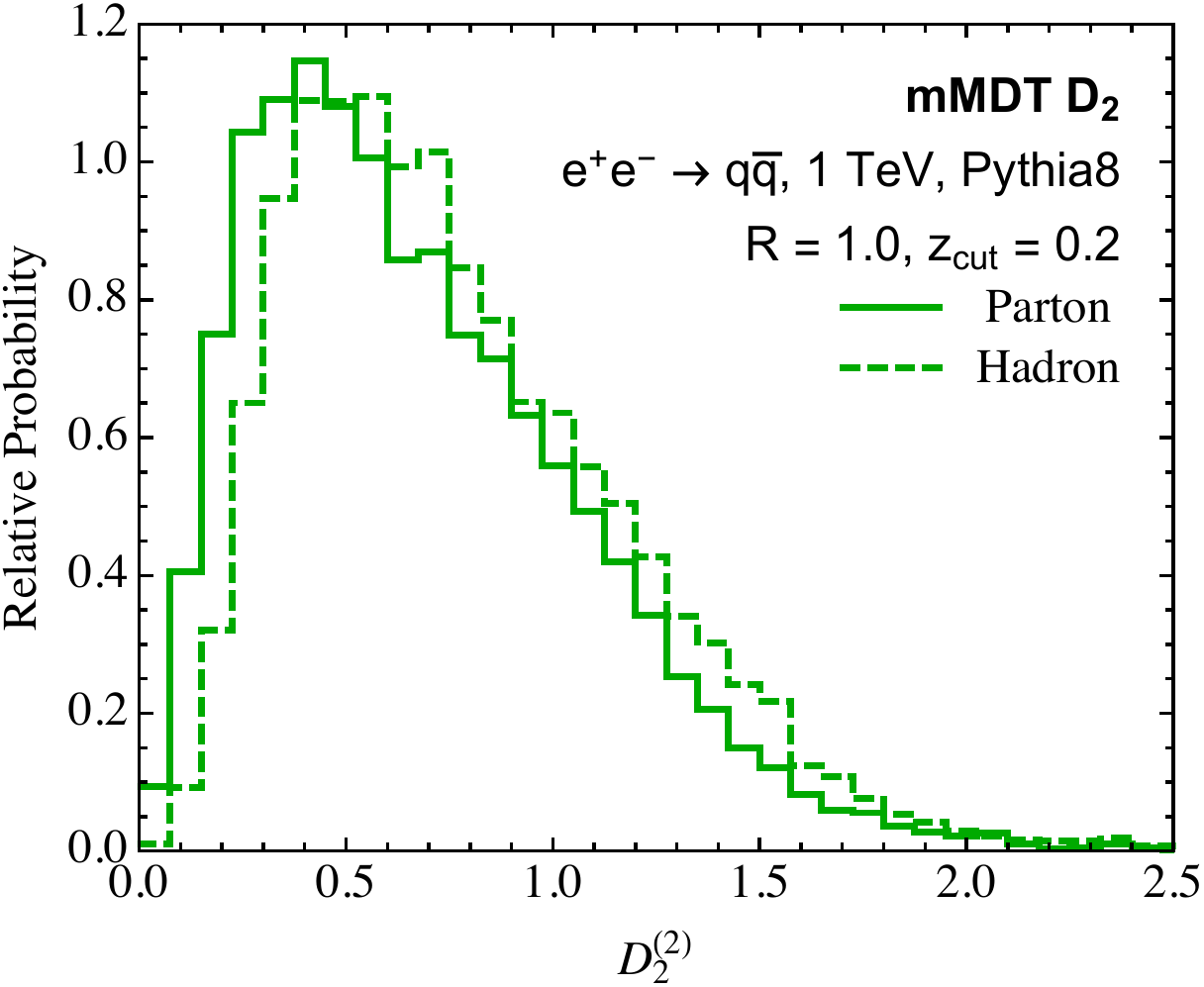}
}\ \ 
\subfloat[]{\label{fig:eegcomp2}
\includegraphics[width = 7.5cm]{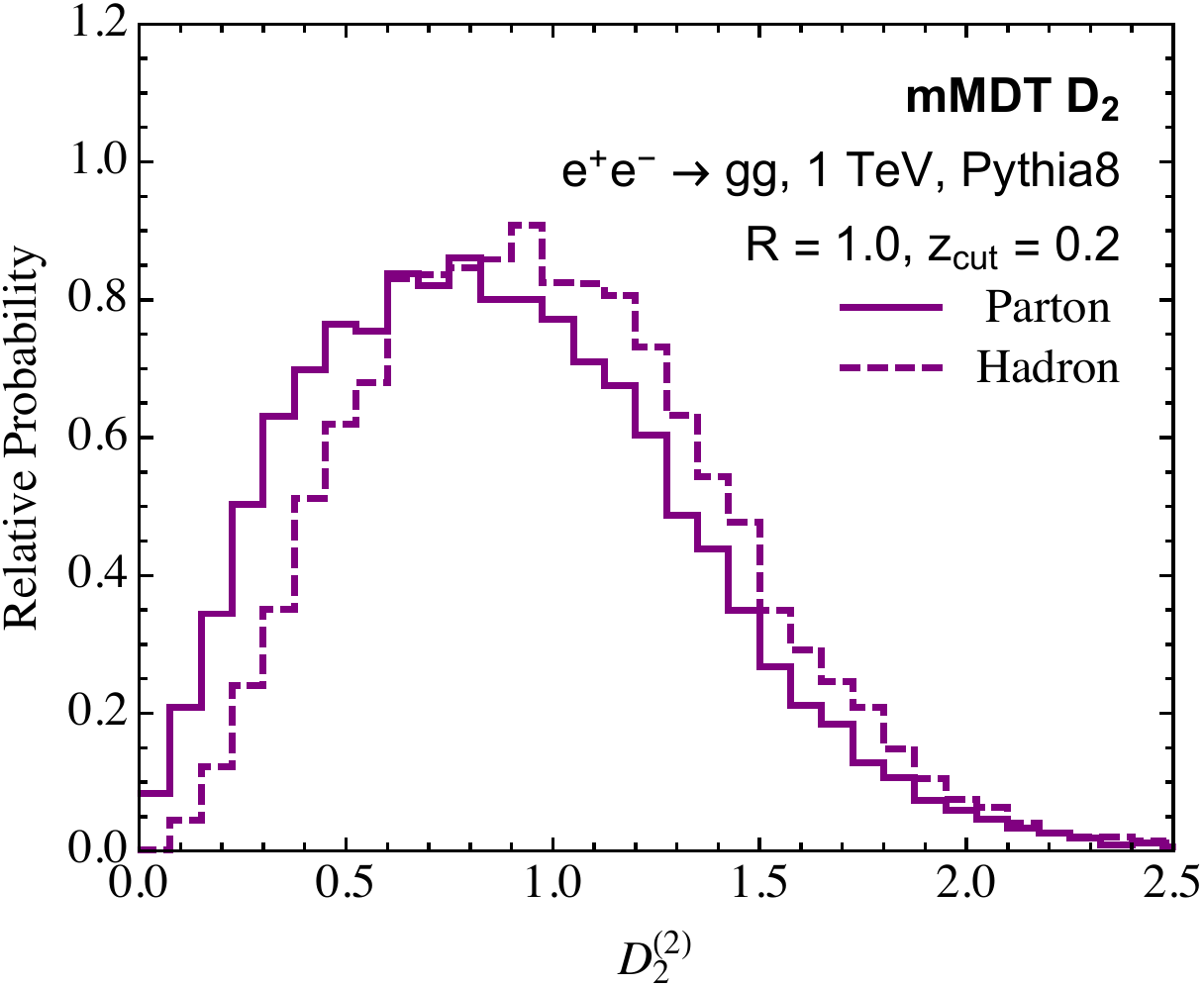}
}
\end{center}
\caption{The effect of hadronization on the $D_2$ distribution for quark (left column) and gluon (right column) jets. Non-perturbative corrections are suppressed by the grooming procedure. A mass cut of $m_J \in [80,100]$ GeV has been applied.
}
\label{fig:eecomphad}
\end{figure}

It is interesting to consider the behavior as a function of $\zcut$, and in particular the differences between the parton shower and analytic results as the value of $\zcut$ is increased, corresponding to a more aggressive grooming. In \Fig{fig:eecompplots} we see that for smaller values of $\zcut$, where the grooming has a smaller effect, better agreement is observed between the analytic and parton shower results. We believe that this arises primarily due to two effects. First, since the endpoint of the distribution scales as $1/\zcut$, for smaller values of $\zcut$, there is a less rapid transition between the resummation and fixed order regime, ensuring that there are well separated resummation and fixed orders regions. Secondly,  we observe that for more aggressive grooming the LO fixed order prediction undershoots the Monte Carlo distribution at large values of $D_2$. Since we are considering normalized distributions, this then translates into a large difference in the peak region, as seen most clearly for $\zcut=0.2$. We believe that this could be remedied by including higher order perturbative corrections, which would fill out the phase space better, more similar to the parton shower. Indeed, the LO predictions are known to undershoot in other grooming or $D_2$ studies \cite{Larkoski:2015uaa,Dasgupta:2015lxh,Frye:2016aiz,Marzani:2017mva,CMS:2017tdn}. It would be very interesting to include higher order corrections, but this is beyond the scope of the current paper.  Although our focus in this paper is primarily on the calculation of groomed $D_2$ for QCD jets, in \Fig{fig:eezcompplots} we show a comparison of our analytic results with parton shower Monte Carlo for a hadronically decaying boosted Z. Excellent agreement is observed.

Finally, it is important to address the impact of hadronization corrections to the distribution. Although it was shown in \Sec{sec:hadr_suppress} that hadronization corrections are suppressed significantly by the grooming procedure, they still have a non-negligible impact on the $D_2$ distribution. In \Fig{fig:eecomphad} we show the effect of hadronization on the groomed $D_2$ spectrum in parton shower Monte Carlo for both quark and gluon jets, and for the different values of $\zcut$ considered above. In \Fig{fig:eepythiaz} the effect of hadronization is shown for the boosted $Z$ distribution. We see that in all cases, the effect of hadronization is quite minor, and does not dominate the shape of the distribution. It can be included in the analytic calculation using a model shape function \cite{Korchemsky:1999kt,Korchemsky:2000kp,Bosch:2004th,Hoang:2007vb,Ligeti:2008ac}, although we will not pursue this further in this paper. A detailed study of the non-perturbative corrections for the $D_2$ distribution at the LHC, and their incorporation through shape functions, are presented in a companion paper \cite{Larkoski:2017iuy}.

\section{Monte Carlo Results in $pp$ Collisions}\label{sec:pppred}

Analytic predictions of the $D_2$ distribution for $pp$ processes of interest are presented in a companion paper \cite{Larkoski:2017iuy}. In this section we perform a Monte Carlo study demonstrating the consequences of the power counting and factorization analysis presented earlier. For convenience, we recall here the major predictions of our analysis. We emphasize that these are robust predictions of the factorization formula, combined with the power counting analysis, which hold within the region of validity of the factorization formula, namely $\ecf{2}{\alpha}\ll \zcut$. They should therefore be independent of the details of the hadronization model or details of the perturbative shower. The non-trivial predictions are:
\begin{itemize}
\item The endpoint of the $D_2$ distribution is fixed as $1/(2\zcut)$. This is independent of the jet mass, jet energy, hadronization, or the angular exponent used to define the energy correlation functions.
\item The scale at which hadronization corrections become important is independent of the $p_T$ of the jet. It depends only on $\Lambda_{\text{QCD}}$, the jet mass, and the value of the $\zcut$ parameter.
\item The distributions depend only the quark vs.~gluon fraction of the jets in the event, but are otherwise process independent. We have previously argued that the soft drop procedure also reduces the dependence of the distribution on the parton flavor.
\end{itemize}
We will see that each of these predictions is well reproduced by Monte Carlo parton shower in $pp$ collisions. The parton-level samples in this section were generated at the 13 TeV LHC with \madgraph{2.5.5} \cite{Alwall:2014hca} and showered with \pythia{8.226} \cite{Sjostrand:2006za,Sjostrand:2014zea} with default settings.  Jets were clustered with the anti-$k_T$ algorithm \cite{Cacciari:2008gp} in \fastjet{3.1.2}  \cite{Cacciari:2011ma} and the \texttt{EnergyCorrelator} and \texttt{SoftDrop} \fastjet{contrib}s \cite{Cacciari:2011ma,fjcontrib} for jet analysis.

\subsection{Signal Distributions}

\begin{figure}
\begin{center}
\subfloat[]{\label{fig:ppzzpt}
\includegraphics[width=7.5cm]{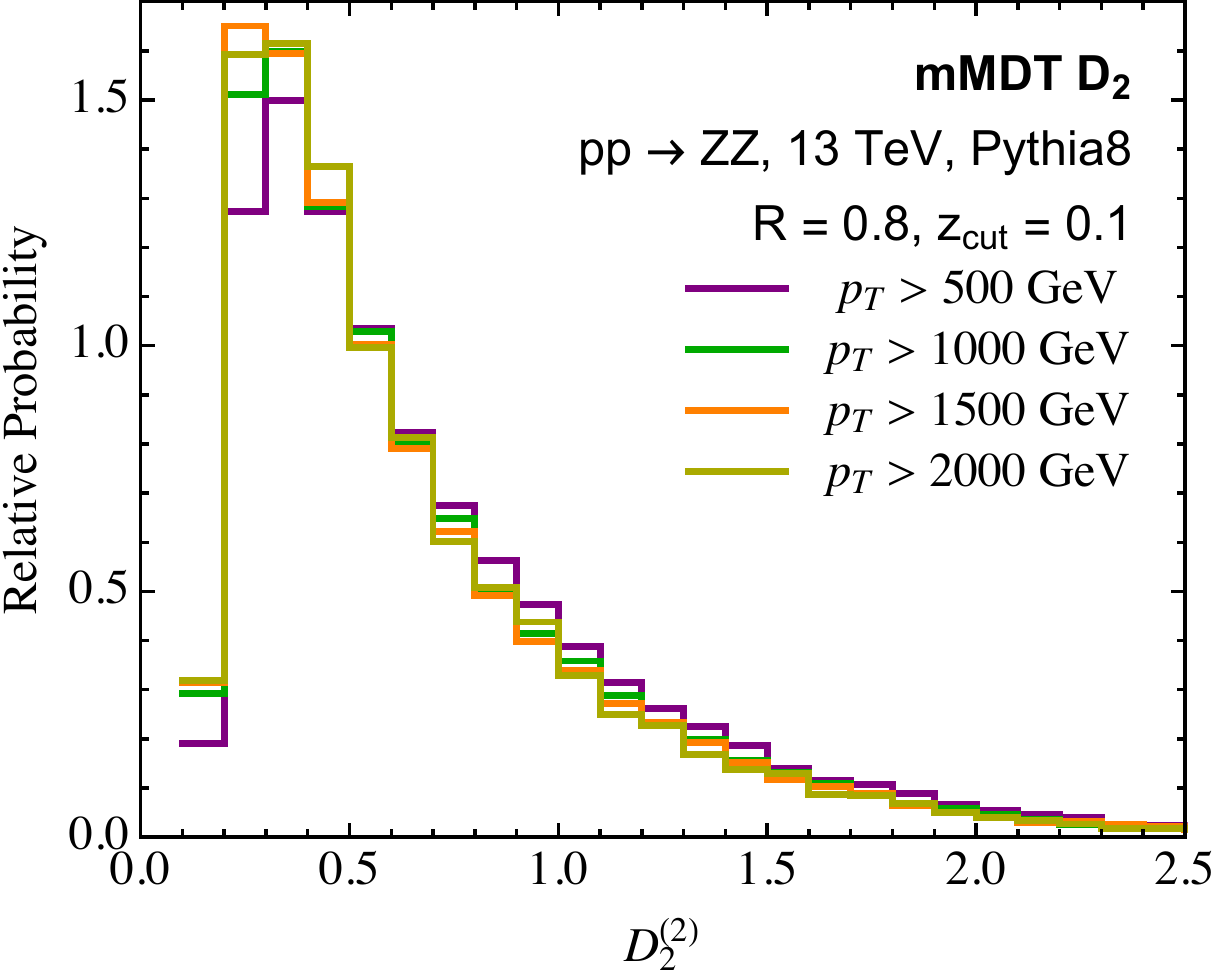}
}\ \ 
\subfloat[]{\label{fig:ppzzR}
\includegraphics[width = 7.5cm]{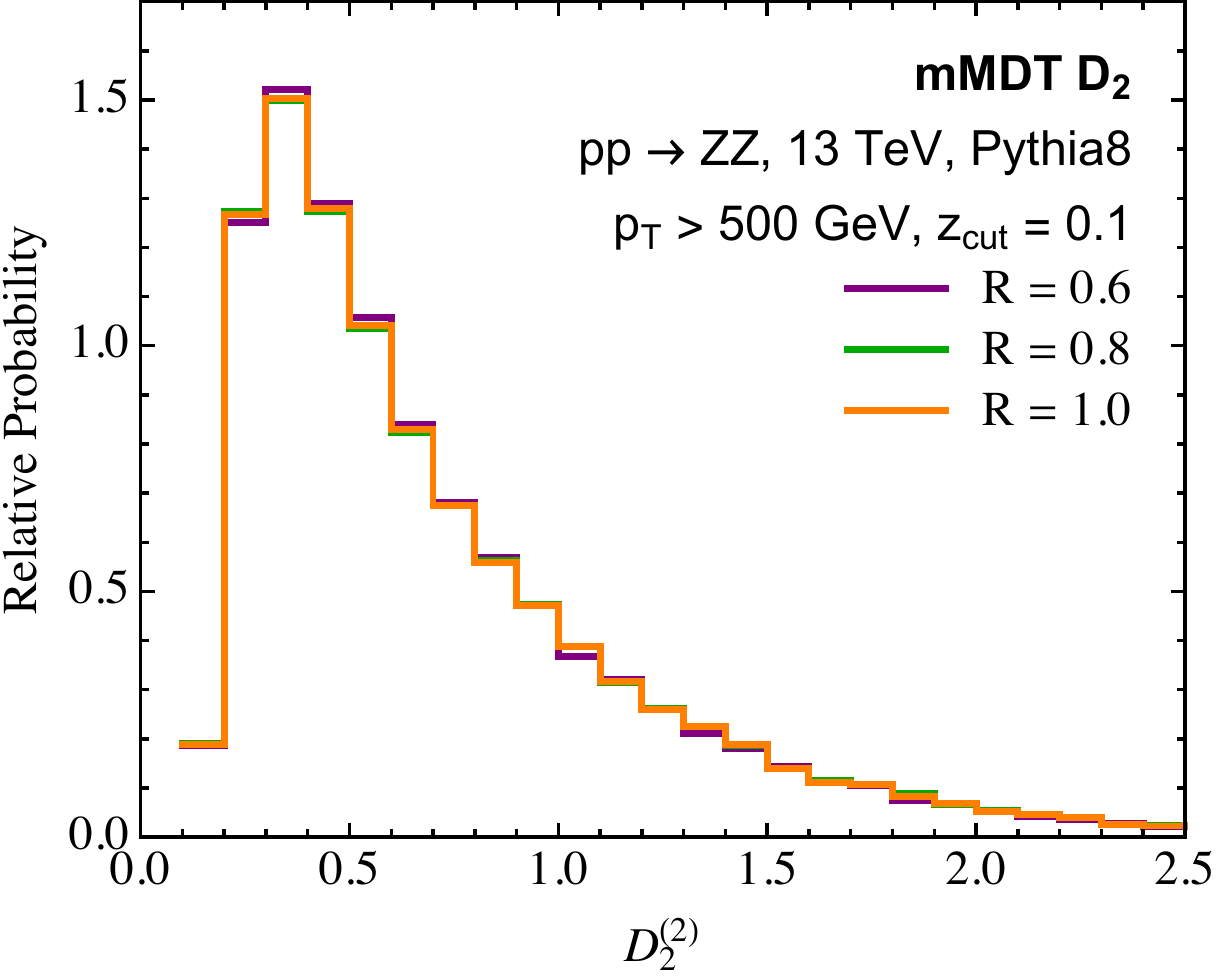}
}
\end{center}
\caption{Distributions of $\Dobs{2}{2}$ measured on mMDT groomed hadronically-decaying $Z$ jets with $\zcut = 0.1$ from the process $pp\to ZZ$ at the 13 TeV LHC.  The groomed jet mass is restricted to lie in the range  $80\text{ GeV} < m_J < 100\text{ GeV}$.  (a) Distributions of $\Dobs{2}{2}$ for various $p_T$ cuts, with the jet radius fixed to $R= 0.8$. (b) Distributions of $\Dobs{2}{2}$ for various jet radii $R$, with the jet $p_T$ cut fixed to $p_T > 500$ GeV.
}
\label{fig:ppzzplots}
\end{figure}

We first show the distributions for signal jets.  The signal jets are clustered from $pp\to ZZ$ events in which one $Z$ boson is forced to decay to neutrinos, and the other to hadrons.  In \Fig{fig:ppzzplots}, we plot the groomed jet $\Dobs{2}{2}$ distributions of the hadronically-decaying $Z$ boson.  In these plots, the mMDT groomer (soft drop with $\beta = 0$) is used, with the parameter $\zcut = 0.1$.  A mass cut of $80\text{ GeV} < m_J < 100\text{ GeV}$ is also imposed on the groomed jet.  First, in \Fig{fig:ppzzpt}, we plot the $\Dobs{2}{2}$ distribution for various $p_T$ cuts on the jets.  The signal distributions are stable with respect to $p_T$, which might be expected as the $Z$ boson is a color-singlet and has an intrinsic, Lorentz-invariant energy scale, $m_Z$.  Importantly, the grooming appears to successfully remove contamination radiation at wide angles in the jet that would distort the distribution, and potentially become more important at larger $p_T$.   

In \Fig{fig:ppzzR}, we fix the transverse momentum range to $p_T > 500$ GeV, and vary the clustered jet radius from $R=0.6$ to $R=1.0$.  Essentially no effect on the distribution is observed in changing the jet radius, corroborating the performance of the mMDT groomer in removing contamination radiation.  Note also that with this $p_T$ range and these jet radii, the Z boson decay products are well-contained within the jet.  For this $p_T$ range, the angular scale of the $Z$ boson decay products $R_Z$ is approximately
\begin{equation}\label{eq:jetrscale}
R_Z \simeq \frac{2m_Z}{p_T}\simeq 0.4\,.
\end{equation}

\subsection{Background Distributions}

It may have been expected that the signal distributions were robust under changes of jet parameters, both due to grooming as well as the intrinsic mass scale.  Here, we will show that mMDT/soft drop grooming also renders the background distributions extremely robust to jet parameters.  We study jets produced in two processes for our background: $pp\to Zj$ and $pp\to Hj$.  The production of the $Z$ or $H$ bosons in association with the jet enables a handle on the quark and gluon jet fractions.  Because soft drop grooming formally makes quark and gluon jet definitions infrared and collinear safe, we could in principle extract the individual quark and gluon jet distributions of $\Dobs{2}{2}$ from these two samples; however, since separate quark and gluon distributions were studied in the context of $e^+e^-$ in \Sec{sec:eepred}, here we will focus only on the mixed distributions.  To easily isolate the hadronic jet in these events, we force the $Z$ boson to decay to neutrinos and the $H$ boson to decay to photons.  As with the signal events, a mass cut of $80\text{ GeV} < m_J < 100\text{ GeV}$ is imposed on the groomed jet.

\begin{figure}
\begin{center}
\subfloat[]{\label{fig:ppzjpt}
\includegraphics[width=7.5cm]{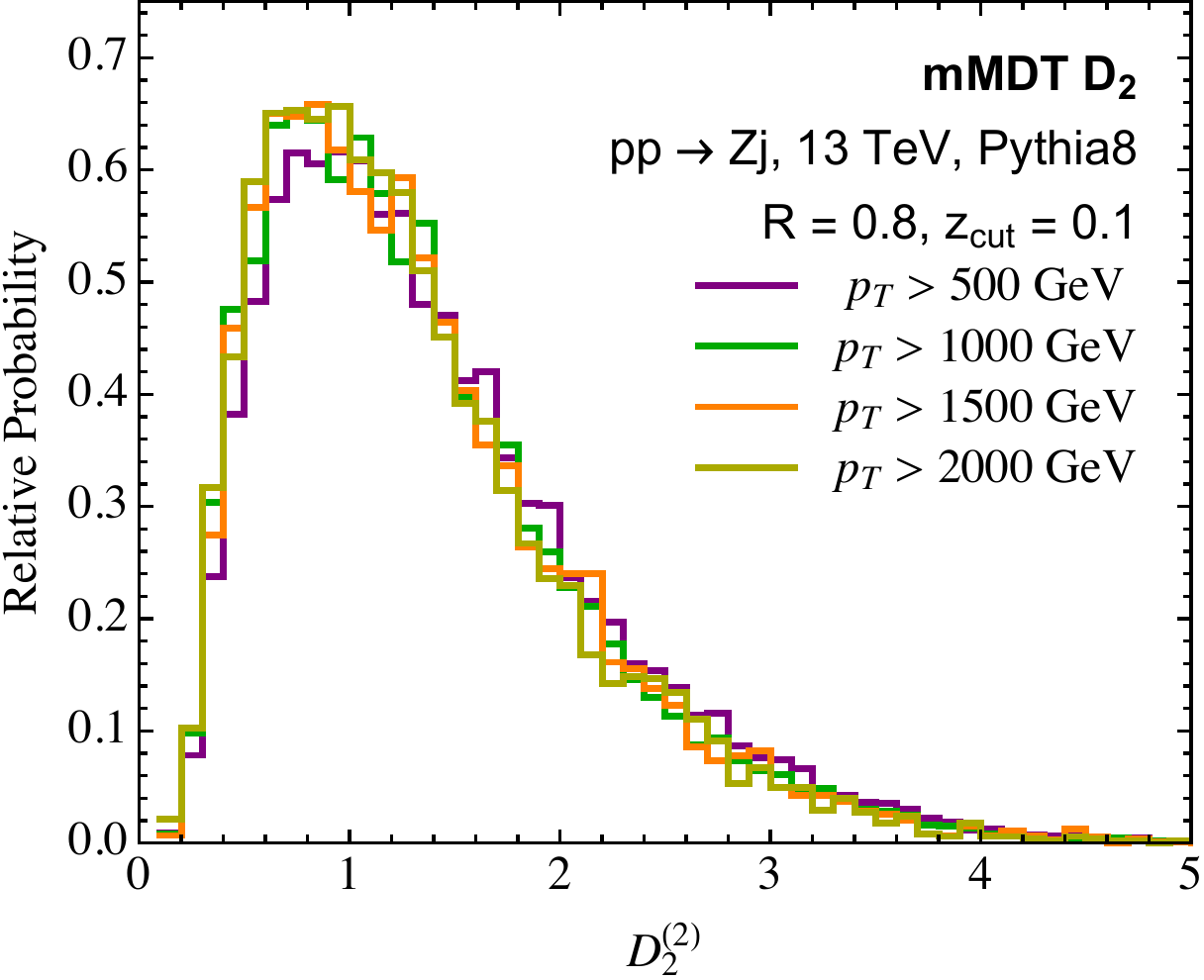}
}\ \ 
\subfloat[]{\label{fig:pphjpt}
\includegraphics[width = 7.5cm]{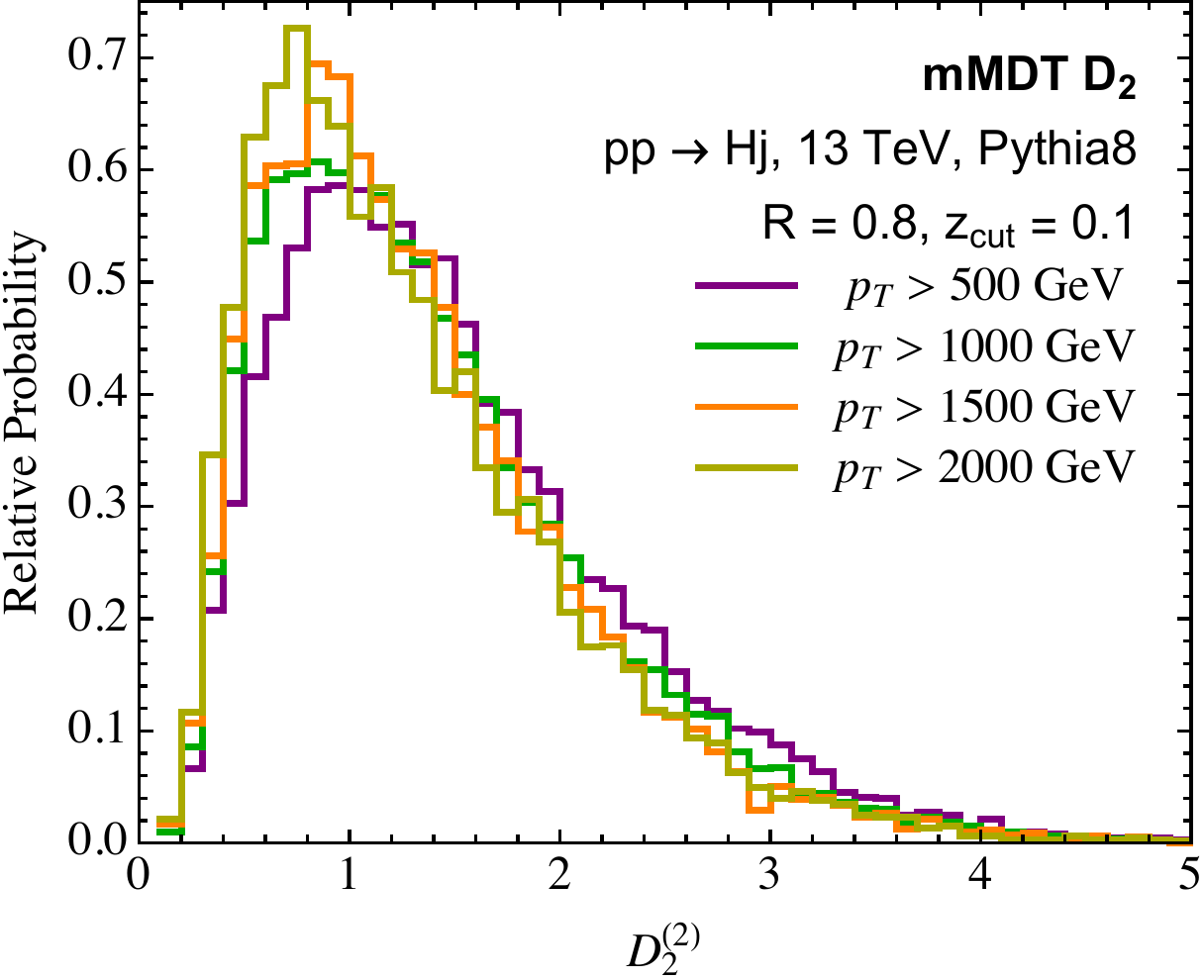}
}
\end{center}
\caption{
Distributions of $\Dobs{2}{2}$ measured on mMDT groomed QCD jets with $\zcut = 0.1$ from the processes $pp\to Zj$ (left) and $pp\to Hj$ (right) at the 13 TeV LHC.  The groomed jet mass is restricted to lie in the range  $80\text{ GeV} < m_J < 100\text{ GeV}$.  Various $p_T$ cuts are shown, with the jet radius fixed to $R= 0.8$.
}
\label{fig:ppjptplots}
\end{figure}

In \Fig{fig:ppjptplots}, we plot the mMDT $\Dobs{2}{2}$ distributions for various jet $p_T$ cuts, for both the $pp\to Zj$ and $pp\to Hj$ samples.  As predicted from our factorization formula, the background distributions are very weakly dependent on the jet $p_T$.  This is a consequence of the facts that the endpoint of the distributions are fixed at $D_2^{\max} = 1/(2\zcut) = 5$ and that non-perturbative effects become important at a scale set by the jet mass, and not the jet $p_T$.  Additionally, the distributions in the $Z$ and $H$ samples are very similar, demonstrating that quark vs.~gluon flavor effects are small.  This is a consequence of the constraints on the threshold and endpoint kinematics: the constraints on the bounds of the distribution were derived independent of jet flavor, and so the distribution must be only weakly-dependent on the jet flavor.

\begin{figure}
\begin{center}
\subfloat[]{\label{fig:ppzjR}
\includegraphics[width=7.5cm]{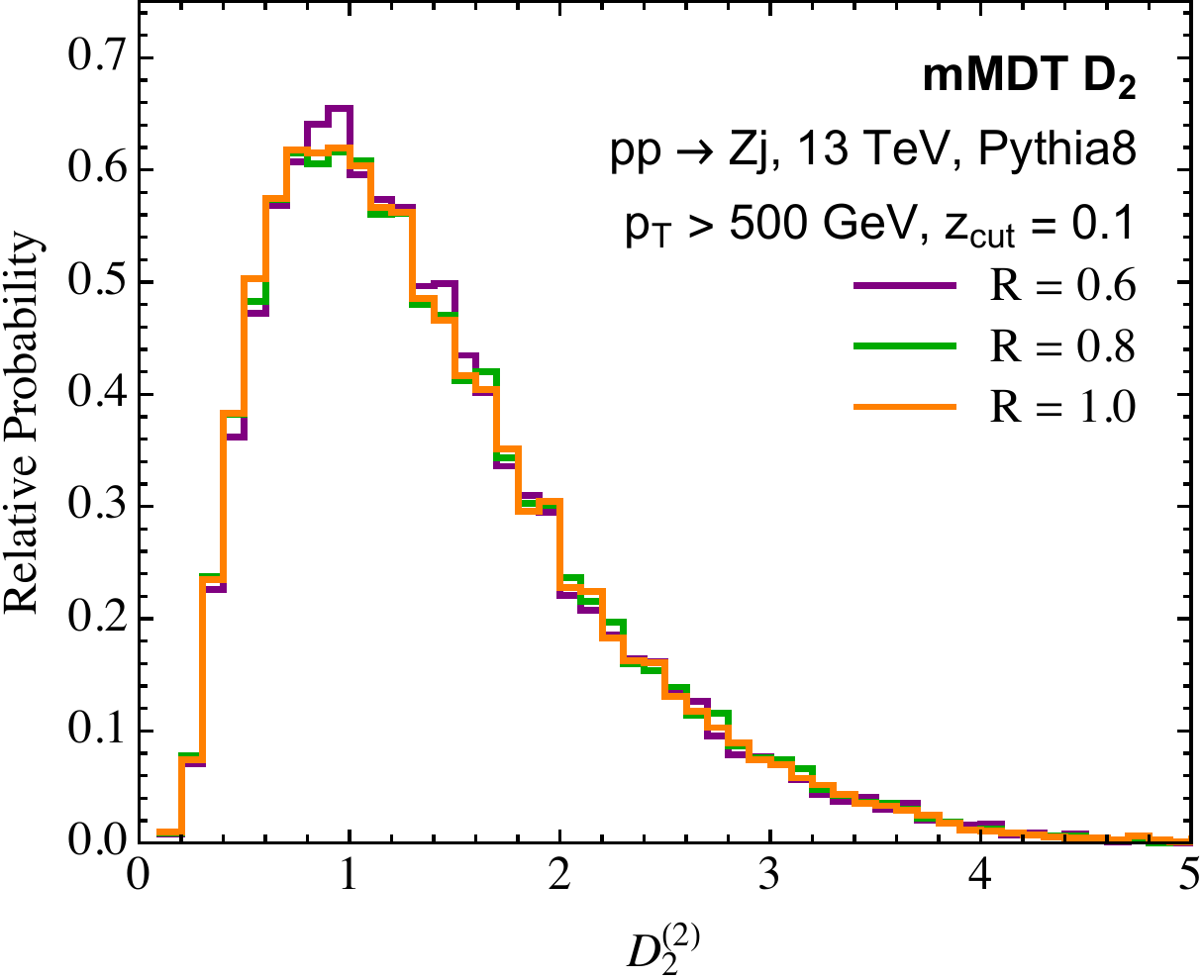}
}\ \ 
\subfloat[]{\label{fig:pphjR}
\includegraphics[width = 7.5cm]{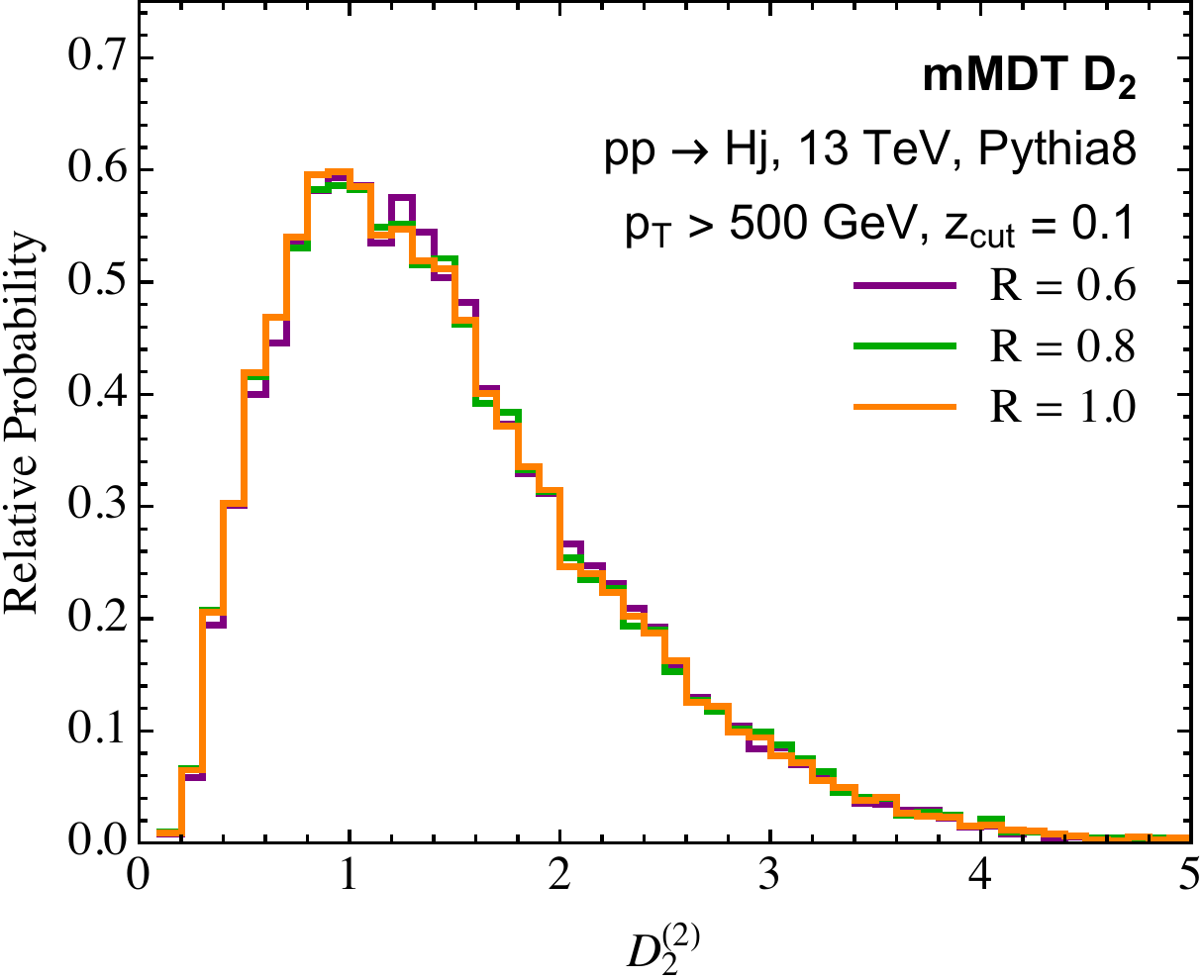}
}
\end{center}
\caption{
Distributions of $\Dobs{2}{2}$ measured on mMDT groomed QCD jets with $\zcut = 0.1$ from the processes $pp\to Zj$ (left) and $pp\to Hj$ (right) at the 13 TeV LHC.  The groomed jet mass is restricted to lie in the range  $80\text{ GeV} < m_J < 100\text{ GeV}$.  Various jet radii $R$ are shown, with the jet $p_T$ cut fixed to $p_T> 500$ GeV.
}
\label{fig:ppjRplots}
\end{figure}

Plots of the groomed $\Dobs{2}{2}$ distributions on the background jet samples with different jet radii are shown in \Fig{fig:ppjRplots}.  Here, the jet $p_T$ cut is fixed to $p_T > 500$ GeV, while the jet radius ranges from $R = 0.6$ to $R = 1.0$.  As observed with the signal distributions, there is extremely weak dependence on the jet radius, demonstrating that mMDT/soft drop is efficient at removing wide-angle radiation in the jet that would be sensitive to the jet radius.  After grooming, all radiation in the jet is collinear, and the relevant angular scales are set by the ratio of the groomed mass to the jet $p_T$.  By construction, this angular scale is always less than the jet radius (see \Eq{eq:jetrscale}), and so the jet radius is never relevant.

\section{Conclusions}\label{sec:conc}

In this paper we have performed a detailed study of the factorization properties of groomed multi-prong observables, focusing in particular on the $D_2$ observable with mMDT or soft drop $\beta=0$ grooming. We derived factorization formulae which describe the observable to all orders in $\alpha_s$, and allow us to make powerful statements of phenomenological relevance about the behavior of the groomed $D_2$ observable.  Most interesting are the fixed endpoint of the distribution at $1/(2\zcut)$, and the independence of non-perturbative corrections on the jet energy scale. Combined, these imply a remarkable robustness of the groomed $D_2$ observable, which is important for jet substructure applications.

We have introduced factorization formulae describing each region of phase space relevant for groomed boosted boson discrimination. Some of these factorization formulae follow by combining those which previously existed in the literature, however some are new. In particular, we derived a novel factorization describing the production of a soft subjet with energy the scale of the soft drop parameter $\zcut$. This factorization has the interesting property that clustering effects enter into the hard matching coefficient for the production of the soft subjet. We computed the functions entering the factorization at one-loop, and showed renormalization group consistency of the factorization formulae. While we have focused on applying these factorization formulae to the particular case of soft dropped $D_2$, we believe that they will be more generally applicable for describing groomed jet substructure observables, particularly those based on the energy correlation functions. This includes, for example, the $N_2$ \cite{Moult:2016cvt} observable used by CMS \cite{CMS-PAS-EXO-17-001,CMS-PAS-HIG-17-010,CMS:2017cbv}, or more ambitiously, energy correlation based observables for boosted top tagging \cite{Larkoski:2013eya,Moult:2016cvt,Larkoski:2014zma}.

We performed a numerical study for $e^+e^-\to$ dijets at NLL order, considering both the case of $e^+e^- \to Z\to q\bar q$ and $e^+e^- \to H\to gg$, allowing us to understand the differences between the $D_2$ distributions for quark and gluon jets. Non-perturbative effects were found to be small, and good agreement was found between the predictions of the Monte Carlo parton shower, and the analytic calculation. Since the $D_2$ observable probes multi-particle splittings, it may prove useful for testing Monte Carlo generators that implement $1\to3$ \cite{Hoche:2017iem,Hoche:2017hno} or $2\to4$ splittings \cite{Li:2016yez}. 

Perhaps most interestingly, we have shown that due to the grooming procedure, our calculations extend straightforwardly to proton-proton collisions at the LHC. This allows for experimentally realistic jet substructure observables currently used at the LHC to be calculated with theoretical precision. In this paper we performed a Monte Carlo study, showing that the features predicted by the factorization formulae are reproduced by Monte Carlo parton shower generators. In particular, we focused on the robustness of the groomed $D_2$ distribution as a function of the jet $p_T$, the scaling of non-perturbative hadronization effects, and the partonic content of the jet. We also showed that the groomed $D_2$ distribution is process independent up to the quark-gluon fraction of the jet. These are experimentally desirable features, which can be derived from a first principles theoretical description, and make groomed $D_2$ a promising observable for QCD studies at the LHC, as well as putting its theoretical understanding as a jet substructure tagger on firm theoretical footing. Analytic results for $D_2$ distributions for relevant processes at the LHC are presented in a companion publication \cite{Larkoski:2017iuy}.

\begin{acknowledgments}
We thank Ben Nachman, Stefan Prestel and Matt Schwartz for helpful discussions. This work is supported by the U.S. Department of Energy (DOE) under cooperative research agreements DE-FG02-05ER-41360, and DE-SC0011090 and by the Office of High Energy Physics of the U.S. DOE under Contract No. DE-AC02-05CH11231, and the LDRD Program of LBNL. This work was performed in part at Aspen Center for Physics, which is supported by National Science Foundation grant PHY-1607611. D.N. thanks both the Berkeley Center for Theoretical Physics and LBNL for hospitality while portions of this work were completed, as well as support from the Munich Institute for Astro- and Particle Physics (MIAPP) of the DFG cluster of excellence ``Origin and Structure of the Universe'', and support from DOE contract DE-AC52-06NA25396 at LANL and through the LANL/LDRD Program.
\end{acknowledgments}

\appendix

\section{Ingredients for Collinear Subjets}\label{app:collinear_sub}

In this appendix we present the one-loop calculation of all the functions appearing in the collinear subjets factorization formula.

\subsection{Kinematics and Notation}
We begin by briefly describe the kinematics and notation. We follow closely \cite{Larkoski:2015kga}.
We let $Q$ be the center of mass energy of the $e^+e^-$ collisions. The energy deposited in each hemisphere is therefore $Q/2$, and the four-momenta of the hemispheres are
\begin{align}
p_{\text{hemisphere}_1}=\left( \frac{Q}{2},\vec p_1   \right)\,, \qquad p_{\text{hemisphere}_2}=\left( \frac{Q}{2},-\vec p_1   \right)\,,
\end{align}
so that we have $s=Q^2\,.$ For the kinematics of the subjets, we will use the following notation
\begin{align}
&\text{Subjet a,b momenta: }& &p_a,p_b\,,\\
&\text{Subjet a,b spatial directions: }& &\hat{n}_{a}, \hat{n}_b\,,\\
&\text{Thrust axis: }& &\hat{n}=\frac{\hat{n}_{a}+ \hat{n}_b}{|\hat{n}_{a}+ \hat{n}_b|}\,,\\
&\text{Light-cone vectors: }& &n=(1,\hat{n}),\,\bar{n}=(1,-\hat{n}),\nonumber\\
& & &n_{a,b}=(1,\hat{n}_{a,b}),\,\bar{n}_{a,b}=(1,-\hat{n}_{a,b})\,.
\end{align}
These satisfy the relations
\begin{align} \label{eq:cs_kinematics1}
n\cdot n_a&=n\cdot n_b= \frac{n_a \cdot n_b}{4}\,, \qquad
\bar{n}\cdot n_a=\bar{n}\cdot n_b=2\,, \\
n_{\perp a,b}\cdot \bar{n}_{\perp a,b}&=-n_{\perp a,b}\cdot n_{\perp a,b}=\hat{n}_{\perp a,b}\cdot\hat{n}_{\perp a,b}=\frac{n_a\cdot n_b}{2}\,. \label{eq:cs_kinematics2}
\end{align}
For a particle in the collinear sector $a$ or $b$, we have
\begin{align}\label{energy_in_diverse_coor}
p_a\sim\frac{1}{2}(\bar{n}\cdot p_a) n_a,\quad p_b\sim\frac{1}{2}(\bar{n}\cdot p_b) n_b\,,\qquad
p_a^0\sim\frac{1}{2}(\bar{n}\cdot p_a),\quad p_b^0\sim\frac{1}{2}(\bar{n}\cdot p_b)\,.
\end{align}
We will label the energy fractions carried in each subjet by
\begin{align}\label{energy_fractions}
z_{a,b}&=\frac{2p_{a,b}^0}{Q}=\frac{\bar n\cdot p_{a,b}}{Q}\,,
\end{align}
where the second relation is true to leading power.

We can now compute the leading power expressions for the observables in these kinematics.
The value of $\ecf{2}{\alpha}$ is given by
\begin{align}\label{eq:lp_e2}
\ecf{2}{\alpha}&=\frac{1}{E_J^2} E_a E_b \left( \frac{2p_a\cdot p_b}{E_a E_b}  \right)^{\alpha/2}  \nn \\
&=2^{\alpha/2}  z_az_b\left(n_a\cdot n_b\right)^{\alpha/2}\,.
\end{align}
For three emissions, with momenta $k_1, k_2, k_3$, the expression for the three point energy correlation function is
\begin{align}
\ecf{3}{\alpha}&=\frac{1}{E_J^3}k_1^0 k_2^0 k_3^0 \left(\frac{2 k_1 \cdot k_2}{k_1^0 k_2^0}  \right)^{\alpha/2}     \left(\frac{2k_1\cdot k_3}{k_1^0 k_3^0}  \right)^{\alpha/2}      \left(\frac{2k_2 \cdot k_3}{k_2^0 k_3^0}  \right)^{\alpha/2}\,.
\end{align}
For an emission collinear with one of the subjets, where we have the splitting $p_{a,b}\rightarrow k_1+k_2$, we have
\begin{align}\label{eq:collinear_limits_eec3}
\left.\ecf{3}{\alpha}\right|_{k_1,k_2\parallel n_a}&=   2^{5\alpha/2}z_b (n_a\cdot n_b)^{\alpha}\left(\frac{k_1\cdot k_2}{Q^2}\right)^{\frac{\alpha}{2}}\left(\frac{\bar n_a\cdot k_1}{Q}\right)^{1-\frac{\alpha}{2}}\left(\frac{\bar n_a\cdot k_2}{Q}\right)^{1-\frac{\alpha}{2}}\,,\\
\left.\ecf{3}{\alpha}\right|_{k_1,k_2\parallel n_b}&= 2^{5\alpha/2}z_a (n_a\cdot n_b)^{\alpha}\left(\frac{k_1\cdot k_2}{Q^2}\right)^{\frac{\alpha}{2}}\left(\frac{\bar n_b\cdot k_1}{Q}\right)^{1-\frac{\alpha}{2}}\left(\frac{\bar n_b\cdot k_2}{Q}\right)^{1-\frac{\alpha}{2}}\,.
\end{align}
For a third collinear-soft emission $k$ off of the $p_{a,b}$ partons, we have
\begin{align}\label{eq:collinear_soft_limits_eec3}
\left.\ecf{3}{\alpha}\right|_{k\to \text{c-soft}}&=2^{3\alpha/2+1}z_az_b\left(n_a\cdot n_b\right)^{\alpha/2}\left(\frac{\bar{n}\cdot k}{2Q}\right)^{1-\alpha}\left(\frac{n_a\cdot k}{Q}\right)^{\alpha/2}\left(\frac{n_b\cdot k}{Q}\right)^{\alpha/2}\,.
\end{align}

The factorization formula in this region of phase space, which we repeat here for convenience, is:
{\small\begin{align}\label{eq:ninjafact_repeat}
\frac{d^3\sigma}{dz d\ecf{2}{\alpha}\, d\ecf{3}{\alpha} }&= H(Q^2) S(\zcut) S_c(\zcut,\theta>\theta_{ab}) H_2(z,\ecf{2}{\alpha})\nn \\
&\hspace{2cm}\cdot J_1(\ecf{3}{\alpha})\otimes J_2(\ecf{3}{\alpha})\otimes C_s(\ecf{3}{\alpha},\theta<\theta_{ab})\otimes C^{\text{NG}}_s(\ecf{3}{\alpha},\zcut) \,.
\end{align}}
The convolutions over $\ecf{3}{\alpha}$ can be turned into products by Laplace transforming with respect to the variable $\ecf{3}{\alpha}$. We denote the Laplace transformed variables and functions with a tilde.  When Laplace transformed, the cross section can be written as
{\small\begin{align}
\frac{d^2\sigma}{dz d\ecf{2}{\alpha}\, d\ecflp{3}{\alpha} }&= H(Q^2) S(\zcut) S_c(\zcut,\theta>\theta_{ab}) H_2(z,\ecf{2}{\alpha}) \nn \\
&\hspace{2cm} \cdot \tilde J_1(\ecflp{3}{\alpha}) \tilde J_2(\ecflp{3}{\alpha}) \tilde C_s(\ecflp{3}{\alpha},\theta<\theta_{ab}) C^{\text{NG}}_s(\ecflp{3}{\alpha},\zcut) \,.
\end{align}}
In this appendix, we will present the expressions for these Laplace transformed functions.

\subsection{Matrix Element Definitions}
We give the matrix element definitions of all functions in the factorization formula in \Eqs{eq:ninjafact_repeat}{eq:sc_sj_fact}, and \Eq{eq:signal_fact}. The factorization formulae presented in this paper are formulated in the language of SCET \cite{Bauer:2000yr,Bauer:2001ct,Bauer:2001yt,Bauer:2002nz}. We refer readers unfamiliar with SCET to the reviews \cite{iain_notes,Becher:2014oda}. The jet functions are identical to \cite{Larkoski:2015kga}, so we only give the soft matrix elements. They are defined in terms of soft Wilson lines 
\begin{align}\label{eq:soft_wilson_line}
S_q&={\bf P} \exp \left( ig \int\limits_0^\infty ds\, q\cdot A(x+sq)    \right)\,.
\end{align}
Explicitly, we have
\begin{align}
C_s(\ecflp{3}{\alpha},\theta<\theta_{ab})&=\frac{1}{N_{f}}\text{tr}\Big[\mathbf{T}_{f}\langle 0|T\{S_{a}S_{b}S_{\nbar}\}\exp\Big(\hspace{-0.2cm}-\ecflp{3}{\alpha}\ecfop{3}{\alpha}\big|_{\text{SD} a,b}\Big)\bar T\{S_{a}S_{b}S_{\nbar}\}|0\rangle\mathbf{T}^{\dagger}_{f}\Big]\,, \\
S_c(\zcut,\theta>\theta_{ab})&=\frac{1}{N_{f}}\text{tr}\Big[\mathbf{T}_{f}\langle 0|T\{S_{a}S_{b}S_{\nbar}\}\Theta_{\text{SD}}(a,b,\zcut)\bar T\{S_{a}S_{b}S_{\nbar}\}|0\rangle\mathbf{T}^{\dagger}_{f}\Big]\,,\\
C^{\text{NG}}_s(\ecflp{3}{\alpha},\zcut)&=  \frac{1}{N_{f}}\frac{\text{tr}\Big[\mathbf{T}_{f}\langle 0|T\{S_{a}S_{b}S_{\nbar}\}\Theta_{\text{SD}}(a,b,\zcut)\exp\Big(-\ecflp{3}{\alpha}\ecfop{3}{\alpha}\big|_{\text{SD} a,b}\Big)\bar T\{S_{a}S_{b}S_{\nbar}\}|0\rangle\mathbf{T}^{\dagger}_{f}\Big]}{C_s(\ecflp{3}{\alpha},\theta<\theta_{ab})S_c(\zcut,\theta>\theta_{ab})}\,.
\end{align}
Here $a,b$ are the light-cone directions of the subjets, the operator $\Theta_{\text{SD}}(a,b,\zcut)$ imposes an energy cut of $Q\zcut/2$ on any emission that is not clustered into legs $a,b$ before they themselves are clustered, and $\ecfop{3}{\alpha}\big|_{\text{SD} a,b}$ returns only the energy-correlation function computed on all momenta that are clustered into $a$ or $b$ \emph{before} $a$ and $b$ are clustered. This works out to be a purely geometrical constraint, for both the soft-collinear and collinear-soft functions, since all emissions are at a scale where they cannot pass soft drop on their own.  $\mathbf{T}_f$ are color matrices contracted into the Wilson lines, and in general depend upon the flavor structure of the splitting, as does the representation of the Wilson lines, that is, whether the $1\to2$ subjet splitting was $g\rightarrow gg$, $q\rightarrow qg$, or $g\rightarrow q\bar{q}$.

Throughout these appendices we will refer to the perturbative order of the calculation with a superscript $(L)$, where $L$ denotes the loop order of the calculations. For example, for a function $F$, we have
\begin{align}
F&=F^{(T)}+F^{(1)}+F^{(2)}+...\,,
\end{align}
where $F^{(L)}$ denotes the $L$-th loop correction to the function $F$. The superscript $(T)$ stands for ``tree".

\subsection{Hard Function}

The hard function for $e^+e^-\to$ dijets can be found in \cite{Bauer:2003di,Manohar:2003vb,Ellis:2010rwa,Bauer:2011uc}. To $\mathcal{O}(\alpha_s)$ it is given by
\begin{equation}
    H(Q^2) = 1 + \frac{\alpha_s\,C_F}{2\pi} \left(-\log^2 \frac{\mu^2}{Q^2} - 3\log \frac{\mu^2}{Q^2} - 8+\frac{7}{6}\pi^2\right)\,,
\end{equation}
and its anomalous dimension is given by
\begin{equation}
\gamma_H = \frac{\alpha_sC_F}{\pi}\left(
-2\,\log\frac{\mu^2}{Q^2}-3
\right)\,.
\end{equation}
For notational simplicity, throughout these appendices we do not explicitly write the arguments of the anomalous dimensions.

\subsection{Hard Splitting Function}

The hard splitting function describing the $q\to qg$ splitting into two collinear subjets was first derived in the SCET$_+$ context in \cite{Bauer:2011uc} using results from \cite{Ellis:1980wv,Kosower:1999rx}, while the $g\rightarrow gg$ splitting and $g\rightarrow q\bar{q}$ was derived in \cite{Larkoski:2015kga}, where all the coefficients were given in terms of the variables most useful for the current study. Taking as an example initial quarks splitting to $\mathcal{O}(\alpha_s)$, the hard splitting function is given by
\begin{align}
H_2^{q\to qg}(\ecf{2}{2},z_q,\mu)&= \frac{\alpha_s C_F}{2\pi} \frac{1}{  \ecf{2}{2}      } \frac{1+z_q^2}{1-z_q}   
\\
& \hspace{-2.75cm}\times\left \{ 1+\frac{\alpha_s}{2\pi} \left[ \left( \frac{C_A}{2} -C_F \right) \left( 2\, \log\left( \frac{Q^2}{\mu^2} \frac{   \ecf{2}{2}    }{4} \right) \log\, z_q +\log^2 z_q +2\text{Li}_2(1-z_q)  \right) \right. \right. \nonumber \\
&\hspace{-0.5cm} -\frac{C_A}{2} \left (  \log^2 \left( \frac{Q^2}{\mu^2} \frac{   \ecf{2}{2}     }{4} \right)-\frac{7\pi^2}{6}  \left. \left.+2\,\log \left( \frac{Q^2}{\mu^2} \frac{    \ecf{2}{\alpha}      }{4} \right)\log (1-z_q) +\log^2(1-z_q)+2\text{Li}_2 (z_q) \right) \right. \right. \nn \\
&\left. \left. \hspace{-0.5cm}+ \vphantom{ \log^2 \left( \frac{Q^2}{\mu^2} \frac{   \ecf{2}{2}     }{4}  \right) }(C_A-C_F)\frac{1-z_q}{1+z_q^2} \right] \right\}\,,\nonumber
\end{align}
and its anomalous dimension is
\begin{equation}
\gamma_{H_2}^{q\to qg} = -2\frac{\alpha_s}{\pi}\left(
\frac{C_A}{2}-C_F
\right)\log\, z_q-\frac{\alpha_s}{\pi}C_A\log\frac{4\mu^2}{\ecf{2}{2}Q^2}+\frac{\alpha_s}{\pi}C_A \log(1-z_q)-\frac{\alpha_s}{2\pi}\left(
\frac{11}{3}C_A-\frac{2}{3}n_f
\right)\,.
\end{equation}

\subsection{Jet Functions}\label{app:jetcoll}

The jet functions in the collinear subjets region of phase space are identical to the case that no grooming is applied, and therefore are given in \cite{Larkoski:2015kga}. To $\mathcal{O}(\alpha_s)$, the quark jet function in the direction $n_a$ is given by
\begin{align}\label{eq:qjet_final}
\tilde{J}_{q,n_{a}}^{(1)} (Q_J,\ecflp{3}{2})&=\frac{\alpha_s}{2\pi}C_F\left[\frac{2}{\epsilon^2}+\frac{3}{2\epsilon}+\frac{2}{\epsilon}   \log\left( 16(n_a\cdot n_b)^2(1-z_q) \ecflp{3}{2}\frac{\mu^2}{Q^2}  \right)\right.\\
&
\hspace{-1cm}
\left.+\log^2\left( 16(n_a\cdot n_b)^2(1-z_q) \ecflp{3}{2}\frac{\mu^2}{Q^2}  \right)+\frac{3}{2}\log\left( 16(n_a\cdot n_b)^2(1-z_q) \ecflp{3}{2}\frac{\mu^2}{Q^2}  \right)-\frac{\pi^2}{3}+\frac{7}{2}\right] \,,\nonumber
\end{align}
and its anomalous dimension is 
\begin{equation}
\gamma_{J_q} = \frac{\alpha_s}{2\pi}C_F\left[
4\,  \log\left( 4\frac{\left(\ecf{2}{2}\right)^2}{z_q^2(1-z_q)}\ecflp{3}{2}\frac{\mu^2}{Q^2}  \right)+3
\right]\,.
\end{equation}
The gluon jet function is given by
\begin{align}\label{eq:gjet_final}
\tilde{J}_{g,n_{a}}^{(1)}(Q_J,\ecflp{3}{2})&=\frac{\alpha_s}{2\pi}C_A\left[\frac{2}{\epsilon^2}   +\frac{2 }{ \epsilon}\log\left( 16(n_a\cdot n_b)^2 z_q \ecflp{3}{2}\frac{\mu^2}{Q^2}  \right)   +\frac{1}{ \epsilon} \frac{11C_A-2n_f}{6C_A}  \right. \nn\\
   &
    \hspace{-0.0cm}
    + \log^2\left( 16(n_a\cdot n_b)^2 z_q \ecflp{3}{2}\frac{\mu^2}{Q^2}  \right)+\left(
    \frac{11}{6}-\frac{n_f}{3C_A}
    \right)\log\left( 16(n_a\cdot n_b)^2 z_q \ecflp{3}{2}\frac{\mu^2}{Q^2}  \right)
    \nonumber \\
   &
   \left.
   \hspace{-0.0cm}
   +\frac{67}{18}-\frac{\pi^2}{3}-\frac{5 n_f}{9 C_A }\right]\,,
\end{align}
and its anomalous dimension is
\begin{equation}
\gamma_{J_g} = \frac{\alpha_s}{2\pi}C_A\left[
4\,  \log\left( 4\frac{\left(\ecf{2}{2}\right)^2}{z_q(1-z_q)^2} \ecflp{3}{2}\frac{\mu^2}{Q^2}  \right)+\frac{11C_A-2n_f}{3C_A}
\right]\,.
\end{equation}

\subsection{Wide-Angle Soft Function}\label{app:softanom}

The wide angle soft function in the collinear subjets region of phase space does not resolve the collinear splitting, and is therefore identical to that for the soft dropped mass in $e^+e^-$, which was derived in \cite{Frye:2016aiz}. At $\mathcal{O}(\alpha_s)$ it can be written as a sum over dipoles, and is given by
\begin{align}
S(\zcut) = 1+\frac{\alpha_s C_F}{2\pi}\left(
\log^2\frac{\mu^2}{\zcut^2 Q^2}-\frac{\pi^2}{6}
\right)\,.
\end{align}
Its anomalous dimension is
\begin{equation}
\gamma_S=2\frac{\alpha_s C_F}{\pi}\log\frac{\mu^2}{\zcut^2 Q^2}\,.
\end{equation}

\subsection{Collinear-Soft Function}

The collinear-soft function, $C_s(\ecf{3}{\alpha},\theta<\theta_{ab})$ is new, and we therefore calculate it to $\mathcal{O}(\alpha_s)$ in this appendix. At $\mathcal{O}(\alpha_s)$, it can be written as a sum over dipoles
{\small\begin{align}
C^{(1)}_{s}(\ecf{3}{\alpha},\theta<\theta_{ab})&=\frac{1}{N_f}\text{tr}\Big[\mathbf{T}_f^{\dagger}\Big(C_{s, ab}(\ecf{3}{\alpha},\theta<\theta_{ab})+C_{s, a\nbar}(\ecf{3}{\alpha},\theta<\theta_{ab})+C_{s, b\nbar}(\ecf{3}{\alpha},\theta<\theta_{ab})\Big)\mathbf{T}_f\Big]\,,
\end{align}}
where the contribution from a given dipole is given by
{\small\begin{align}
C_{s, ij}(\ecf{3}{\alpha},\theta<\theta_{ab}) &= -2g^2\mu^{2\epsilon}{\bf T}_i\cdot {\bf T}_j \int\frac{d^dk}{(2\pi)^{d-1}}\frac{n_i\cdot n_j}{(n_i\cdot k)(n_j\cdot k)}\delta(k^2) \Theta_\text{SD}\delta_{\ecf{3}{2}}\,.
\end{align}}
The soft drop constraint in this region of phase space is given by
{\small\begin{align}\label{eq:two_prong_soft_drop_clustered}
\Theta_\text{SD} &= \Theta\left(n_a\cdot n_b-\frac{n_b\cdot k}{\nbar\cdot k}\right)\Theta\left(\frac{n_a\cdot k}{\nbar\cdot k}-\frac{n_b\cdot k}{\nbar\cdot k}\right)+\Theta\left(n_a\cdot n_b-\frac{n_a\cdot k}{\nbar\cdot k}\right)\Theta\left(\frac{n_b\cdot k}{\nbar\cdot k}-\frac{n_a\cdot k}{\nbar\cdot k}\right)\,.
\end{align}}
The soft drop constraint is easily understood: the emission is to be clustered into direction $a$ or direction $b$ of the splitting before the two legs are themselves clustered. The measurement function is
\begin{equation}
\delta_{\ecf{3}{2}} = \delta\left(
\ecf{3}{2} - 16\ecf{2}{2}\frac{(n_a\cdot k)(n_b\cdot k)}{Q\nbar\cdot k }
\right)\,.
\end{equation}
We may take the light-cone direction of the parent jet to be $n=\frac{1}{2}(n_{a}+n_{b})$, conjugate to $\nbar$. Decomposing $k$ into the light-cone basis formed by $n$-$\nbar$, we have
\begin{align}
n_{a}\cdot k &=\frac{n_a\cdot n_b}{4}\nbar\cdot k+n\cdot k+n_a\cdot k_\perp\,,\\
n_{b}\cdot k &=\frac{n_a\cdot n_b}{4}\nbar\cdot k+n\cdot k-n_a\cdot k_\perp\,.
\end{align}
Here we have used $n_a\cdot\nbar=n_b\cdot\nbar\approx 2$. We can write the dot product in the transverse plane as
\begin{align}
n_a\cdot k_\perp&=-\sqrt{n_a\cdot n_b}~k_\perp\text{cos}\phi\,.
\end{align} 
Here we choose a fixed but arbitrary direction in the transverse plane for the projection of the direction $n_a$, defining the angle $\phi$. Solving the on-shell condition and rescaling, we have
\begin{align} 
\nbar\cdot k&\rightarrow Q z\,,\qquad
k_{\perp}\rightarrow \, Q\, z\,\sqrt{n_a\cdot n_b}\,k_{\perp},
\end{align}
and the Lorentz invariants become
\begin{align}
n_{a}\cdot k &=\frac{Q}{4}n_a\cdot n_b\,z\Big(1+4k_\perp^2-4k_\perp\text{cos}\phi\Big)=\frac{Q}{4}n_a\cdot n_b\,z(1+4\vec{k}^{\,2}-4\vec{k}\cdot\hat{n})\,,\\
n_{b}\cdot k &=\frac{Q}{4}n_a\cdot n_b\,z\Big(1+4k_\perp^2+4k_\perp\text{cos}\phi\Big)=\frac{Q}{4}n_a\cdot n_b\,z(1+4\vec{k}^{\,2}+4\vec{k}\cdot\hat{n})\,.
\end{align}
We can now see that the integration can be efficiently represented by an integral in the transverse plane. The integration measure, having solved the on-shell conditions with these rescalings, is then given by
{\small\begin{align}
\int\frac{d^dk}{(2\pi)^{d-1}}\delta(k^2)&=\frac{1}{2\pi}\Big(Q^2n_a\cdot n_b\Big)^{1-\epsilon}\int_0^{\infty}dz z^{1-2\epsilon}\int\frac{d^{2-2\epsilon}\vec{k}}{(2\pi)^{2-2\epsilon}}\,.
\end{align}}
We can now perform a shift $\vec{k}\rightarrow \vec{k}+\frac{1}{2}\hat{n}$, to get the compact form
\begin{align}
n_{a}\cdot k &=Q n_a\cdot n_b\,z \vec{k}^{\,2}\,,\\
n_{b}\cdot k &=Q n_a\cdot n_b\,z(\vec{k}+\hat{n})^2\,.
\end{align}
The measurement function then becomes
\begin{align}
\delta_{\ecf{3}{2}} = \delta\left(
\ecf{3}{2} - \frac{\ecf{2}{2}(n_a\cdot n_b)^2}{16}z\vec{k}^{\,2}(\vec{k}+\hat{n})^2
\right)\,,
\end{align}
and the soft drop condition becomes
{\small\begin{align}\label{eq:soft_drop_condition_compact}
\Theta_\text{SD} &=\Theta\Big(1-\vec{k}^{\,2}\Big)\Theta\Big((\vec{k}+\hat{n})^2-\vec{k}^{\,2}\Big)+\Theta\Big(1-(\vec{k}+\hat{n})^2\Big)\Theta\Big(\vec{k}^{\,2}-(\vec{k}+\hat{n})^2\Big)\,.
\end{align}}
We note that the reduction of the collinear-soft function to integrals over a transverse plane is not entirely unexpected, given the duality between time-like and space-like soft processes, see Refs. \cite{Marchesini:2003nh,Hatta:2008st,Avsar:2009yb}.

\subsubsection{Soft Integrals in the Transverse Plane}

We solve the measurement constraint using the $z$ integral, and go to Laplace space. We note that the roles of $a$ and $b$ are interchangeable (this amounts to mapping $\vec{k}\rightarrow-\vec{k}-\hat{n}$), so that we have
{\small\begin{align}\label{eq:collinear_soft_integrals_transverse}
C_{s, ab}(\ecflp{3}{2},\theta<\theta_{ab}) &= -C_{\epsilon}^{\text{in}}{\bf T}_a\cdot {\bf T}_b \int\frac{d^{2-2\epsilon}{k}}{(2\pi)^{2-2\epsilon}}\frac{\Theta\Big(1-\vec{k}^{\,2}\Big)\Theta\Big((\vec{k}+\hat{n})^2-\vec{k}^{\,2}\Big)}{(\vec{k}^{\,2}(\vec{k}+\hat{n})^2)^{1-2\epsilon}}\,,\\
C_{s, a\nbar}(\ecflp{3}{2},\theta<\theta_{ab}) &= -C_{\epsilon}^{\text{in}}{\bf T}_a\cdot {\bf T}_{\nbar} \int\frac{d^{2-2\epsilon}{k}}{(2\pi)^{2-2\epsilon}}\frac{\Theta\Big(1-\vec{k}^{\,2}\Big)\Theta\Big((\vec{k}+\hat{n})^2-\vec{k}^{\,2}\Big)}{(\vec{k}^{\,2}(\vec{k}+\hat{n})^2)^{1-2\epsilon}}\Big((\vec{k}^{\,2})+(\vec{k}+\hat{n})^2)\Big)\,,\nonumber\\
C_{s, b\nbar}(\ecflp{3}{2},\theta<\theta_{ab}) &= -C_{\epsilon}^{\text{in}}{\bf T}_b\cdot {\bf T}_{\nbar} \int\frac{d^{2-2\epsilon}{k}}{(2\pi)^{2-2\epsilon}}\frac{\Theta\Big(1-\vec{k}^{\,2}\Big)\Theta\Big((\vec{k}+\hat{n})^2-\vec{k}^{\,2}\Big)}{(\vec{k}^{\,2}(\vec{k}+\hat{n})^2)^{1-2\epsilon}}\Big((\vec{k}^{\,2})+(\vec{k}+\hat{n})^2)\Big)\,. \nn
\end{align}}
Here we have introduced the prefactor
\begin{align}
C_{\epsilon}^{\text{in}}=8\alpha_s\left(\frac{\mu}{Q}\ecf{2}{2}\ecflp{3}{2}(n_a\cdot n_b)^{3/2}\right)^{2\epsilon}\Gamma(-2\epsilon)\,.
\end{align}

\subsubsection{Isolating Divergent Contributions}

We note that all integrals in \Eq{eq:collinear_soft_integrals_transverse} only have a divergence at $\vec{k}^2=0$, so that we may add and subtract to each color structure the integral
\begin{align}\label{eq:collinear_soft_subtraction}
-C_{\epsilon}^{\text{in}}\int\frac{d^{2-2\epsilon}{k}}{(2\pi)^{2-2\epsilon}}\frac{\Theta\left(1-\vec{k}^{\,2}\right)}{(\vec{k}^{\,2})^{1-2\epsilon}}&=-\frac{C_{\epsilon}^{\text{in}}}{(4\pi)^{1-\epsilon}\epsilon\,\Gamma(1-\epsilon)}\,.
\end{align}
The resulting subtracted integrals are all finite, and can be evaluated numerically as a Taylor series in $\epsilon$ in terms of two-dimensional integrals. We have found an analytic expression for all the divergent contributions. For instance, in the $a$-$b$ dipole, we first perform the integral over the magnitude of the vector in the transverse plane, which after some algebra, gives the angular integral
{\small\begin{align}
\int\frac{d^{2-2\epsilon}{k}}{(2\pi)^{2-2\epsilon}}&\frac{\Theta\Big(1-\vec{k}^{\,2}\Big)\Theta\Big((\vec{k}+\hat{n})^2-\vec{k}^{\,2}\Big)}{(\vec{k}^{\,2}(\vec{k}+\hat{n})^2)^{1-2\epsilon}}-\int\frac{d^{2-2\epsilon}\vec{k}}{(2\pi)^{2-2\epsilon}}\frac{\Theta\left(1-\vec{k}^{\,2}\right)}{(\vec{k}^{\,2})^{1-2\epsilon}}\nonumber\\
&=\frac{1}{2\pi^2}\Bigg(\int_{0}^{\frac{\pi}{3}}d\phi\,\phi\text{cot}\phi+\frac{1}{2}\int_{\frac{\pi}{3}}^{\pi}d\phi\Big\{(\pi-\phi)\text{cot}\phi-\log(1-\text{cos}\phi)-\log\,2\Big\}\Bigg)+O(\epsilon) \nn \\
&=-\frac{\text{Cl}_2(\frac{\pi}{3})}{4\pi^2}+O(\epsilon)\,.
\end{align}}
Here $\text{Cl}_2(x)$ is the Clausen function, which has a value of approximately $1.01494$ at its maximum $x=\frac{\pi}{3}$. The other dipoles are handled similarly. Putting in the appropriate factors for $\overline{\text{MS}}$ scheme, we find
{\small\begin{align}
C_{s, ab}(\ecflp{3}{2},\theta<\theta_{ab}) &= \frac{\alpha_s}{\pi}{\bf T}_a\cdot {\bf T}_b\Bigg(\frac{1}{\epsilon^2}+\frac{2}{\epsilon}\left(L_{\theta<\theta_{ab}}-2\frac{\text{Cl}_2(\frac{\pi}{3})}{\pi}\right)+2L_{\theta<\theta_{ab}}^2-8\frac{\text{Cl}_2(\frac{\pi}{3})}{\pi}L_{\theta<\theta_{ab}}\Bigg)+O(\epsilon^0) \,,\nonumber\\
C_{s, a\nbar}(\ecflp{3}{2},\theta<\theta_{ab}) &= \frac{\alpha_s}{\pi}{\bf T}_a\cdot {\bf T}_{\nbar}\Bigg(\frac{1}{\epsilon^2}+\frac{2}{\epsilon}\left(L_{\theta<\theta_{ab}}+2\frac{\text{Cl}_2(\frac{\pi}{3})}{\pi}\right)+2L_{\theta<\theta_{ab}}^2+8\frac{\text{Cl}_2(\frac{\pi}{3})}{\pi}L_{\theta<\theta_{ab}}\Bigg)+O(\epsilon^0)\,,\nonumber\\
C_{s, b\nbar}(\ecflp{3}{2},\theta<\theta_{ab}) &= \frac{\alpha_s}{\pi}{\bf T}_b\cdot {\bf T}_{\nbar}\Bigg(\frac{1}{\epsilon^2}+\frac{2}{\epsilon}\left(L_{\theta<\theta_{ab}}+2\frac{\text{Cl}_2(\frac{\pi}{3})}{\pi}\right)+2L_{\theta<\theta_{ab}}^2+8\frac{\text{Cl}_2(\frac{\pi}{3})}{\pi}L_{\theta<\theta_{ab}}\Bigg)+O(\epsilon^0)\,,
\end{align}}
where the logarithm is defined as
{\small\begin{align}
L_{\theta<\theta_{ab}}&=\log\,\left(\frac{16\mu(\ecf{2}{2})^{5/2}\,\ecflp{3}{2}e^{\gamma_E}}{(z(1-z))^{3/2}Q}\right)\,.
\end{align}}
From these expressions is is straightforward to extract the anomalous dimension and perform the renormalization group evolution using standard techniques.

\subsection{Soft-Collinear Function}\label{sec:soft_drop_fails}
Now we compute how radiation that is not clustered into the two hard prongs gets groomed. Such radiation is described by the function $S_c(\zcut,\theta>\theta_{ab})$. At one-loop order, the contribution from the generic dipole $i,j$ is given by
{\small\begin{align}\label{eq:soft_drop_fails_soft_collinear_one_loop}
S^{(1)}_{c}(\zcut,\theta>\theta_{ab})&=\frac{1}{N_f}\text{tr}\Big[\mathbf{T}_f^{\dagger}\Big(S_{c, ab}(\zcut,\theta>\theta_{ab})+S_{c, a\nbar}(\zcut,\theta>\theta_{ab})+S_{c, b\nbar}(\zcut,\theta>\theta_{ab})\Big)\mathbf{T}_f^{\dagger}\Big]\,,\\
S_{c, ij}(\zcut,\theta>\theta_{ab}) &= -2g^2\mu^{2\epsilon}{\bf T}_i\cdot {\bf T}_j \int\frac{d^dk}{(2\pi)^{d-1}}\frac{n_i\cdot n_j}{(n_i\cdot k)(n_j\cdot k)}\delta(k^2) (1-\Theta_\text{SD})\Theta_{\zcut}\,,
\end{align}}
where $\Theta_\text{SD}$ was defined in \Eq{eq:two_prong_soft_drop_clustered}, and the constraint on the energy fraction of the groomed radiation is given by
\begin{align}\label{eq:soft_drop_fails}
\Theta_{\zcut}&=\Theta\left(\zcut-\frac{\nbar\cdot k}{Q}\right)\,.
\end{align}
We use the same coordinate system as above, and arrive at the following representation for each dipole
{\small\begin{align}
S_{c, ab}(\zcut,\theta>\theta_{ab}) &= -C_{\epsilon}^{\text{out}}{\bf T}_a\cdot {\bf T}_b\int\frac{d^{2-2\epsilon}{k}}{(2\pi)^{2-2\epsilon}}\Bigg(1-\Theta\Big(1-\vec{k}^{\,2}\Big)\Theta\Big((\vec{k}+\hat{n})^2-\vec{k}^{\,2}\Big)\Bigg)\frac{1}{\vec{k}^{\,2}(\vec{k}+\hat{n})^2}\,,\nonumber\\
S_{c, a\nbar}(\zcut,\theta>\theta_{ab}) &= C_{\epsilon}^{\text{out}}{\bf T}_a\cdot {\bf T}_{\nbar} \int\frac{d^{2-2\epsilon}{k}}{(2\pi)^{2-2\epsilon}}\frac{\Theta\Big(1-\vec{k}^{\,2}\Big)\Theta\Big((\vec{k}+\hat{n})^2-\vec{k}^{\,2}\Big)}{\vec{k}^{\,2}(\vec{k}+\hat{n})^2}\Big((\vec{k}^{\,2})+(\vec{k}+\hat{n})^2)\Big)\,,\nonumber\\
S_{c, b\nbar}(\zcut,\theta>\theta_{ab}) &= C_{\epsilon}^{\text{out}}{\bf T}_b\cdot {\bf T}_{\nbar} \int\frac{d^{2-2\epsilon}{k}}{(2\pi)^{2-2\epsilon}}\frac{\Theta\Big(1-\vec{k}^{\,2}\Big)\Theta\Big((\vec{k}+\hat{n})^2-\vec{k}^{\,2}\Big)}{\vec{k}^{\,2}(\vec{k}+\hat{n})^2}\Big((\vec{k}^{\,2})+(\vec{k}+\hat{n})^2)\Big)\,,
\end{align}}
where we have defined the constant
\begin{align}
C_{\epsilon}^{\text{out}}&=-\frac{4\alpha_s}{\epsilon}\left(\frac{\mu}{\zcut Q(n_a\cdot n_b)^{1/2}}\right)^{2\epsilon}\,.
\end{align}
To isolate the $\vec{k}=0$ divergence, we now add and subtract the integral
\begin{align}\label{eq:collinear_soft_subtraction2}
C_{\epsilon}^{\text{out}}\int\frac{d^{2-2\epsilon}{k}}{(2\pi)^{2-2\epsilon}}\frac{\Theta\Big(1-\vec{k}^{\,2}\Big)}{\vec{k}^{\,2}}&=\frac{2C_{\epsilon}^{\text{out}}}{(4\pi)^{1-\epsilon}\epsilon\,\Gamma(1-\epsilon)}\,.
\end{align}
We then find that the divergences and $\mu$-dependent logs have the structure
{\small\begin{align}
S_{c, ab}(\zcut,\theta>\theta_{ab}) &= -2\frac{\alpha_s}{\pi}{\bf T}_a\cdot {\bf T}_b\Bigg(\frac{4}{\epsilon}\frac{\text{Cl}_2(\frac{\pi}{3})}{\pi}+8\frac{\text{Cl}_2(\frac{\pi}{3})}{\pi}L_{\theta>\theta_{ab}}\Bigg)+O(\epsilon^0) \,,\\
S_{c, a\nbar}(\zcut,\theta>\theta_{ab}) &= -2\frac{\alpha_s}{\pi}{\bf T}_a\cdot {\bf T}_{\nbar}\Bigg(\frac{1}{\epsilon^2}+\frac{2}{\epsilon}\left(L_{\theta>\theta_{ab}}-2\frac{\text{Cl}_2(\frac{\pi}{3})}{\pi}\right)-2L_{\theta>\theta_{ab}}^2+8\frac{\text{Cl}_2(\frac{\pi}{3})}{\pi}L_{\theta>\theta_{ab}}\Bigg)+O(\epsilon^0)\,,\nonumber\\
S_{c, b\nbar}(\zcut,\theta>\theta_{ab}) &= -2\frac{\alpha_s}{\pi}{\bf T}_b\cdot {\bf T}_{\nbar}\Bigg(\frac{1}{\epsilon^2}+\frac{2}{\epsilon}\left(L_{\theta>\theta_{ab}}-2\frac{\text{Cl}_2(\frac{\pi}{3})}{\pi}\right)-2L_{\theta>\theta_{ab}}^2+8\frac{\text{Cl}_2(\frac{\pi}{3})}{\pi}L_{\theta>\theta_{ab}}\Bigg)+O(\epsilon^0)\,, \nn
\end{align}}
where the logarithm is defined as
{\small\begin{align}
L_{\theta>\theta_{ab}}&=\log\,\left(\frac{\mu(z(1-z))^{1/2}}{\zcut\left(\ecf{2}{2}\right)^{1/2}Q}\right)\,.
\end{align}}
It is again straightforward to extract the anomalous dimensions from the above results, and perform the renormalization group evolution.

\subsection{Anomalous Dimensions}\label{app:anom_dim}

Here we show that the anomalous dimensions sum to zero, showing the consistency of the factorization formula at the one-loop level. We label the anomalous dimension's flavor structure to distinguish different channels.  For $q\to qg$ splitting, the anomalous dimensions calculated above are given by
\begin{align}
\gamma_H^{q} &= \frac{\alpha_s C_F}{\pi}\left(
-2\,\log\frac{\mu^2}{Q^2}-3
\right)\,,\\
\label{eq:collinear_subjets_hard_matching_anom_dim}\gamma_{H_2}^{q\to qg} &= -2\frac{\alpha_s}{\pi}\left(
\frac{C_A}{2}-C_F
\right)\log\, z_q-\frac{\alpha_s}{\pi}C_A\log\frac{4\mu^2}{\ecf{2}{2}Q^2}\nonumber\\
&\hspace{0.4cm}+\frac{\alpha_s}{\pi}C_A \log(1-z_q)-\frac{\alpha_s}{2\pi}\left(
\frac{11}{3}C_A-\frac{2}{3}n_f
\right)\,,\\
\gamma_{J_q} &= \frac{\alpha_s}{2\pi}C_F\left[
4\,  \log\left( 4\frac{\left(\ecf{2}{2}\right)^2}{z_q^2(1-z_q)}\ecflp{3}{2}\frac{\mu^2}{Q^2}  \right)+3
\right]\,,\\
\gamma_{J_g} &= \frac{\alpha_s}{2\pi}C_A\left[
4\,  \log\left( 4\frac{\left(\ecf{2}{2}\right)^2}{z_q(1-z_q)^2} \ecflp{3}{2}\frac{\mu^2}{Q^2}  \right)+\frac{11C_A-2n_f}{3C_A}
\right]\,,\\
\gamma_S^q&=2\frac{\alpha_s C_F}{\pi}\log\frac{\mu^2}{\zcut^2 Q^2}\,,\\
\gamma_{C_s}^{q\to qg} &= -\frac{\alpha_s}{\pi}C_A \left[
\log \left( \frac{4(\ecf{2}{2})^5(\ecflp{3}{2})^2\mu^2}{z_q^3(1-z_q)^3Q^2}\right)-\frac{\text{Cl}_2(\frac{\pi}{3})}{\pi}
\right]\nonumber\\
&\hspace{0.4cm}-\frac{\alpha_s}{\pi}C_F\left[
\log \left(\frac{4(\ecf{2}{2})^5(\ecflp{3}{2})^2\mu^2}{z_q^3(1-z_q)^3Q^2}\right)+\frac{\text{Cl}_2(\frac{\pi}{3})}{\pi}
\right]\,,\\
\gamma_{S_c}^{q\to qg} &= -\frac{\alpha_s}{\pi}C_A \left(
\frac{\text{Cl}_2(\frac{\pi}{3})}{\pi}
\right)-\frac{\alpha_s}{\pi}C_F\left[
\log \left( \frac{4 z_q(1-z_q)\mu^2}{\zcut^2\ecf{2}{2}Q^2}\right)-\frac{\text{Cl}_2(\frac{\pi}{3})}{\pi}
\right]\,.
\end{align}
For $g\to gg$ splitting, we have
\begin{align}
\gamma_H^{q} &= \frac{\alpha_s C_A}{\pi}\left(
-2\,\log\frac{\mu^2}{Q^2}-\beta_0
\right)\,,\\
\label{eq:collinear_subjets_hard_matching_anom_dim_glue}\gamma_{H_2}^{g\to gg} &= -\frac{\alpha_s}{\pi}\left[-C_A\log \left( \frac{4\mu^2}{\ecf{2}{2}z(1-z)Q^2} \right)+\frac{11}{6}C_A-\frac{2}{6}n_f\right]\,,\\
\gamma_S^g&=\gamma_S^q\Big|_{C_F\rightarrow C_A}\,,\\
\gamma_{C_s}^{g\to gg}&=\gamma_{C_s}^{q\to qg}\Big|_{C_F\rightarrow C_A}\,,\\
\gamma_{S_c}^{g\to gg}&=\gamma_{S_c}^{q\to qg}\Big|_{C_F\rightarrow C_A}\,.
\end{align}
One can deduce the anomalous dimension structure for $g\rightarrow q\bar{q}$ using the appropriate color generators in the collinear-soft matrix elements, and the matching for the splitting given in \cite{Larkoski:2015kga}. To achieve NLL accuracy, one must include the contribution from the two-loop cusp anomalous dimension to the coefficient multiplying logarithmic terms in each of the anomalous dimensions. For $e^+e^-\to $ hadrons events in which we divide the event into hemispheres, groom each hemisphere, and then measure $\ecf{2}{2}$ and $\ecf{3}{2}$ on each hemisphere, the anomalous dimensions must satisfy
\begin{align}
0= \frac{\gamma_H+\gamma_S}{2}+\gamma_{H_2}+\gamma_{J_q}+\gamma_{J_g}+\gamma_{C_s}+\gamma_{S_c}\,.
\end{align}
One can verify that indeed this is satisfied, demonstrating consistency of the factorization at one-loop, or NLL accuracy. Due to the highly non-trivial combinations of scales appearing in the different functions, this provides a strong cross-check of our calculation and factorization formula.

\subsection{Description of Resummation, Scale Choices and Profiles}\label{app:scales_resummation}

Since we are resumming to NLL, the contribution to the cross section from each factorized function is given by the formula
\begin{align}
F(\mu)&=F(\mu_f)\exp\left(\int^{\mu}_{\mu_f}\frac{d\mu'}{\mu'}\gamma^F(\alpha_s(\mu'),\mu')\right)\,,
\end{align}
where $\gamma^F$ is the appropriate anomalous dimension for the considered function, and $\mu_f$ is the scale where we run the function to. For canonical scale setting, $\mu_f$ is where all large logarithms are minimized. To perform the resummation, we substitute in \Eq{eq:ninjafact_repeat} the resummed expression for each function. In general, so that we may make use of profiles to control the precise value of the resummation scale where needed, and so that we can match to the fixed order result, we keep all terms in the renormalized functions that explicitly contribute to the anomalous dimensions. Thus, when we turn off resummation, we will recover explicitly the differential cross section. When we resum we always scale set in the cumulative distribution. That is, we exponentiate all anomalous dimensions in Laplace space, perform the inverse Laplace transform with generic endpoints to the RG evolution, and integrate to get the cumulative distribution. \emph{Then} we set all scales to their canonical values (which are given below), and take the derivative to get the differential distribution. For a detailed discussion of how to implement the resummation procedure in SCET, see \cite{Almeida:2014uva}. 

The canonical scales, given as functions of $D_2$, are
\begin{align}
\mu_{H}&=Q\,,\\
\mu_{H_2}&=\frac{1}{2}Q\sqrt{\ecf{2}{2}z(1-z)}\,,\\
\mu_{C_s}&=D_2\,\sqrt{\ecf{2}{2}(z(1-z))^{3}}\frac{Q}{2}\,,\\
\mu_{S_c}&=\zcut \frac{Q}{2} \sqrt{\frac{\ecf{2}{2}}{z(1-z)}}\,,\\
\mu_{J_a}&=\sqrt{D_2\,\ecf{2}{2}z(1-z)^2}\frac{Q}{2}\,,\\
\mu_{J_b}&=\sqrt{D_2\,\ecf{2}{2}z^2(1-z)}\frac{Q}{2}\,,\\
\mu_{S}&=\zcut Q\, ,\\
\mu&=\sqrt{\ecf{2}{2}}\frac{Q}{2}.\label{eq:factorization_scale}
\end{align}

When performing the resummation, we take the wide-angle soft scale as the common scale where we factorize. When assessing resummation uncertainties, we vary all scales up and down by a factor of two. To handle the Landau pole in the running coupling, we smoothly ``freeze out'' the running coupling as a function of its scale at $1$ GeV, so that it is simply a constant function below this value, and vary this freeze out scale up and down by $0.5$ GeV. In general we have very little sensitivity to the freezing scale of the running coupling. To achieve NLL accuracy, we also promote the coefficient of the logarithmic terms in the anomalous dimensions in \App{app:anom_dim} to two-loop accuracy, as given by the perturbative expansion of the cusp anomalous dimension, and the running of the coupling (including when integrating the anomalous dimensions).

To turn off the resummation, we use the simple profile
\begin{align}
p(x;t,s)=\Theta(x-t)\left[1-\exp\left(-s(x-t)^2\right)\right]\,.
\end{align}
This function is zero below $t$, and asymptotes to one as $x-t\gg\frac{1}{s}$. Using this function, the scale choice for a function $F$ in the factorization formula can be profiled as
\begin{align}
\mu_F^{\text{profile}}(D_2)&=\mu_F\left(1-p(D_2;t,s)\right)+\mu\cdot p(D_2;t,s)\,,
\end{align}
where $\mu$ is the common factorization scale in \Eq{eq:factorization_scale}, and $\mu_F$ is the canonical choice. We chose as default value for the profile transition points
\begin{align}
s=4\zcut,\qquad t=\zcut\,.
\end{align}
This choice ensures that canonical resummation effectively dominates at $D_2\sim 1$, and that the resummation is turned off at the endpoint $D_2=1/(2\zcut)$. As part of the uncertainty estimate (in addition to scale variations of the canonical resummation scales), these choices were varied by $50\%$. We found that these profiles were sufficient to numerically cancel any zero that developed by dividing by the singular result in \Eq{eq:matching_to_fo}, when the singular result developed a zero in the physical range. A more sophisticated profile would ensure that such a cancellation would happen exactly, but we found negligible differences between the matched result and the fixed-orded cross section even in a small neighborhood about the zero, given our numerical accuracy and sampling of the matched spectra.

\section{Ingredients for Collinear-Soft Subjets}\label{app:csoft_sub}

In this appendix, we present the calculations of the functions in the collinear-soft subjets factorization formula.  The notation and formulation of the calculations will be identical to that presented in \App{app:collinear_sub}.  The factorization formula is
\begin{align}
\frac{d^3\sigma}{dz d\ecf{2}{\alpha} \, d\ecf{3}{\alpha}}= H(Q^2) S(\zcut) H_2^{sj}(z, \ecf{2}{\alpha},\zcut)  C_s(\ecf{3}{\alpha},\theta<\theta_{12}) \otimes J_{sc}(\ecf{3}{\alpha})\otimes J(\ecf{3}{\alpha}) \,.
\end{align}
The convolutions can be removed by Laplace transforming in $\ecf{3}{2}$, after which we find
\begin{align}
\frac{d^3\sigma}{dz d\ecflp{2}{\alpha} \, d\ecflp{3}{\alpha}}= H(Q^2) S(\zcut) H_2^{sj}(z, \ecf{2}{\alpha},\zcut)  \tilde C_s(\ecflp{3}{\alpha},\theta<\theta_{12}) \tilde J_{sc}(\ecflp{3}{\alpha})\tilde J(\ecflp{3}{\alpha}) \,.
\end{align}
We again denote the Laplace transformed variables and functions with a tilde.
The calculation of the low scale functions has been presented in the previous section, so we will focus on the hard function calculation.

\subsection{Hard Function Calculation}\label{app:hardfunc_sd}

The hard function for the collinear-soft subjet region of phase space has two parts at one-loop: there is the pure virtual term, which is familiar, and also the term with two emissions, where one emission is removed by the soft drop groomer. We focus on the case where $\alpha=2$. The tree level integrand is given by integrating over the square of the tree-level soft current for the $n$ and $\nbar$ dipole
{\small\begin{align} 
H_{cs}^{(T)}(z,\ecf{2}{2},\zcut) &= \int \frac{d^dk_1}{(2\pi)^d} 2\pi\delta(k_1^2)\delta\left(z-\frac{\nbar\cdot k_1}{Q}\right)\delta\left(\ecf{2}{2}-\frac{n\cdot k_1}{Q}\right)|{\cal M}^{R}_{n\nbar}(k_1)|^2\,, \nn\\
&=H_{2}^{(T)}(z,\ecf{2}{2},\zcut)\Big|_{z\rightarrow 0}\,,
\end{align}}
and we can see that it exactly matches the soft limit of the splitting function.
The one-loop contribution to the hard function can be calculated from
{\small\begin{align}\label{eq:one_loop_collinear_soft_subjets}
H_{cs}^{(1)}(z,\ecf{2}{2},\zcut) &= \int \frac{d^dk_1}{(2\pi)^d} 2\pi\delta(k_1^2)\delta\left(z-\frac{\nbar\cdot k_1}{Q}\right)\delta\left(\ecf{2}{2}-\frac{n\cdot k_1}{Q}\right)\Bigg\{|{\cal M}^{RV}_{n\nbar}(k_1)|^2\nonumber\nonumber\\
&\hspace{3cm}+2\int \frac{d^dk_2}{(2\pi)^d} 2\pi\delta(k_2^2) |{\cal M}^{RR}_{n\nbar}(k_1,k_2)|^2(1-\Theta_{\text{SD}})\Theta_{\zcut}\Bigg\}\,.
\end{align}}
The matrix elements are given by squaring the one soft emission plus virtual current, and two soft emission currents, where the hard directions are given by the $n$ and $\nbar$ directions. The double real emission is obtained from \cite{Catani:2000pi}. The soft drop geometrical constraints are again given by Eq. \eqref{eq:two_prong_soft_drop_clustered}, with the substitutions $n_a\rightarrow n$, $n_b\rightarrow\frac{k_1}{\nbar\cdot k_1}$, and $k\rightarrow k_2$. The constraint on the momentum fraction of $k_2$ is given by Eq. \eqref{eq:soft_drop_fails}, with $k\rightarrow k_2$. Critically, we note that the soft drop conditions are \emph{purely geometrical}: we can rephrase the soft drop constraint of Eq. \eqref{eq:two_prong_soft_drop_clustered} in terms of angles, as is done in Eq. \eqref{eq:clustering_one_loop}. Thus regardless of how one power counts the relative momentum fractions of emissions $k_1$ and $k_2$, the exact same soft drop condition applies to an emission which is not clustered. Then all that can change is the expansion of the matrix element $|{\cal M}^{RR}_{n\nbar}(k_1,k_2)|^2$, whether we take it to be strongly-ordered in the energy of the emissions, or not. If we strongly order, then we reproduce the calculation of \App{sec:soft_drop_fails}, once we factor out the tree-level result for the $k_1$ emission that forms a leg of the hard $1\rightarrow 2$ splitting and take the limit that $z\rightarrow 0$. With this observation we may write Eq. \eqref{eq:one_loop_collinear_soft_subjets} as
{\small\begin{align}\label{eq:one_loop_collinear_soft_subjets_II}
H_{cs}^{(1)}(z,\ecf{2}{2},\zcut) &= \int \frac{d^dk_1}{(2\pi)^d} 2\pi\delta(k_1^2)\delta\left(z-\frac{\nbar\cdot k_1}{Q}\right)\delta\left(\ecf{2}{2}-\frac{n\cdot k_1}{Q}\right)\Bigg\{|{\cal M}^{RV}_{n\nbar}(k_1)|^2\nonumber\nonumber\\
&+2\int \frac{d^dk_2}{(2\pi)^d} 2\pi\delta(k_2^2)\left(|{\cal M}^{RR}_{n\nbar}(k_1,k_2)|^2\Big |_{\text{s.o.:}\, k_1^0\gg k_2^0}\right)(1-\Theta_{\text{SD}})\Theta_{\zcut}\nonumber\\
&+2\int \frac{d^dk_2}{(2\pi)^d} 2\pi\delta(k_2^2)\left( |{\cal M}^{RR}_{n\nbar}(k_1,k_2)|^2-|{\cal M}^{RR}_{n\nbar}(k_1,k_2)|^2\Big |_{\text{s.o.:}\, k_1^0\gg k_2^0}\right)(1-\Theta_{\text{SD}})\Theta_{\zcut}\Bigg\}\nn\\
&=H_{cs}^{(T)}(z,\ecf{2}{2},\zcut)\left(S^{(1)}_{c}(\zcut,\Theta>\Theta_{ab})+H^{(1)}_{2}(z,\ecf{2}{2})\Big|_{z_g\rightarrow 0}+\text{const}\right)\,,
\end{align}}
where the subscript ``$\text{s.o.}$" denotes strongly ordered.
Here $S^{(1)}_{c}(\zcut,\theta>\theta_{ab})$ is given by Eq. \eqref{eq:soft_drop_fails_soft_collinear_one_loop}, and corresponds to the strongly ordered contribution, and the virtual correction is fully captureed by taking the energy fraction of the gluon to zero ($z_g\rightarrow 0$) in the collinear-splitting function calculation. The residual constant is formally given as
{\small\begin{align}
\text{const}&=\frac{2}{H_{cs}^{(T)}(z,\ecf{2}{2},\zcut)}\int \frac{d^dk_1}{(2\pi)^d} \pi\delta(k_1^2)\delta\Big(z-\frac{\nbar\cdot k_1}{Q}\Big)\delta\Big(\ecf{2}{2}-\frac{n\cdot k_1}{Q}\Big)\nonumber\\
&\qquad\int \frac{d^dk_2}{(2\pi)^d} 2\pi\delta(k_2^2)\left( |{\cal M}^{RR}_{n\nbar}(k_1,k_2)|^2-|{\cal M}^{RR}_{n\nbar}(k_1,k_2)|^2\Big |_{\text{s.o.:}\, k_1^0\gg k_2^0}\right)(1-\Theta_{\text{SD}})\Theta_{\zcut}\,.
\end{align}}
We label this as a constant, since this contribution is easily found to be ultraviolet and infrared finite. We may therefore set $\epsilon=0$ in its calculation, and it will give no contribution to the anomalous dimension. We emphasize that the ability to simplify the matching in the collinear-soft subjet region is purely because the soft drop constraint is geometrical, and thus is not sensitive to the exact relative power counting of the two emissions.

\subsection{Hard Function Anomalous Dimension}\label{app:hardfunc_cs_anom_dim}
We give the collinear-soft subjet anomalous dimensions for both quark and gluon initiated jets. We have
{\small\begin{align}\label{eq:collinear_soft_subjet_hard_matching_anom_dim}
\gamma_{H_{cs}}^{q\rightarrow gq}=& -\frac{\alpha_s}{\pi}C_A\log\frac{4\mu^2}{\ecf{2}{2}(1-z_q)Q^2}-\frac{\alpha_s}{2\pi}\left(
\frac{11}{3}C_A-\frac{2}{3}n_f\right)\nonumber\\
&-\frac{\alpha_s}{\pi}C_A \left(
\frac{\text{Cl}_2(\frac{\pi}{3})}{\pi}
\right)-\frac{\alpha_s}{\pi}C_F\left[
\log \left( \frac{4 (1-z_q)\mu^2}{\zcut^2\ecf{2}{2}Q^2} \right )-\frac{\text{Cl}_2(\frac{\pi}{3})}{\pi}
\right]\,,\\
\gamma_{H_{cs}}^{g\rightarrow gg}=& -\frac{\alpha_s}{\pi}C_A\log\frac{4\mu^2}{\ecf{2}{2}zQ^2}-\frac{\alpha_s}{2\pi}\left(
\frac{11}{3}C_A-\frac{2}{3}n_f\right)-\frac{\alpha_s}{\pi}C_A
\log\frac{4 z\mu^2}{\zcut^2\ecf{2}{2}Q^2}
\,.
\end{align} }
We note that we do not have large logs of $z$ or $1-z_q$ over $\zcut$, since in this region of phase space these scales are parametrically the same.

\section{Ingredients for Signal Factorization}\label{app:signal}

In this appendix, we present the results for the signal factorization formula. The factorization formula is
\begin{align}
\frac{d^2\sigma}{dz d\ecf{2}{\alpha}d\ecf{3}{\alpha}}=H(Q^2)H_2^{Z\rightarrow q \bar q}(z,\ecf{2}{\alpha},m_Z^2) J_1(\ecf{3}{\alpha})\otimes J_2(\ecf{3}{\alpha})\otimes S(\ecf{3}{\alpha},\zcut) \,. 
\end{align}
Laplace transforming in $\ecf{3}{\alpha}$ removes the convolutions, and factoring out the global contributions, we have
{\small\begin{align}
\frac{d^2\sigma}{dz d\ecf{2}{\alpha}d\ecflp{3}{\alpha}}&=H(Q^2)H_2^{Z\rightarrow q \bar q}(z,\ecf{2}{\alpha},m_Z^2) \tilde J_1(\ecflp{3}{\alpha}) \tilde J_2(\ecflp{3}{\alpha}) \tilde S(\ecflp{3}{\alpha},\zcut)\,\\
&\hspace{-1cm}= H(Q^2)H_2^{Z\rightarrow q \bar q}(z,\ecf{2}{\alpha},m_Z^2) \tilde J_1(\ecflp{3}{\alpha}) \tilde J_2(\ecflp{3}{\alpha})C_s(\ecflp{3}{\alpha},\theta<\theta_{ab})S_c(\zcut,\theta>\theta_{ab})C^{\text{NG}}_s(\ecflp{3}{\alpha},\zcut)\,. \nn
\end{align}}
Results in this appendix will be expressed in Laplace space.  Note that the anomalous dimension of the hard function $H(Q^2)$ is zero in QCD because the production of the $Z$ boson occurs in the QCD vacuum.

\subsection{Soft Matrix Elements}
The soft matrix elements are simplified relative to the earlier case of QCD splittings, since we have no Wilson line in the direction of the recoil. We therefore have
\begin{align}
C_s(\ecflp{3}{\alpha},\theta<\theta_{ab})&=\frac{1}{N}\text{tr}\Big[\langle 0|T\{S_{a}S_{b}\}\exp\Big(-\ecflp{3}{\alpha}\ecfop{3}{\alpha}\big|_{\text{SD} a,b}\Big)\bar T\{S_{a}S_{b}\}|0\rangle\Big]\,, \\
S_c(\zcut,\theta>\theta_{ab})&=\frac{1}{N}\text{tr}\Big[\langle 0|T\{S_{a}S_{b}\}\Theta_{\text{SD}}(a,b,\zcut)\bar T\{S_{a}S_{b}\}|0\rangle\Big]\,,\\
C^{\text{NG}}_s(\ecflp{3}{\alpha},\zcut)&=\frac{1}{N}\frac{\text{tr}\Big[\langle 0|T\{S_{a}S_{b}\}\Theta_{\text{SD}}(a,b,\zcut)\exp\Big(-\ecflp{3}{\alpha}\ecfop{3}{\alpha}\big|_{\text{SD} a,b}\Big)\bar T\{S_{a}S_{b}\}|0\rangle\Big]}{C_s(\ecflp{3}{\alpha},\theta<\theta_{ab})S_c(\zcut,\theta>\theta_{ab})}\,.
\end{align}
As before, $S_q$ are soft Wilson lines, as defined in \Eq{eq:soft_wilson_line}.

\subsection{Hard Decay Function}

The anomalous dimension of the hard decay function, $H_2^{Z\rightarrow q \bar q}(\ecf{2}{\alpha},m_Z^2)$, is identical to the anomalous dimension for the hard function of $e^+e^-\to q\bar q$, at $Q^2=m_Z^2$. We therefore have
\begin{equation}
\gamma_{H_2} = \frac{\alpha_sC_F}{\pi}\left(
-2\,\log\frac{\mu^2}{m_Z^2}-3
\right)\,.
\end{equation}
Identifying the mass of the jet with the two-point energy correlation function as
\begin{equation}
m_Z^2 =\ecf{2}{2}E_J^2 =\frac{\ecf{2}{2}Q^2}{4}\,,
\end{equation}
this anomalous dimension can also be expressed as
\begin{equation}
\gamma_{H_2} = \frac{\alpha_sC_F}{\pi}\left(
-2\,\log\frac{4\mu^2}{\ecf{2}{2}Q^2}-3
\right)\,.
\end{equation}
The tree-level matrix element for $Z$ boson decay to quarks is well-known and was first calculated in \Ref{Altarelli:1979ub}.

\subsection{Jet Functions}

The subjets of the decay of a $Z$ boson are quarks.  The jet functions have the corresponding anomalous dimensions
\begin{align}
\gamma_{J_1} = \frac{\alpha_s C_F}{2\pi}\left[
4\,  \log\left( 4\frac{\left(\ecf{2}{2}\right)^2}{z_a^2z_b}\ecflp{3}{2}\frac{\mu^2}{Q^2}  \right)+3
\right]\,,\\
\gamma_{J_2} = \frac{\alpha_s C_F}{2\pi}\left[
4\,  \log\left( 4\frac{\left(\ecf{2}{2}\right)^2}{z_az_b^2}\ecflp{3}{2}\frac{\mu^2}{Q^2}  \right)+3
\right]\,.
\end{align}
Here $z_a$ and $z_b$ are the energy fractions of the two subjets, with $z_a+z_b = 1$.

\subsection{Collinear-Soft Function}

The soft radiation from the dipole of the subjets is constrained by the same restrictions as for background jets.  In this case, however, there is only one non-trivial dipole formed from the subjets with color factor ${\bf T}_a\cdot {\bf T}_b = -C_F$.  Therefore, the anomalous dimension of this function is just
\begin{equation}
\gamma_{C_s} = -2\frac{\alpha_s C_F}{\pi} \left[
\log \left( \frac{4(\ecf{2}{2})^5(\ecflp{3}{2})^2\mu^2}{z_a^3z_b^3Q^2} \right)-\frac{\text{Cl}_2(\frac{\pi}{3})}{\pi}
\right]\,.
\end{equation}

\subsection{Soft-Collinear Function}

As with the collinear-soft function, there is only one contribution to the soft-collinear function, namely from emissions that are emitted at a wide angle. We find that the anomalous dimension of the soft-collinear function is
\begin{equation}
\gamma_{S_c}=-2\frac{\alpha_s C_F}{\pi} \left(\frac{\text{Cl}_2(\frac{\pi}{3})}{\pi} \right)\,.
\end{equation}

\subsection{Consistency}

The anomalous dimensions of the functions in the factorization formula are
\begin{align}
\gamma_{H_2} &= \frac{\alpha_sC_F}{\pi}\left(
-2\,\log\frac{4\mu^2}{\ecf{2}{2}Q^2}-3
\right)\,,\\
\gamma_{J_1}& = \frac{\alpha_s C_F}{2\pi}\left[
4\,  \log\left( 4\frac{\left(\ecf{2}{2}\right)^2}{z_a^2z_b}\ecflp{3}{2}\frac{\mu^2}{Q^2}  \right)+3
\right]\,,\\
\gamma_{J_2} &= \frac{\alpha_s C_F}{2\pi}\left[
4\,  \log\left( 4\frac{\left(\ecf{2}{2}\right)^2}{z_az_b^2}\ecflp{3}{2}\frac{\mu^2}{Q^2}  \right)+3
\right]\,,\\
\gamma_{C_s} &= -2\frac{\alpha_s C_F}{\pi} \left[
\log \left( \frac{4(\ecf{2}{2})^5(\ecflp{3}{2})^2\mu^2}{z_a^3z_b^3Q^2} \right)-\frac{\text{Cl}_2(\frac{\pi}{3})}{\pi}
\right]\,,\\
\gamma_{S_c}&=-2\frac{\alpha_s C_F}{\pi}\left(\frac{\text{Cl}_2(\frac{\pi}{3})}{\pi}\right)\,.
\end{align}
It is straightforward to verify that the sum of anomalous dimensions is $0$, namely
\begin{equation}
0=\gamma_{H_2}+\gamma_{J_1}+\gamma_{J_2}+\gamma_{C_s}+\gamma_{S_c}\,.
\end{equation}
This demonstrates that the factorization formula is consistent at one-loop or NLL accuracy.

\section{$1\rightarrow 3$ Splitting Function Integration}\label{app:1to3}

In this appendix we describe the calculation of the $D_2$ distribution using the $1\to 3$ splitting functions.
The splitting function calculation on the soft dropped jet is given by
{\small\begin{align}
\frac{d\sigma}{de_2 dD_2}&\propto(2\pi)^{(3-2\epsilon)}\int[d^dk_a]_+[d^dk_b]_+[d^dk_c]_+\delta(2E_J-\nbar\cdot k_a-\nbar\cdot k_b-\nbar\cdot k_c)\delta^{(d-2)}\Big(k_{a\perp}+k_{b\perp}+k_{c\perp}\Big)\nonumber\\
&\qquad\times\delta\left(e_2-\frac{1}{E_J^2}((2k_a\cdot k_b)+(2 k_a\cdot k_c) +(2k_b\cdot k_c))\right)\delta\Bigg(D_2-\frac{8(2k_a\cdot k_b)\,(2 k_a\cdot k_c) \,(2k_b\cdot k_c)}{E_J^3(\ecf{2}{2})^3\nbar\cdot k_a\nbar\cdot k_b \nbar\cdot k_c}\Bigg)\nonumber\\
&\qquad\times\text{SD}\Big(\zcut,k_a,k_b,k_c\Big)\Theta_{J}\Big(R;,k_a,k_b,k_c\Big)\text{Split}\Big(k_a,k_b,k_c;\nbar\Big)\,.
\end{align}}
The soft drop and jet region constraints are defined as
{\scriptsize\begin{align}
\Theta_{J}\Big(R;,k_a,k_b,k_c\Big)&=\Theta\left(\text{tan}^2\frac{R}{2}-\frac{n\cdot k_a}{\nbar \cdot k_a}\right)\Theta\left(\text{tan}^2\frac{R}{2}-\frac{n\cdot k_b}{\nbar \cdot k_b}\right)\Theta\left(\text{tan}^2\frac{R}{2}-\frac{n\cdot k_c}{\nbar \cdot k_c}\right) \,,\\
\text{SD}\Big(\zcut,k_a,k_b,k_c\Big)&= \\
&\hspace{-1.9cm}\left[\Theta\left(\frac{1}{2E_J}\text{min}\big(\nbar\cdot(k_b+k_c),\nbar\cdot k_a \big)-\zcut\right)\Theta\left(\frac{k_a\cdot k_b}{\nbar\cdot k_a\nbar\cdot k_b}-\frac{k_b\cdot k_c}{\nbar\cdot k_b\nbar\cdot k_c}\right)\Theta\left(\frac{k_a\cdot k_c}{\nbar\cdot k_a\nbar\cdot k_c}-\frac{k_b\cdot k_c}{\nbar\cdot k_b\nbar\cdot k_c}\right) \right.\nonumber\\
&\hspace{-1.9cm}+\Theta\left(\frac{1}{2E_J}\text{min}\big(\nbar\cdot(k_a+k_b),\nbar\cdot k_c \big)-\zcut\right)\left(\frac{k_a\cdot k_c}{\nbar\cdot k_a\nbar\cdot k_c}-\frac{k_a\cdot k_b}{\nbar\cdot k_a\nbar\cdot k_b}\right)\left(\frac{k_b\cdot k_c}{\nbar\cdot k_b\nbar\cdot k_c}-\frac{k_a\cdot k_b}{\nbar\cdot k_a\nbar\cdot k_b}\right)\nonumber\\
&\hspace{-1.9cm}\left.+\Theta\left(\frac{1}{2E_J}\text{min}\big(\nbar\cdot(k_a+k_c),\nbar\cdot k_b \big)-\zcut\right)\Theta\left(\frac{k_a\cdot k_b}{\nbar\cdot k_a\nbar\cdot k_b}-\frac{k_a\cdot k_c}{\nbar\cdot k_a\nbar\cdot k_c}\right)\Theta\left(\frac{k_b\cdot k_c}{\nbar\cdot k_b\nbar\cdot k_c}-\frac{k_a\cdot k_c}{\nbar\cdot k_a\nbar\cdot k_c}\right)\right]\,. \nn
\end{align}}
We note that the jet region constraint is that of a simple cone algorithm, where we have drawn a cone of radius $R$ about the direction $n$. Further, we have demanded that the total transverse momentum of all emissions inside this cone is zero. Because of the soft drop constraint, we could actually take the effective cone radius to be infinite, that is, the jet be the whole sphere. However, in our numerical integration of the splitting functions, we keep the jet region constraint so that we can explicitly test the $R$ and $\ecf{2}{2}$ independence of the fixed order result. 

\subsection{Solving Observable Constraints}

To integrate the above phase space, we express the kinematic variables as
\begin{align}
\nbar\cdot k_a&=2 E_Jz_a,\qquad  n\cdot k_a=\frac{k_{a\perp}^2}{2 E_Jz_a}\,, \qquad
\nbar\cdot k_b=2 E_Jz_b,\qquad n\cdot k_b=\frac{k_{b\perp}^2}{2 E_Jz_b}\,,\nn\\
\nbar\cdot k_c&=2 E_J(1-z_a-z_b),\qquad n\cdot k_c=\frac{k_{c\perp}^2}{2 E_J(1-z_a-z_b)}\,,\qquad
k_{c\perp}=-k_{a\perp}-k_{b\perp}\,.
\end{align}
We can then shift $k_{a\perp}$ as
\begin{align}
k_{a\perp}\rightarrow k_{a\perp}-\frac{z_b}{1-z_a}k_{b\perp}\,.
\end{align}
This allows us to solve the $\ecf{2}{2}$ delta function constraint as
\begin{align}
k_{a\perp}^2&=\frac{z_a(1-z_a-z_b)}{z_b(1-z_b)}\Big(\ecf{2}{2}E_J^2z_b(1-z_b)-k_{b\perp}^2\Big)\,,\qquad k_{b\perp}^2<\ecf{2}{2}E_J^2z_b(1-z_b)\,.
\end{align}
We again rescale $k_{b\perp}^2$ as
\begin{align}
k_{b\perp}^2&=\ecf{2}{2}E_J^2z_b(1-z_b)k^2\,.
\end{align}
Our left over variables are $z_a,z_b,k^2$ and $\phi$, where $\phi$ is a relative angle in the transverse plane. The Lorentz invariants have the functional form
{\small\begin{align}
k_{a}\cdot k_{b}&=E_J^2\ecf{2}{2}\Bigg(\frac{k^2(z_a-(1-z_a)z_b+z_b^2)+z_b(1-z_a-z_b)-2\sqrt{k^2(1-k^2)z_az_b(1-z_a-z_b)}\text{cos}\phi}{2(1-z_b)}\Bigg)\,, \nn\\
k_{b}\cdot k_{c}&=E_J^2\ecf{2}{2}\Bigg(\frac{k^2(1-z_b-z_a(1+z_b))+z_az_b+2\sqrt{k^2(1-k^2)z_az_b(1-z_a-z_b)}\text{cos}\phi}{2(1-z_b)}\Bigg)\,,\nn\\
k_{a}\cdot k_{c}&=E_J^2\ecf{2}{2}\Bigg(\frac{(1-k^2)(1-z_b)}{2}\Bigg)\,.
\end{align}}
It is convenient to introduce the variables
\begin{align}
s_{ij}&=\frac{k_i\cdot k_j}{E_J^2\ecf{2}{2}}\,,\qquad a\leq i<j\leq c\,.
\end{align}
Further we use the fact that if we rescale the invariants $k_i\cdot k_j$ in the splitting function by a factor $\lambda$, this simply induces an overall factor of $\lambda^{-2}$
\begin{align}
\text{Split}\Big( k_a,k_b,k_c;\nbar\Big)\rightarrow\frac{1}{\lambda^2}\text{Split}\Big( k_a,k_b,k_c;\nbar\Big)\,.
\end{align}
The final phase space then has the form
{\small\begin{align}
\frac{d\sigma}{d\ecf{2}{2} dD_2}&\propto \frac{g^4}{\ecf{2}{2}}(2\pi)^{-6+4\epsilon}\left(\frac{2\pi^{1-\epsilon}}{\Gamma(1-\epsilon)}\right)\left(\frac{\mu^2}{\ecf{2}{2}E_J^2}\right)^{\epsilon}\left(\frac{\mu^2}{\ecf{2}{2}E_J^2}\right)^{\epsilon} \nn \\
&\qquad \cdot \int_0^1dz_a\int_0^1dz_b\int_0^1dk^2\int_0^{\pi}d\phi\,\text{sin}^{-2\epsilon}\phi\,\Theta(1-z_a-z_b)\nonumber\\
&\qquad \cdot\Big(z_az_b(1-z_a-z_b)k^2(1-k^2)\Big)^{-\epsilon}\delta\Bigg(D_2-\frac{8s_{ab}s_{bc}s_{ac}}{z_az_b(1-z_a-z_b)}\Bigg)\nonumber\\
&\qquad\cdot\text{SD}\Big(\zcut,z_a,z_b,k^2,\phi\Big)\Theta_{J}\Big(R;\ecf{2}{2},z_a,z_b,k^2,\phi\Big)\text{Split}\Big(z_a,z_b,k^2,\phi\Big)\,.
\end{align}}
Note that the only dependence on $\ecf{2}{2}$ enters through the jet constraints. Had we expanded the jet region constraints, we would have formally found the cross section to depend on $\ecf{2}{2}$ only via an overall scaling.  That is, the shape of the differential spectrum is independent of $\ecf{2}{2}$. Retaining the jet region constraints we can explicitly test the independence on the jet radius at all values of $D_2$ by varying either $\ecf{2}{2}$ or $R$. At finite $D_2$, the cross section is finite, and so we can set the dimensional regularization parameter to zero, and numerically compute the result.

\subsection{Numerical Integration}

To perform the numerical integration, we first rescale $z_b$ as
\begin{align}
z_b&\rightarrow (1-z_a)z_b\,.
\end{align}
This gives the integral
{\small\begin{align}
\frac{d\sigma}{d\ecf{2}{2} dD_2}&\propto \frac{2g^4}{\ecf{2}{2}}(2\pi)^{-6+4\epsilon}\left(\frac{2\pi^{1-\epsilon}}{\Gamma(1-\epsilon)}\right)\left(\frac{\mu^2}{\ecf{2}{2}E_J^2}\right)^{\epsilon}\left(\frac{\mu^2}{\ecf{2}{2}E_J^2}\right)^{\epsilon}\int_0^1dz_a\int_0^1dz_b\int_0^1dk^2\int_0^{\pi}d\phi\,\text{sin}^{-2\epsilon}\phi\,\nonumber\\
&\qquad\cdot\Big(z_a(1-z_a)^2z_b(1-z_b)k^2(1-k^2)\Big)^{-\epsilon}\delta\Bigg(D_2-\frac{8s_{ab}s_{bc}s_{ac}}{z_a(1-z_a)^2z_b(1-z_b)}\Bigg)\nonumber\\
&\qquad\cdot \text{SD}\Big(\zcut,z_a,(1-z_a)z_b,k^2,\phi\Big)\Theta_{J}\Big(R;\ecf{2}{2},z_a,(1-z_a)z_b,k^2,\phi\Big)(1-z_a)\nn \\
&\qquad\cdot\text{Split}\Big(z_a,(1-z_a)z_b,k^2,\phi\Big)\,.
\end{align}}
We can then transform to the rapidity-like variables
\begin{align}
z_a=\frac{1}{2}\left[1+\text{tanh}\Big(\frac{y_a}{\delta}\Big)\right]\,,\qquad
z_b=\frac{1}{2}\left[1+\text{tanh}\Big(\frac{y_b}{\delta}\Big)\right]\,,\qquad
k^2=\frac{1}{2}\left[1+\text{tanh}\Big(\frac{y_{k^2}}{\delta}\Big)\right]\,.
\end{align}
This transformation has the Jacobian
\begin{align}\label{eq:rapidity_jacobin}
\int_0^1dz_a\,\int_0^1dz_b\,\int_0^1dk^2&=\int_{-\infty}^{\infty}dy_a\int_{-\infty}^{\infty} dy_b \int_{-\infty}^{\infty}dy_{k^2} J(y_a,y_b,y_{k^2})\,,
\end{align}
with
\begin{align}
J(y_a,y_b,y_{k^2})&=\frac{1}{16\delta^3}\text{sech}^2\Big(\frac{y_a}{\delta}\Big)\text{sech}^2\Big(\frac{y_b}{\delta}\Big)\text{sech}^2\Big(\frac{y_{k^2}}{\delta}\Big)\,.
\end{align}
The numerical integral is then performed at $\epsilon=0$. To perform the integral, we randomly sample $y_{a}, y_{b}$, and $y_{k^2}$ uniformly on an interval $[-y_{\text{max}},y_{\text{max}}]$, while $\phi$ is uniformly sampled on $[0,2\pi]$. The variable $\delta$ simply controls how often one samples the region where $D_2\sim 1$. The value of $y_{\text{max}}$ sets the minimal $D_2$ that we can integrate down to. For each generated phase space point we compute the value of corresponding value of $D_2$ and the weight
\begin{align}
w(y_a,y_b,y_{k^2},\phi)&=\text{SD}\Big(\zcut,z_a,(1-z_a)z_b,k^2,\phi\Big)\Theta_{J}\Big(R;\ecf{2}{2},z_a,(1-z_a)z_b,k^2,\phi\Big)\nonumber\\
&\qquad\cdot(1-z_a)\text{Split}\Big(z_a,(1-z_a)z_b,k^2,\phi\Big)\,.
\end{align}
Then a histogram $H_{D_2}$, indexed by $D_2$, is updated according to
\begin{align}
H_{D_2}&\rightarrow H_{D_2}+J(y_a,y_b,y_{k^2})\frac{w(y_a,y_b,y_{k^2},\phi)}{\Delta D_2}\,,
\end{align}
where $\Delta D_2$ is the width of bin at position $D_2$.
We then divide all bins in the histogram by the total number of phase space points sampled. In the limit of infinitely narrow bins, infinite statistics, and $y_{\text{max}}\rightarrow\infty$, the final histogram $H_{D_2}$ is proportional to the differential spectrum
\begin{align}
\frac{d\sigma}{d\ecf{2}{2} dD_2}\propto H_{D_2}\,.
\end{align}
Aside from the transformation to the rapidity like variables $y_{a},y_{b}$ and $y_{k^2}$, which serve to smooth out the soft and collinear singularities, no further importance sampling is performed. To actually fix the normalization of the histogram, the simplest procedure is to compare to the analytic predictions in the singular region. For a fixed bin size, we can fit for the singular behavior of the histogram at a specific soft drop parameter $z_{\text{cut}}$. Then we take the ratio to the analytic result in the limit $D_2\rightarrow 0$. This selects for the ratio of the double-logarithmic terms, and gives the relative normalization of the histogram to the singular result. This normalization is the same for all other values of $z_{\text{cut}}$ when using the same bin spacing, so that for all other $z_{\text{cut}}$ we can then \emph{test} that the fixed order result is reproducing the analytic double and single logarithmic structure at small $D_2$.

\subsection{Singular Cross Section}
To find the singular cross section for gluon splitting, $g\to gg$, we sum the collinear-soft and jet function contributions. This gives
{\small\begin{align}
\frac{d\sigma}{d\ecf{2}{2} dD_2}&=\frac{\alpha_s}{\pi}\int_{0}^{1}dz_a\Theta\Big(\text{min}\big(z_a,1-z_a\big)-\zcut\Big)\frac{P_{g\rightarrow gg}(z_a)}{\ecf{2}{2}}  \\
&\hspace{4cm}\cdot\left(\frac{\alpha_sC_A}{\pi\,D_2}\right)\left(-\log\,D_2-\frac{\beta_0}{C_A}-\frac{3}{2}\log\,(z_a(1-z_a))\right)\,.\nn
\end{align}}
Performing a similar calculation for $q\to qg$ gives
{\small\begin{align}
\left.\frac{d\sigma}{d\ecf{2}{2} dD_2}\right |_{C_A}&=\frac{\alpha_s}{\pi}\int_{0}^{1}dz_a\Theta\Big(\text{min}\big(z_a,1-z_a\big)-\zcut\Big) \frac{P_{q\rightarrow qg}(z_a)}{\ecf{2}{2}}   \\
&\qquad \cdot \left(\frac{\alpha_sC_A}{\pi\,D_2}\right)\Bigg(-\frac{1}{2}\log\,D_2-\frac{\beta_0}{2C_A}-\log\,(1-z_a)-\frac{1}{2}\log\,z_a+\frac{\text{Cl}_2(\frac{\pi}{3})}{\pi}\Bigg)\,,\nn\\
\left. \frac{d\sigma}{d\ecf{2}{2} dD_2}\right|_{C_F}&=\frac{\alpha_s}{\pi}\int_{0}^{1}dz_a\Theta\Big(\text{min}\big(z_a,1-z_a\big)-\zcut\Big) \frac{P_{q\rightarrow qg}(z_a)}{\ecf{2}{2}}\\
&\qquad \cdot \left(\frac{\alpha_sC_F}{\pi\,D_2}\right)\Bigg(-\frac{1}{2}\log\,D_2-\frac{3}{8}-\frac{1}{2}\log\,(1-z_a)-\log\,z_a-\frac{\text{Cl}_2(\frac{\pi}{3})}{\pi}\Bigg)\,. \nn
\end{align}}

\section{Collinear Non-Global Logarithms}\label{sec:NGL_alg}

In this appendix we give the appropriate modification of the Dasgupta-Salam Monte Carlo algorithm \cite{Dasgupta:2001sh} for computing NGLs in the large-$N_c$ limit, changed to account for the phase space constraints of the soft drop procedure. The origin of these NGLs was discussed in \Sec{sec:resolved}, and the lowest order diagram that contributes was shown in \Fig{fig:collinear_NGLs}. Throughout this section, we define the eikonal factor as
\begin{align}
W_{x y}(j)&=\frac{x\cdot y}{(x\cdot j)\,(j\cdot y)}\,.
\end{align}
Here the round bracket is defined as $(i\cdot j)\equiv 1-\cos \theta_{ij}$.

We start the algorithm with a list of initial dipoles, $\mathcal{D}_{\text{init}}$, that depends on the flavor structure of the $1\rightarrow 2$ splitting, and we generate a list of active radiating dipoles $\mathcal{D}$ and a list of emissions $E$ that fail soft drop. That is, they are not clustered into the legs $a$ and $b$ of the hard $1\rightarrow 2$ splitting. We introduce the phase space constraint given this list of emissions
{\small\begin{align}\label{eq:clustering}
C_{ab}\Big(j,E\Big)&=\Theta\Big(\theta_{ab}-\theta_{aj}\Big)\Theta\Big(\theta_{bj}-\theta_{aj}\Big)\prod_{i\in E}\Theta\Big(\theta_{ij}-\theta_{aj}\Big)+\Theta\Big(\theta_{ab}-\theta_{bj}\Big)\Theta\Big(\theta_{aj}-\theta_{bj}\Big)\prod_{i\in E}\Theta\Big(\theta_{ij}-\theta_{bj}\Big)\,.
\end{align}} 
Here $\theta_{xy}$ is the angle between $x$ and $y$. In other words, we test emission $j$ to see if it is closer in angle to either direction $a$ or $b$ than any other emission in the list $E$. If it is, it will pass soft drop by virtue of being clustered into the hard splitting. Finally, we introduce the one emission phase space that sets the resummation of the Sudakov or global logarithms
\begin{align}\label{eq:clustering_one_loop}     
C_{ab}(j)&=\Theta\Big(\theta_{ab}-\theta_{aj}\Big)\Theta\Big(\theta_{bj}-\theta_{aj}\Big)+\Theta\Big(\theta_{ab}-\theta_{bj}\Big)\Theta\Big(\theta_{aj}-\theta_{bj}\Big)\,.
\end{align}
Though written explicitly in terms of angles, this is the same phase space as \Eq{eq:two_prong_soft_drop_clustered}, since the Lorentz products can be simplified to an angular constraint once we factor out and cancel the energy scales. 

The idea of the algorithm is that in the leading logarithmic approximation for the NGLs, every emission not clustered into the $a$-$b$ dipole is formally at an energy scale below $Q\zcut$, and the emissions are ordered in energy. We therefore generate the emissions at each step according to a distribution
\begin{align}
\tilde{F}_{\mathcal{D}}(j)&=\sum_{i\in \mathcal{D}}W_{x_iy_i}(j)-\sum_{i\in \mathcal{D}_{\text{init}}}C_{ab}(j)W_{a_i\,b_i}(j)\,.
\end{align}
Here $\mathcal{D}$ is the set of decaying dipoles, and $\mathcal{D}_{\text{init}}$ the set of initial dipoles. After generating the emission, we update the list of real emissions $E$ and the dipole list $\mathcal{D}$, so long as the emission qualifies as a real emission, rather than a virtual process. The virtual subtraction is implemented via a veto algorithm following the original Dasgupta-Salam algorithm. We allow the initial dipoles to decay until we have an emission that is clustered into the $a$-$b$ dipole, before it is clustered into any other emission, that is, $C_{ab}(j,E)=1$. This replaces the hemisphere condition to start a new event. Once this condition is met, we end the event and book the histogram. Thus, for each event we must track all the real emissions that have been created so far, and check for each new emission whether it clusters into the $a$-$b$ dipole rather than any other emission generated so far.\footnote{The clustering constraint produced by the soft drop procedure introduces a non-Markovian evolution. Whether or not we terminate the dipole evolution depends on the complete emission history up to that point. However, the generation of emissions still proceeds in a Markovian fashion, as the decay of each dipole is independent and universal.} The virtual subtraction condition, that is whether or not we treat the emission as a virtual correction and so reweight the event, is triggered when $C_{ab}(j)>0$ and $j$ was emitted from a dipole containing an initial leg. We take the high scale for the NGL evolution to be $\mu_{S_c}$, and evolve down to the scale $\mu_{C_s}$, see \App{app:scales_resummation}.

\bibliography{sd_D2_bib}
\end{document}